\documentclass[sn-mathphys,Numbered]{sn-jnl-LB}

\usepackage{graphicx}%
\usepackage{multirow}%
\usepackage{amsmath,amssymb,amsfonts}%
\usepackage{amsthm}%
\usepackage{mathrsfs}%
\usepackage[title]{appendix}%
\usepackage{xcolor}%
\usepackage{textcomp}%
\usepackage{manyfoot}%
\usepackage{booktabs}%
\usepackage{algorithm}%
\usepackage{algorithmicx}%
\usepackage{algpseudocode}%
\usepackage{listings}%
\usepackage{hyperref} 
\hypersetup{ colorlinks=true,  linkcolor=blue,  filecolor=magenta,       urlcolor=cyan, citecolor=black}  
\usepackage{graphicx}        
\usepackage{multicol} 
\usepackage{geometry}
\usepackage{amsmath}           
\usepackage[normalem]{ulem} 
\usepackage {ulem} 
\usepackage{inputenc}
  \usepackage{verbatim} 
\usepackage{cleveref}
\usepackage{xcolor}
\usepackage{framed}
\graphicspath{ {./figures/} }
\def\W{Wider\o e}

\def\BT{Bruno Touschek}
\def\RW{Rolf Wider\o e}
\def\W{Wider\o e}
\def\PM{Pierre Marin}
\def\LAL{Laboratoire de l'Acc\'el\'erateur Lin\'eaire}

\def\FL{Fran\c cois Lacoste}
\def\JH{Jacques Ha\"issinski}

\def\WW2{Second World War II}

\def\RLM{Reichsluftfahrtministerium}

\def\Gott{G\"ottingen}

\def\LNF{Laboratori Nazionali di Frascati}

\def\RLM{Reichsluftfahrtministerium}

\def\Gott{G\"ottingen}

\def\tm{\textcolor{magenta}}

\def\tb{\textcolor{blue}}
\def\bef{\begin{figure}}
\def\enf{\end{figure}}
\def\befoot{\begin{footnotesize}}
\def\enfoot{\end{footnotesize}}

\def\RSUPD{Rome Sapienza University, Physics Department Archives}

\def\tm{\textcolor{magenta}}

\def\tb{\textcolor{blue}}

\def\ep{e^+e^-}

\usepackage{url}

\begin{document}

\title[Two theoretical physicists for AdA]{
Raoul Gatto and Bruno Touschek's joint legacy in the rise of electron-positron physics \footnote{INFN – 24-05-LNF, 
21/5/2024}}

\author[1]{\fnm{Luisa} \sur{Bonolis}}

\author[2]{\fnm{Franco} \sur{Buccella}}

\author[3,4]{\fnm{Giulia} \sur{Pancheri}}

\affil[1]{\orgdiv{} \orgname{Max Planck Institut f\"ur Wissenschaftsgeschichte}, \orgaddress{\street{Boltzmannstrasse 22}, \city{Berlin}, \postcode{14195},  \country{Germany --}} lbonolis@mpiwg-berlin.mpg.de}

\affil[2]{\orgdiv{} \orgname{INFN Sezione di Roma1, c/o Dipartimento di Fisica, Sapienza Universit\`a di Roma}, \orgaddress{\street{Piazzale Aldo Moro 5}, \city{Rome}, \postcode{00184},  \country{Italy -- }} franco.buccella@roma1.infn.it}

\affil[3]{\orgdiv{} \orgname{INFN Frascati Laboratories}, \orgaddress{\street{Via Enrico Fermi 56}, \city{Frascati}, \postcode{00044},  \country{Italy -- }}giulia.pancheri@lnf.infn.it}

\affil[4]{\orgdiv{} \orgname{CREF, Enrico Fermi Center for Study and Research}, \orgaddress{\street{Via  Panisperna, 89a}, \city{Rome}, \postcode{00184},  \country{Italy}}}








\abstract{
Raoul Gatto and Bruno Touschek’s collaboration in the establishment of electron-positron colliders as a fundamental discovery tool in particle
physics will be illustrated. In particular, we will tell the little-known story of how Gatto and Touschek's pioneering visions combined to provide
the theoretical foundation for AdA, the first matter-antimatter collider, and how their friendship with Wolfgang Pauli and Gerhard L\"uders was crucial
to their understanding of the CPT theorem, the basis for AdA's success. We will see how these two exceptional scientists shaped physics between Rome
and Frascati, from the proposal to build AdA and soon after the larger machine ADONE in 1961, to the discovery of the $J/\Psi$ particle in 1974. We
will also highlight Gatto and Touschek's contribution in mentoring an extraordinary cohort of students and collaborators whose work contributed to
the renaissance of Italian theoretical physics after the Second World War and to the establishment of the Standard Model of particle physics.}



\maketitle

\newpage
\begin{center}
\bf{Table of Contents }
\end{center}
\begin{enumerate}
\item Introduction Sec.~\ref{sec:intro}, Figs.~\ref{fig:SRpag2}, \ref{fig:Gatto-Geneva}

\item About Gatto and Touschek: an overview Sec.~\ref{sec:BTMLgatto}, Fig~\ref{fig:GattoBTML}
		\begin{itemize}
		\item From Catania to Rome Subsec.~\ref{ssec:catania}, Fig.~\ref{fig:youngGatto}
		\item Touschek came from Vienna Subsec.~\ref{ssec:Vienna}, Figs.~\ref{fig:BTyoung}, \ref{fig:wardrawings}, \ref{fig:mentors-new}, \ref{fig:Pauli-Touschek-1953-Cagliari-lowres096}
		\end{itemize}
\item The particle physics scenario before 1959 Sec.~\ref{sec:particlephysics1950}
	\begin{itemize}
	\item Antiprotons and new discoveries Subsec.~\ref{ssec:antiprotons-etc}
	\item  Proposals for Center-of-mass collisions Subsec.~\ref{ssec:1956cmcollisions}, 
	 \ref{fig:kerstOneill1956}, \ref{fig:Budkertouschek-catania} 	
	\end{itemize}
	
 \item AdA Sec.~\ref{sec:AdA}
	 	\begin{itemize}
		 \item Different  roads to AdA Subsec.~\ref{ssec:AdAroad}, Fig.~\ref{fig:Venice1957}
		\item The birth of AdA between October 1959 and March 1960 Subsec.~\ref{ssec:AdAbirth}, Figs.~\ref{fig:cabibbocalogero}, \ref{fig:panofskySeminar}, \ref{fig:BohrConversiRubbia}, \ref{fig:AdAscheme}
\item Making AdA work Subsec.~\ref{ssec:AdAmaking}, Fig.~\ref{fig:Ada-synchro}
  		\begin{itemize}
		\item AdA and \LAL, Fig.~\ref{fig:JH-these}
     		\item The thesis calculation which opened the way to electron-positron colliders, Figs.~\ref{fig:maianicaponpreparata}, \ref{fig:buccella}
		       		 \end{itemize}
		\end{itemize}
	\item  The  development of the Frascati theory group Sec.~\ref{sec:lnftheory}, Figs.~\ref{fig:persicosalviniamaldiferretti}, \ref{fig:synchrostaffAdA}
\item Physics at ADONE,  Sec.~\ref{sec:ADONE},  Figs.~\ref{fig:adoneandproposal}, \ref{fig:pacini-pancheri-ferrari}
			\begin{itemize}
			\item Beyond perturbation theory Subsec. ~\ref{ssec:resummation}, Fig.~\ref{fig:yogiMarioLia-EtimElspethFrancis}
			\item The unexpected multiple hadron production Subsec.~\ref{ssec:multihadron}, Figs.~\ref{fig:varenna1969}, \ref{fig:1968}, \ref{fig:erice1967}
			\item The discovery of charm Subsec.~\ref{ssec:charm}
  			\end{itemize}
\item Epilogue Sec.~\ref{sec:epilogue}, Figs.~\ref{fig:Lincei}, \ref{fig:BTvideo-AdA}
\item Conclusion Sec.~ \ref{sec:conclusion}
\item What Gatto said of Touschek in 1987,  App.~\ref{app-BTML} 
\item Raoul Gatto to Luisa Bonolis, 2004, App.~\ref{app-LB}
\item Raoul Gatto to Giulia Pancheri, around 2010, App.~ \ref{app-GP}
\item References 
\end{enumerate} 

\section{Introduction} 
\label{sec:intro}
On July 4th, 2012, in the CERN auditorium packed with scientists from all over the world, 
Fabiola Gianotti, for the ATLAS collaboration \cite{Aad:2012tfa},  and Guido Tonelli, for CMS \cite{Chatrchyan:2012xdj},   announced the discovery of the - so called - missing link of the Standard Model of Elementary Particles, the Higgs boson. The discovery had been made in the collision of protons against protons accelerated in  the Large Hadron Collider, a particle accelerator whose ancestry can be traced back to more than 60 years, when the idea of using the kinematic advantage of center of mass collisions was joined to the vision of creating new states of matter in the annihilation of its constituents. This had been  Bruno Touschek's vision, when he proposed 
 to make an experiment of electron-positron collisions and see what could be extracted from the quantum vacuum. 
The  story of how the first 
 matter-antimatter  collider was born, in Italy in 1960, is still partially unknown. It  is  the aim of the present article  
 to describe how it came to be,  in particular at the University of Rome and the nearby INFN Frascati National Laboratories, and the crucial role played by the collaboration between Raoul Gatto and Bruno Touschek, { two theoretical physicists with a deep knowledge of quantum electrodynamics and the phenomenology of elementary particles.}

{In  the 1950's, Italy  was recovering from twenty years of political isolation and the tragic disruptions that  the Second World War had brought to both  society and scientific progress. }Among the many 
protagonists of the  reconstruction of Italian science during this period, there were in Rome two scientists who left important legacies, Raoul Gatto  and Bruno Touschek \tb{\cite{Amaldi:1981,Greco:2004}}. 
Recently new articles and books  have appeared about them
\cite{Bonolis:2011,Bernardini:2015wja,Maiani:2017hol,Casalbuoni:2018smj,Battimelli:2019,Preparata:2020,Casalbuoni:2021,Casalbuoni:2022rzn,Bonolis:2021khz,Pancheri:2022,Bonolis:2023ksv,Pancheri:2023fia}, 
 but  publications about Gatto 
 and the work with Nicola Cabibbo about electron-positron physics \cite{Cabibbo:1960zza,Cabibbo:1961sz},  
do not fully address   his crucial collaboration with Touschek in 
 bringing to success the construction of
  AdA, the first ever electron-positron collider,   and his contribution to the ADONE 
  proposal, {a much higher energy machine to explore electron-positron physics in depth.}    The purpose of this article is  to fill that gap.

In this article we will see the extent of 
 Gatto's  scientific contribution  to the enterprise for which Touschek is famous, on the basis of the available documents and the reconstruction of the creation of AdA in \cite{Pancheri:2022}.  We will also  include Gatto's personal  remembrances of  Bruno Touschek, from  \cite{Gatto:2004zz} and some private  communications from Gatto to the  present authors.
 
\begin{figure}
\centering
 \includegraphics[scale=0.2]{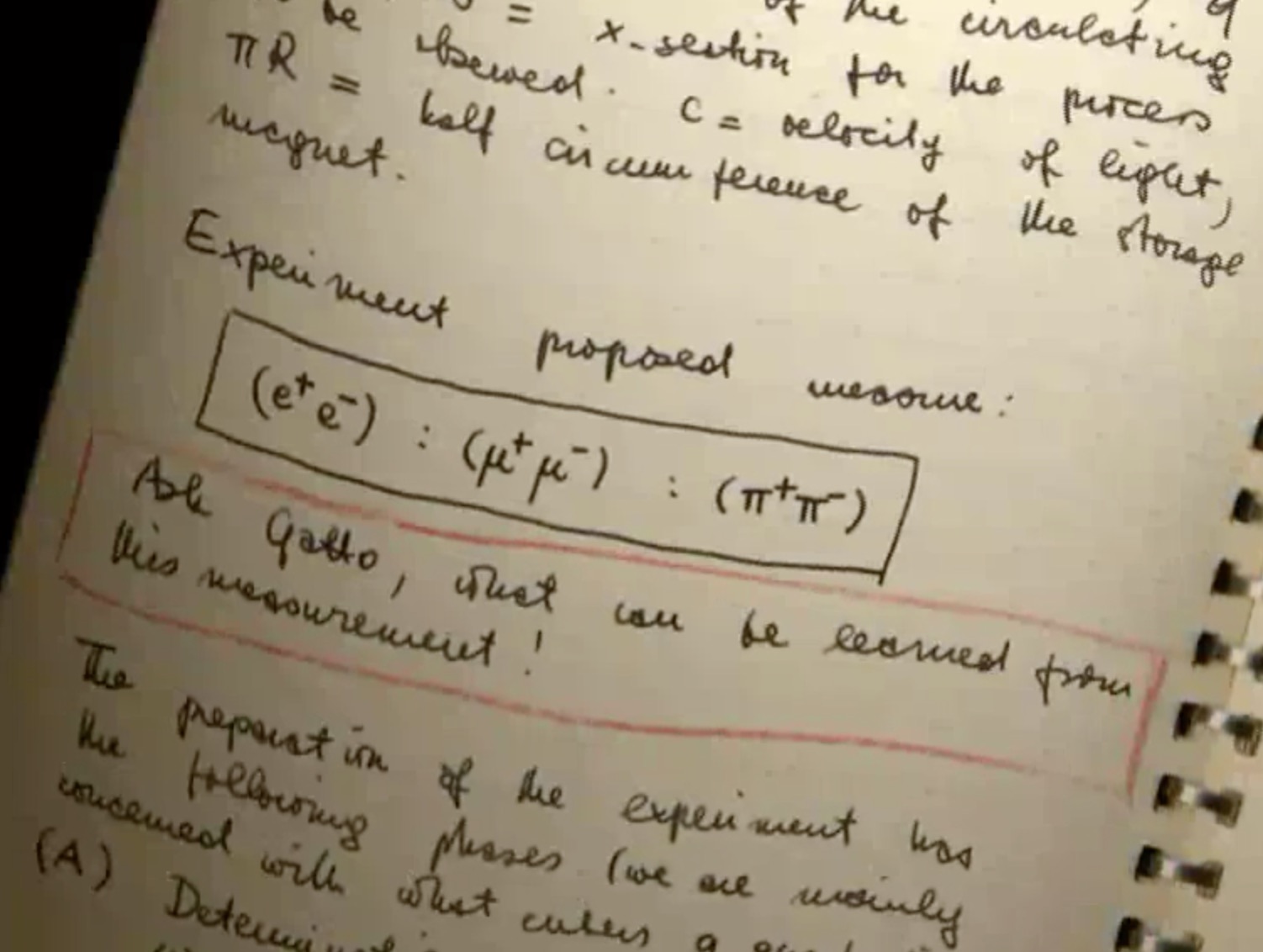} 
\caption{A page from AdA's {\it Storage Ring} Notebook, started by \BT\ on February 18th, 1960, the day after the discussion at the Frascati Scientific Council meeting, where he had proposed an ``experiment worth doing", namely electron-positron collisions and annihilation into new particles. \copyright \ Touschek Family, and  Touschek Papers, \RSUPD, all rights reserved. }
\label{fig:SRpag2}
\end{figure}

  Interestingly, while we have Gatto's memories of Bruno
   \cite{Gatto:2004zz},
   very little was written by Touschek about his friend.  There is however one sentence, which Touschek jotted down  in the first hand-written draft of AdA'a proposal. On February 18th, 1960,  in the famous {\it Storage Ring} Notebook, SR for short,\footnote{The Notebook is 
  {kept among  Bruno Touschek papers in \RSUPD.}}
  Touschek writes : ``Ask Gatto...", as shown in Fig.~\ref{fig:SRpag2}.
These two words testify  to the closeness between these two scientists, as they opened  the road to electron-positron colliders. They discussed together the processes to be studied and needed to prove the feasibility of collider rings as tools for probing the world of elementary particles. Together they assigned these calculations to the students who were asking for a {\it Tesi di laurea in fisica} at University of Rome,
among them Guido Altarelli, Franco Buccella, Etim G. Etim, Giovanni Gallavotti, G. Putzolu,   Paolo di Vecchia, Giancarlo Rossi. 
Together they 
 went to conferences and showed the world what could be gained by the new type of accelerator. 

 After the   proof of feasibility  of  electron-positron colliders in 1964 \cite{Bernardini:1964lqa}, which had followed the  discovery of the Touschek effect in 1963 \cite{Bernardini:1963sc},    Gatto and Touschek's paths
 moved apart, physics-wise and  
geographically.
 \begin{figure}
 \centering
 \includegraphics[scale=0.67]{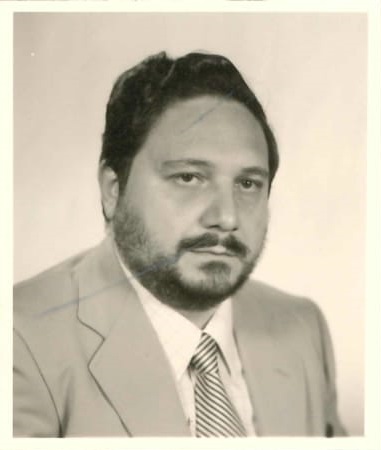}
\includegraphics[scale=0.15]{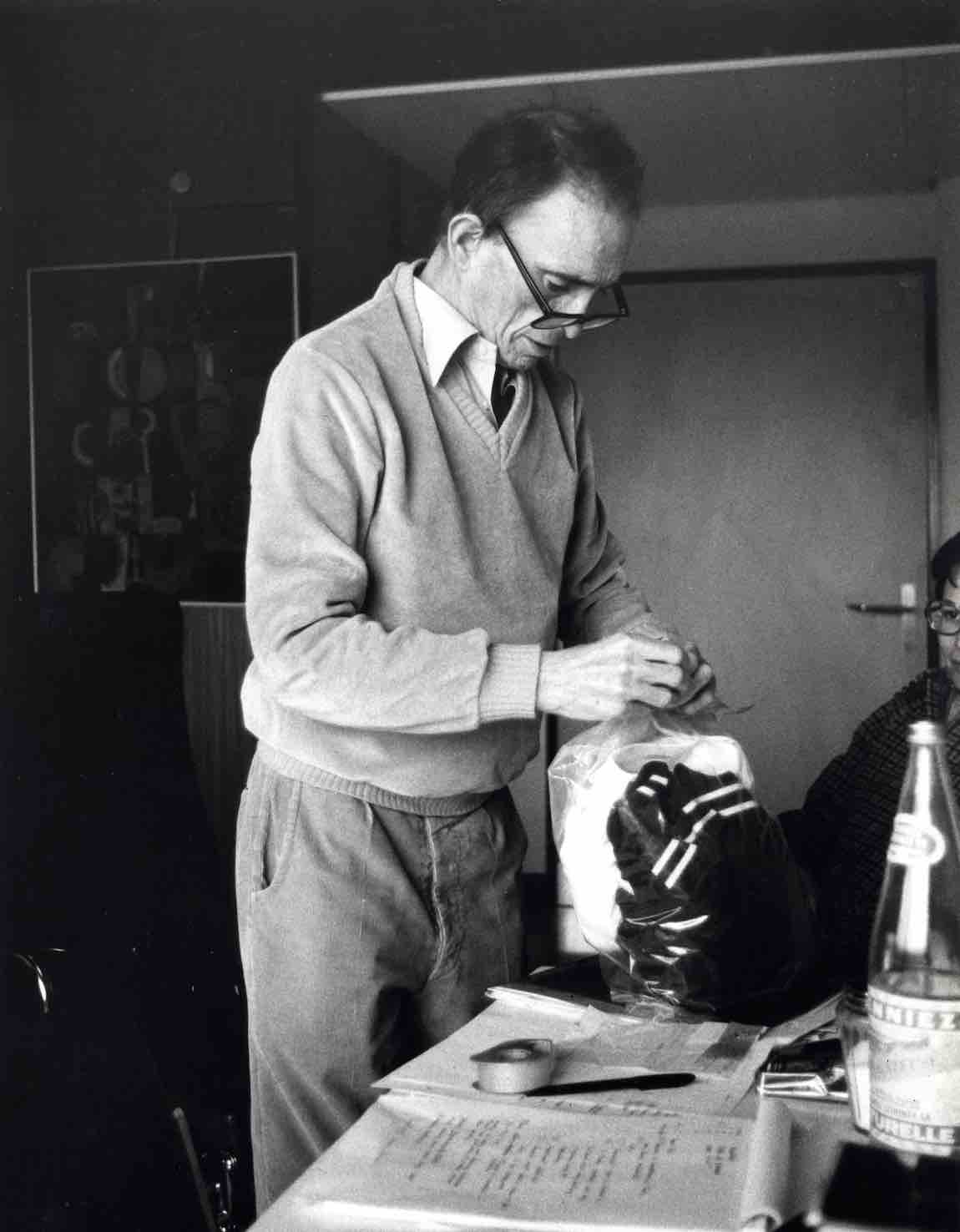}
 \caption{Left:  Raoul Gatto in 1983, from Photot\`eque UNIGE,  \copyright \  University of Geneva; right: Bruno Touschek at CERN in 1978, \copyright Touschek family, all rights reserved;  }
 \label{fig:Gatto-Geneva}
 \end{figure}
Gatto went to Florence to create an extraordinary school in theoretical physics
\cite{Casalbuoni:2018smj},\footnote{Gatto's name, {\it cat} in English, led to his students to be known as the ``gattini", the {\it kittens} in English.}
 then to Padova and  Rome, before finally moving to the  University of Geneva, 
 Fig.~\ref{fig:Gatto-Geneva} 
 left panel, from where he  retired in 1997.  He was for many years 
 editor of {\it Physics Letters B},
which, under his tenure, 
 became 
one of the most influential   journals in particle physics. Touschek remained to work in Rome, except for the time he spent commuting weekly to Orsay,\footnote{See Carlo Bernardini's interview  in the docu-film 
\href{https://www.lnf.infn.it/edu/materiale/video/AdA_in_Orsay.mp4}
{\it Touschek with AdA in Orsay}  by E. Agapito, {L.B.}  and G.P.,  2013.} 
 and 
the time 
at CERN during  the last year of his short life \cite{Amaldi:1981,Rubbia:2023},  Fig.~\ref{fig:Gatto-Geneva} right panel. 
{But 
Gatto and Touschek's  legacy to  physics continued through their students and beyond their separation.}

We will begin  in  Sect.~\ref{sec:BTMLgatto}
with some  highlights   of  Gatto and Touschek's life previous to  the time they 
met  in Rome in 1953. 
An overview of the discoveries in particle physics in the 1950s in Sect. \ref{sec:particlephysics1950} will highlight the scenario  which prepared these two exceptional physicists to join their forces in a collaboration which made the history of
 particle colliders and which is 
described  
 in Sect.~\ref{sec:AdA}.
  Taking a step back in the chronology, 
   we shall  
  present in  Sec.~\ref{sec:lnftheory} a brief history of the Frascati theory group, which was part of the laboratory structure since its very beginning in 1953, and whose later development was fostered by Gatto and Touschek.   The  early years of ADONE, 
 with  the discoveries which followed its commissioning are in 
Sect.~\ref{sec:ADONE}. 

This article is completed by  appendices, 
where  we reproduce three documents about Gatto's participation in the work on electron-positron collisions: the first one, App.~\ref{app-BTML}, 
is courtesy of the \LNF, and  
was  contributed  by Gatto on  the occasion of the  Bruno Touschek Memorial Lectures (BTML) held in Frascati in 1987 \cite{Gatto:2004zz};  the second one is presented    in App.~\ref{app-LB} in English translation from the original  Italian version,  and  is a  private communication to  L.B., via e-mail in 
2003, during the preparation of the docu-film  \href{https://www.youtube.com/watch?v=R2YOjnUGaNY}{\it Bruno Touschek and the art of Physics}; the third one,  App.~\ref{app-GP}, is a letter from Gatto to G.P., around 2010 during the  preparation of Ref.~\cite{Bonolis:2011}.

\section{About  Gatto and  Touschek:  an overview}
\label{sec:BTMLgatto}
\begin{figure}
\centering
\includegraphics[scale=0.081]{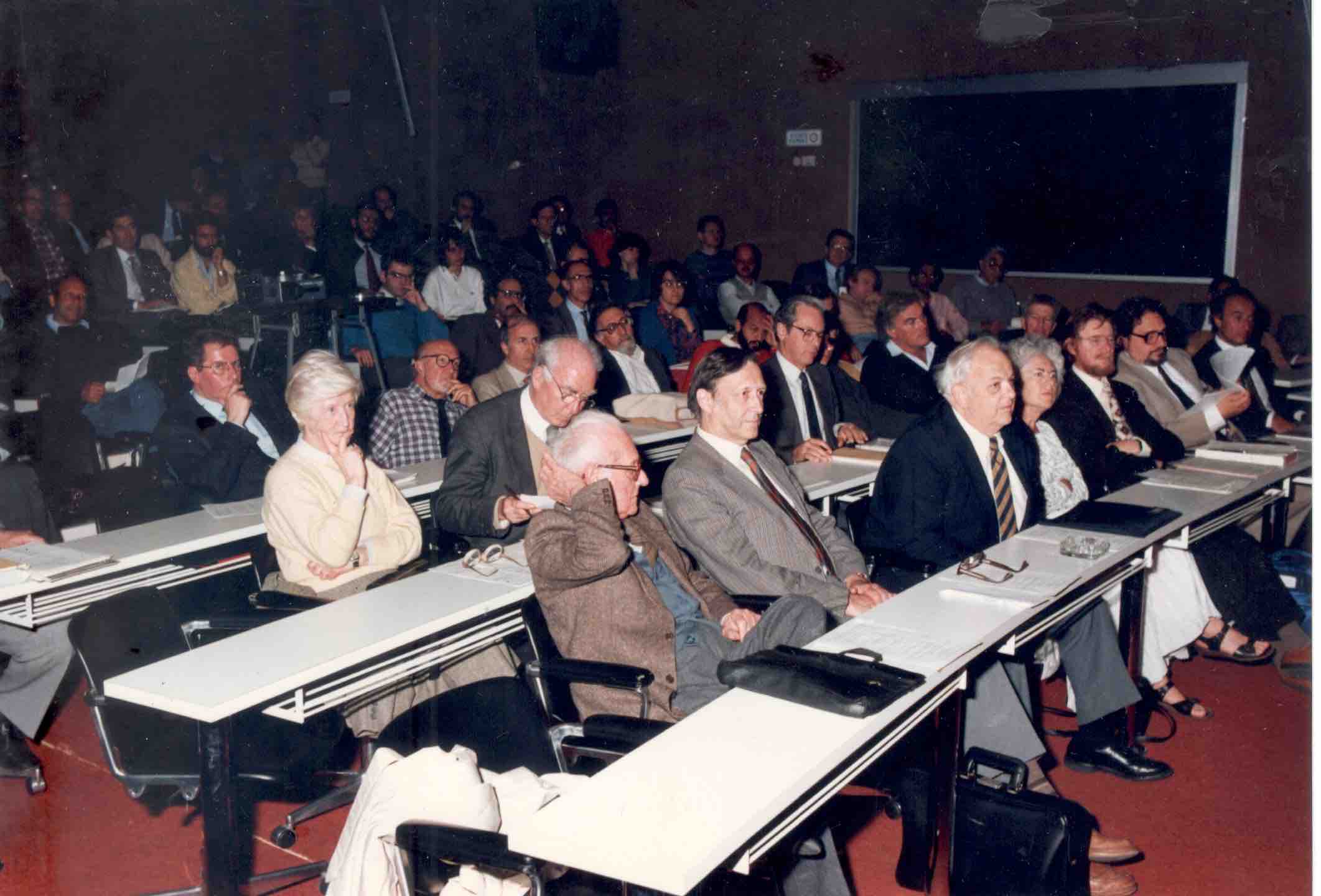}
\includegraphics[scale=0.0443]{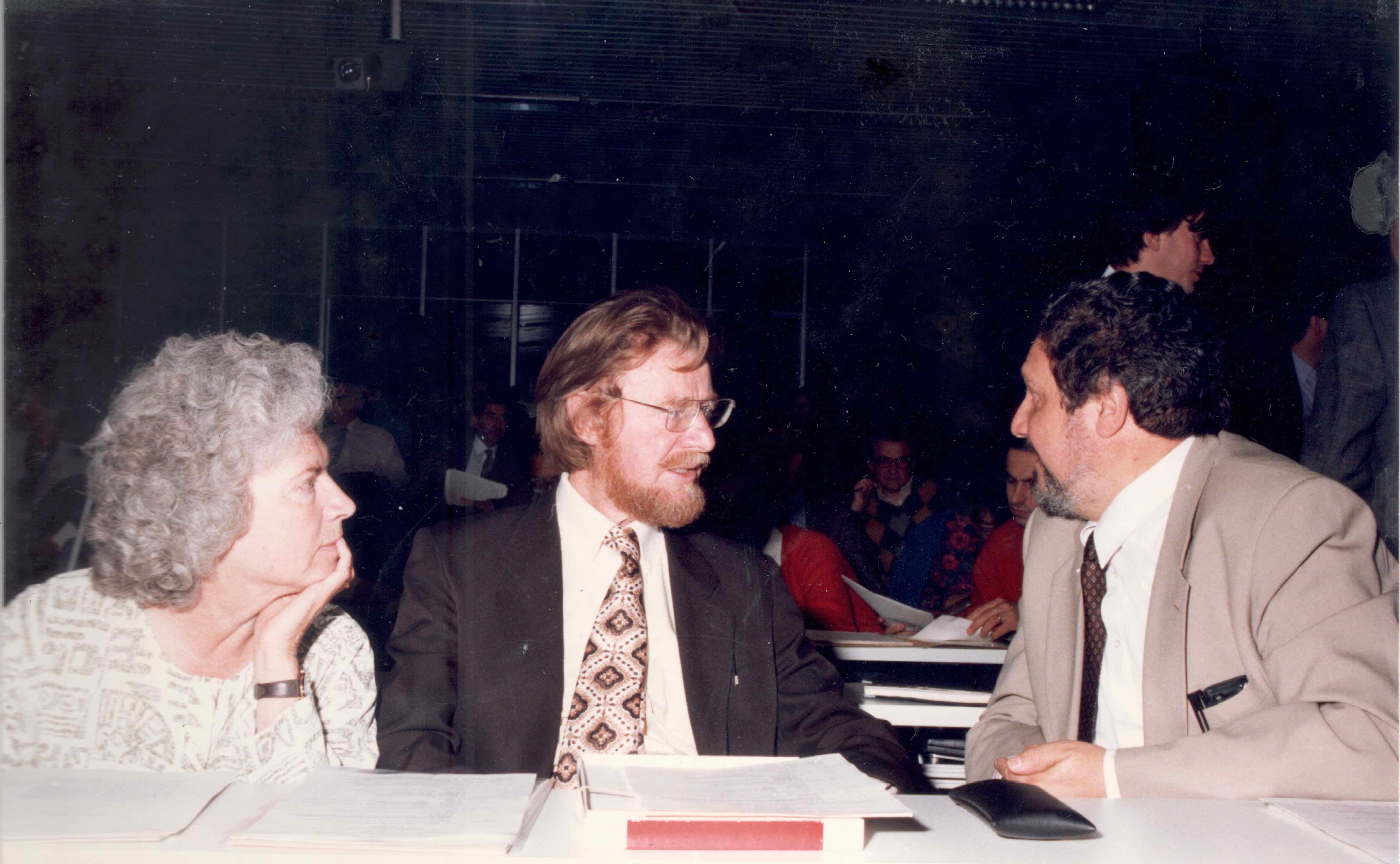}
\caption{
{\LNF \ May 1987,  Bruno Touschek Memorial Lectures: left panel,  seated from  left, first row  Edoardo Amaldi, Simon van der Meer, Burton Richter, Mary Bell, John Bell, Raoul Gatto and Mario Greco,  second row  Ginestra and Giorgio Salvini, Ugo Amaldi;
    right panel,  Raoul Gatto in conversation with Mary and John Bell.  \copyright \  INFN-LNF, all rights reserved.} }
\label{fig:GattoBTML}
\end{figure}

The  Bruno Touschek Memorial Lectures  
were held in Frascati in  May 1987  in memory of \BT,  who 
had passed away 9 years before. Touschek's   friends and colleagues,   were invited to come to Frascati
and contribute their recollections before time erased them, including
 Burton Richter, Carlo Rubbia and Simon van der  Meer.\footnote{Burton Richter shared with Samuel Ting the 1976 Nobel Prize in Physics ``for their pioneering work in the discovery of a heavy elementary particle of a new kind", 
     Carlo Rubbia and Simon van der Meer were awarded the 1984 Nobel Prize in Physics,  ``for their decisive contributions to the large project, which led to the discovery of the field particles W and Z, communicators of weak interaction", 
from 
\url{https://www.nobelprize.org/prizes/physics/1984/summary/.} }  
Due to unforeseen and  unfortunate circumstances,  the written contributions to this event appeared only 17 years later, transcribed from the audio-record, and  published in Ref. \cite{Greco:2004},  after the authors' own review. This gives a unique flavour to the memories presented  in this little volume, which resonates with an immediacy and warmth not always found  in conference proceedings.  Gatto participated in the event which focused on a series of lectures by John Bell on Quantum Mechanics. Fig.~\ref{fig:GattoBTML} shows Gatto with John, Mary Bell and other participants,  in the 
  Conference Hall of the 
  {\it Laboratorio Gas Ionizzati} in the  Frascati Laboratories. 

In Gatto's  words,  
 Bruno comes alive, as a friend and a physicist who played an important role during Gatto's first years in Rome and with whom he shared a strong interest in the physics of the CPT theorem \cite{Luders:1954zz}. His testimony in App.~\ref{app-BTML} constitutes one of the most moving and clear descriptions of their  friendship and the  beginnings of the electron-positron rings story.  Gatto is somewhat  dismissive of his role in the creation and construction of AdA, the first electron-positron storage ring in the world, but we shall argue that  the  close collaboration between Bruno and Raoul  was the key to AdA's success,   much more than what is usually said.

We shall now briefly illustrate  where Gatto and Touschek came from, when they met in Rome in 1953.

\subsection{From Catania   to Rome }
\label{ssec:catania}
 
[Raffaele] Raoul  Gatto was born in Catania on December 8,  1930 and enrolled in physics at the University of Pisa, in 1947, having been awarded a scholarship to the Scuola Normale Superiore (SNS),  a very prestigious institution that accepts students through an extremely selective process.
\footnote{A student accepted by the SNS would have to both graduate from University of Pisa (or Florence) and receive a diploma from the SNS whose courses he had to attend in parallel  with the regular university  studies.}
\begin{figure}
\centering
 \includegraphics[scale=.375]{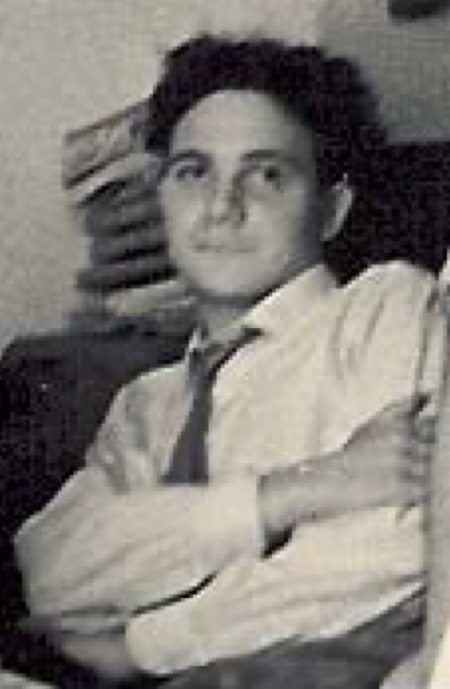}
\includegraphics[scale=0.1]{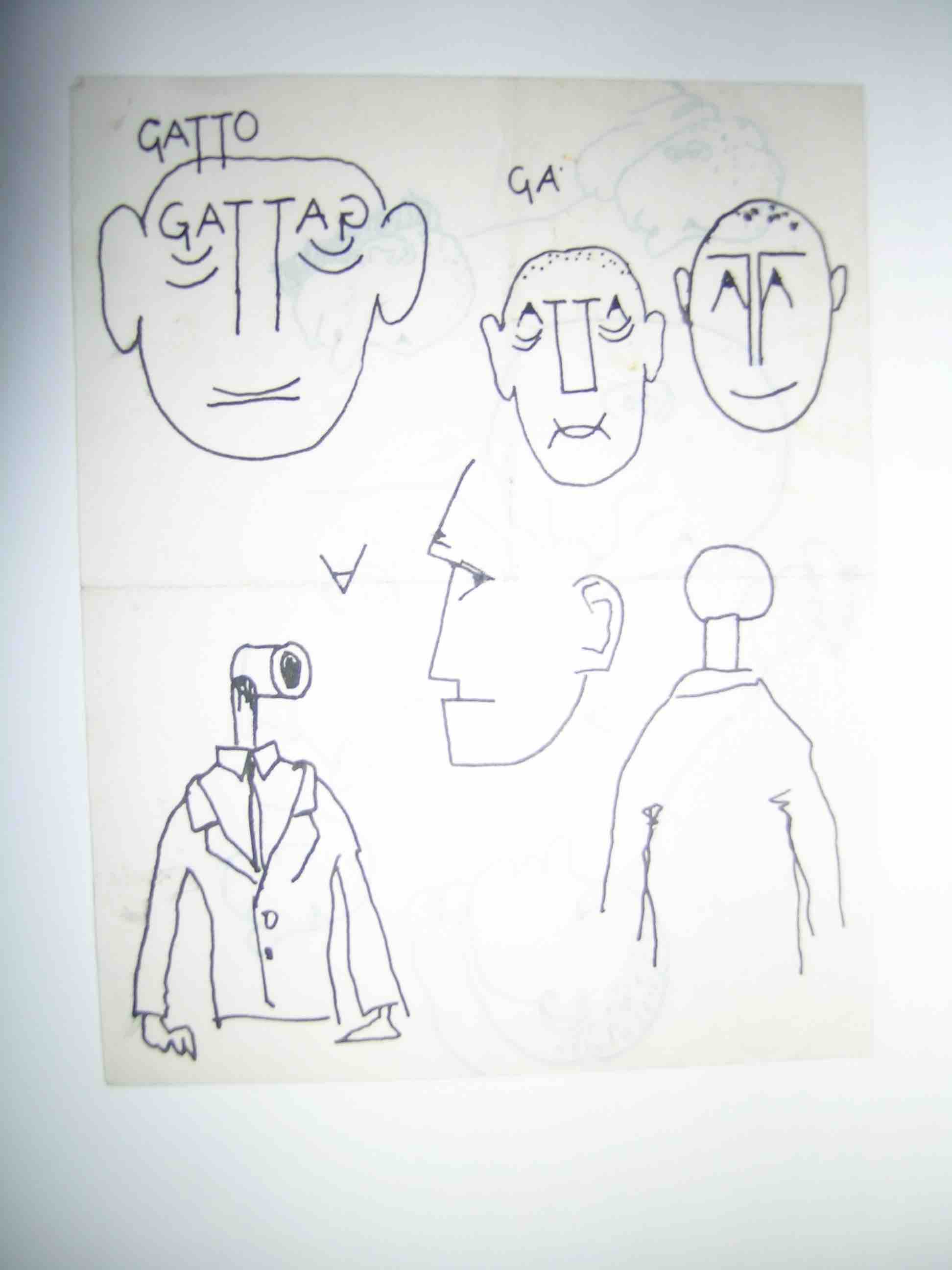}
\caption{A young Raoul Gatto from a photograph  kindly received by L. B., through Carlo Bernardini, courtesy of Gatto's family, all rights reserved; a drawing by Bruno Touschek, where he plays on Gatto's name to make a doodle, \copyright \ Touschek family, all rights reserved. }
\label{fig:youngGatto}
\end{figure}
  In Pisa, among his close friends there was Benedetto De Tollis, with whom he shared a vocation for theoretical physics, 
  and who would later   become Professor 
  in Perugia.\footnote{Recently, a Symposium was held at the University of Perugia to commemorate De Tollis' death in 2018, \url{http://fisgeo.unipg.it/pacetti/nino.}} But in Pisa, at the time, the chair of theoretical physics was vacant, and there was no way to obtain a thesis on the subject. In order to fullfill their dream, a thesis supervisor outside the University of Pisa had to accept them, and the Scuola Normale had to grant special dispensation. The solution may have come from Marcello Conversi, Professor of Experimental Physics in Pisa, who had graduated in Rome with Bruno Ferretti, former assistant to Enrico Fermi, before the latter’s departure for the United States in 1938. Ferretti, a close collaborator with Edoardo Amaldi in the post-war reconstruction of European science, was well known both in Italy and Europe.\footnote{Amaldi  actively discussed with Ferretti ideas for the creation of a large research center for subnuclear physics, the future CERN, where, in 1957, Ferretti would become  the first director of the Theoretical Studies Division.} 
  At the time, he held the Chair of Theoretical Physics in Rome and it was with trepidation that the two friends approached him to ask for his supervision in their thesis work. They were accepted,   the special dispensation was granted,\footnote{In a conversation with Yogendra N. Srivastava, his colleague at University of Perugia, De Tollis used to remember  the epic trip 
when, together with  Gatto, they went to Rome  to ask  Ferretti to be their thesis supervisor: Gatto already confident in his capacities swept easily by the doorman in  charge of letting only deserving students into the Institute, while  De Tollis, affected by a stammering disability, was just as brilliant, but, much unsure of himself,  hesitated for a whole week before picking up enough courage to come  
through the intimidating doors,  personal communication to G.P.}
and Gatto, Fig.~\ref{fig:youngGatto},  graduated with first class honours  {in 1951} with a thesis on the nuclear shell model,\footnote{We cite here from \cite{Casalbuoni:2021}, with   the information directly coming from Raoul Gatto, who had written: ``Tesi di laurea: Modelli a shell dei nuclei".
Private communication by D. Dominici.}
 under Ferretti's guidance, with Conversi as supervisor,
 obtaining  his diploma from the SNS, with Prof. Derenzini with a thesis on statistical
theories of nuclei 
  \cite{Casalbuoni:2021}.

 After graduation, he was offered a position as assistant professor to Ferretti's Chair \cite{Casalbuoni:2018smj,Casalbuoni:2021}, and was in Rome when Touschek arrived at the end of 1952.

\subsection{Touschek came from Vienna}
\label{ssec:Vienna}
\begin{figure}
 \includegraphics[scale=0.194]{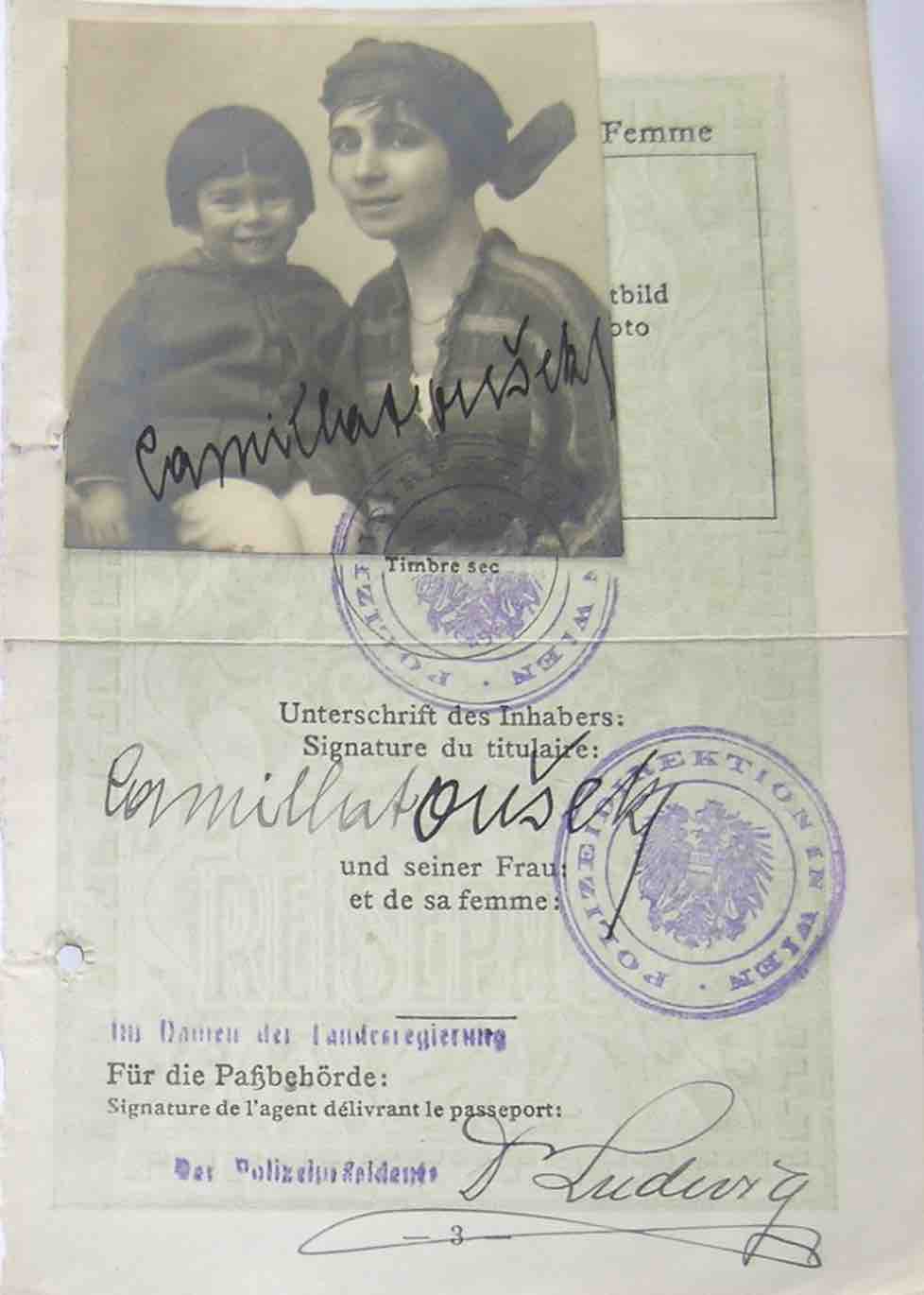}
\includegraphics[scale=0.09]{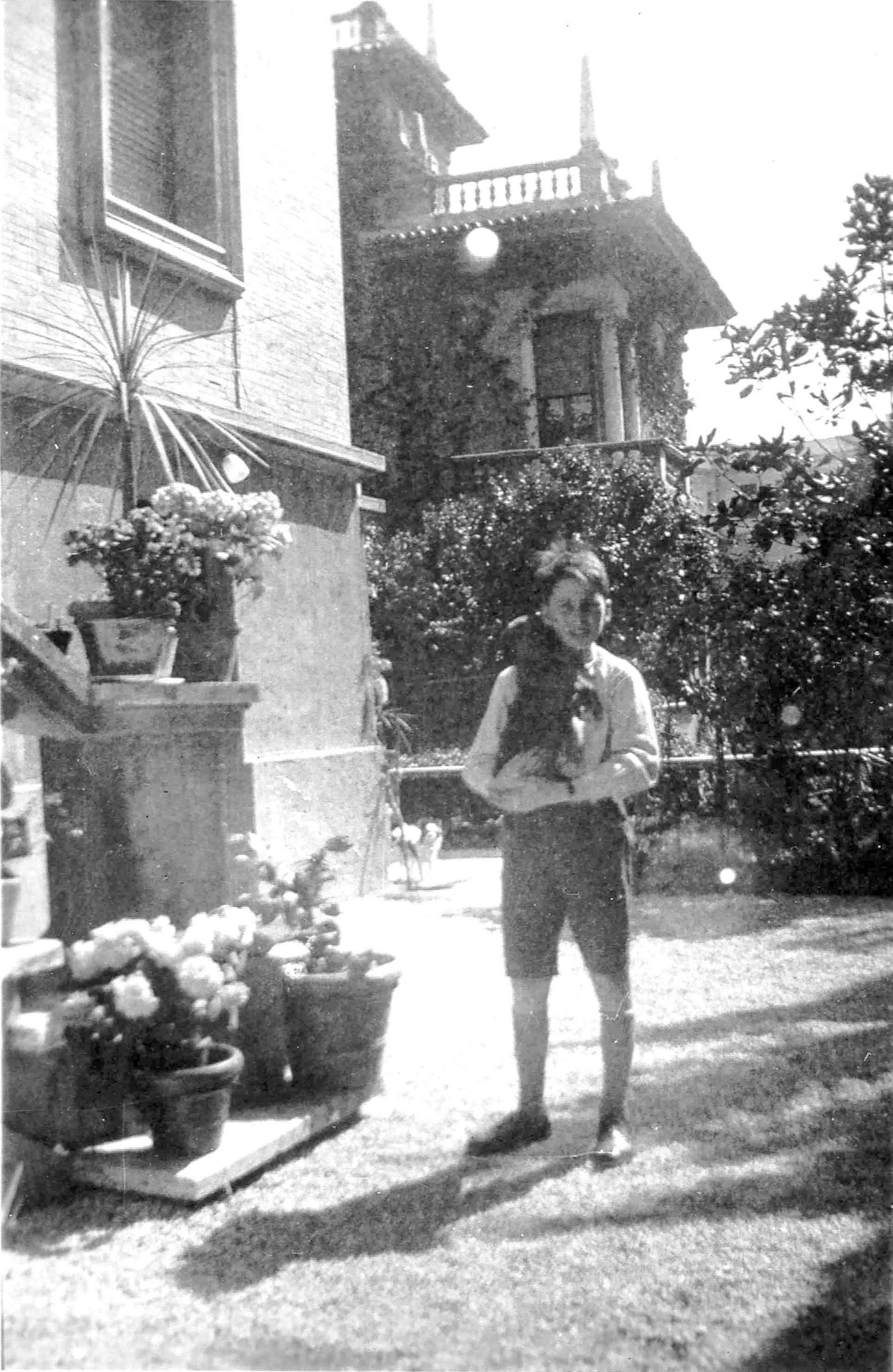}
\caption{Bruno Touschek in a passport photo in his mother's arms, and a later  photograph in his aunt Ada's garden in Rome, \copyright \ Touschek Family, all rights reserved.}
\label{fig:BTyoung}
\end{figure}

Bruno [Francis Xaver] Touschek was born in Vienna on February 3rd 1921. When Touschek arrived in Rome in late December 1952, he was coming to a city he knew well from his pre-war visits to a maternal aunt, Adele Weltmann, married to an Italian industrialist, Gaetano Vannini. Bruno had lost his mother, Camilla n\'ee Weltmann, in 1930,  when only nine years old.  Possible visits to Rome with her can be gathered from a passport photo where he is in his mother's arms, and in another photo which shows him as a young boy, 12 or 13 years old,  in his aunt's garden in Via di Villa Sacchetti, in Rome, Fig.~\ref{fig:BTyoung}.  

In March 1938, the annexation of Austria to Germany, the {\it 
Anschluss}, was declared and everything changed for the Vienna Jews. Suffering  from discrimination in Vienna because of his Jewish origin from the maternal side,  and fearing for worst to come, he thought 
 to leave Austria, and study abroad. From his letters   in March  1939, \cite[Ch. 2]{Pancheri:2022}, and from \cite{Amaldi:1981},  we know that, after having  passed  his high school exams  in  Vienna  in February,  he went to Rome and  visited    aunt Ada (nicknamed from Adele).\footnote{He may have also been in Rome in 1938, since in a March  16th 1939 letters  to his father from Rome,  he
writes ``Rome without F\"uhrer is really a wonderful city \dots", {\it Rom ohne Führer ist wirklich eine wunderbare Stadt},  a possible sarcastic reference  to having been in Rome  during Hitler's visit in 1938.} At this time  he was in Rome expecting  a Visa to emigrate to England, where he had applied to study chemistry in Manchester \cite{Amaldi:1981}.\footnote{A thriving Jewish community lived in Manchester, and Chaim Weizmann, the first President of Israel, had been Senior Lecturer in Chemistry at University of Manchester in 1904, and this may have influenced Bruno in his application to study chemistry there. A concomitant fact is that, after the {\it Anschluss}   in 1938,  many Jewish families  from Vienna looked for their  children to go to Manchester as live-in  with English families, or to be adopted, as recently discussed in \cite{Borger:2024}. See also Williams, Bill. Jews and Other Foreigners: Manchester and the Rescue of the Victims of European Fascism, 1933–40. Manchester University Press, 2011. \url{http://www.jstor.org/stable/j.ctt155j55d.}} 

 It did not happen and he returned in Austria, 
enrolling to study physics at the University of Vienna  in September 1939. After excelling in his first years studies, he was  suspended in June 1940 and then definitely expelled  in January 1942 for not being a ``pure aryan".

In order to continue his studies, Bruno had to leave, and the only place he could go was Germany.  Before leaving, he was given a problem to study by his young mentor Paul Urban, double beta decay,
the process by which two neutrons in an atomic nucleus simultaneously decay through weak interactions into two protons, two electrons, and two antineutrinos $((A,Z) \rightarrow(A,Z+2) + 2e^- + 2\bar{\nu}_e)$, as proposed by Maria Goeppert in 1935 \cite{Goeppert:1935a}.
 \begin{figure}
\centering
\includegraphics[scale=0.4]{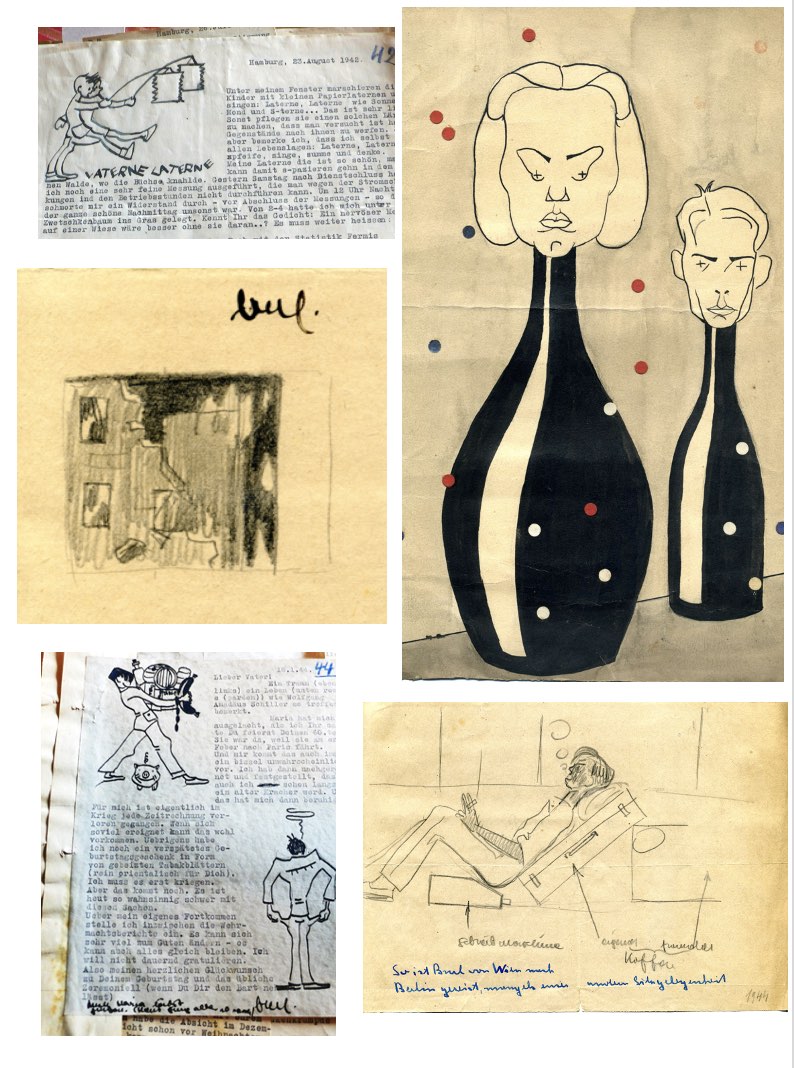}
\caption{A collection of drawings included in Touschek's letters to his father during the war: clockwise from top left,   August 23rd 1942 from Hamburg,  January  2nd 1943 from Berlin, ``Burl travelling from Vienna to Berlin" as from  Bruno's own  caption in German  
placed near a  September   11th 1944 letter,   January  18th 1944,   November  27th 1943 from Berlin; \copyright\ Touschek Family, all rights reserved. }
\label{fig:wardrawings}
\end{figure}
Bruno  left Austria  for Germany at the end of February 1942. Chronicling his movements with drawings and family letters, Fig.~\ref{fig:wardrawings},
he   first  visited  Arnold Sommerfeld, Fig~\ref{fig:mentors-new},  in Munich   and then moved  on to Hamburg and Berlin,  He held various  jobs and  attended university courses, without registering as a student, under  protection   provided by  Sommerfeld's former students and colleagues.  From 1943 until 1945 he worked on building an advanced type of electron accelerator, a betatron, in a secret project directed by the Norwegian scientist \RW\  \cite{Waloschek:1994,Waloschek:2004,Waloschek:2012,Sorheim:2020},  financed by the Aviation Ministry of the Reich, the \RLM, until he was arrested by the Gestapo in mid-March 1945 
\cite[chp.4 and 5]{Pancheri:2022}. 
The events which followed Bruno's arrest on March 1945  include a forced march to the concentration camp in Kiel, and a narrow escape from death, when Bruno was shot by an SS guard and left for dead.  The  Gestapo arrest and the weeks which followed,  between mid March  and April 30th,  are described in \cite{Amaldi:1981}, and have been fully reconstructed in \cite{Bonolis:2011},  from the translation  in English of two letters sent by Touschek to his family in June and October 1945.These letters are part of a regular correspondence which Bruno kept with his father in Vienna, 
occasionally including  drawings  such those  
 displayed in Fig.~\ref{fig:wardrawings}, which reflect both traumas and  joyful times.
 
Liberated  when the war ended, Bruno  was taken in charge by the British  Allied Forces, who   valued the  unique  experience  in  accelerator physics he had acquired with \W , Fig~\ref{fig:mentors-new}. 
The war years left  the young man with a burden of suffering,  as later events in his life would reveal, but at the same time reinforced his determination to survive and became a scientist, in fact a {\it physicist} as he would later write to his father.\footnote{Letter from Touschek to his father, from \Gott, on May 9th, 1946.} 

\begin{figure}[h]
 \centering
  \includegraphics[scale=0.1075]{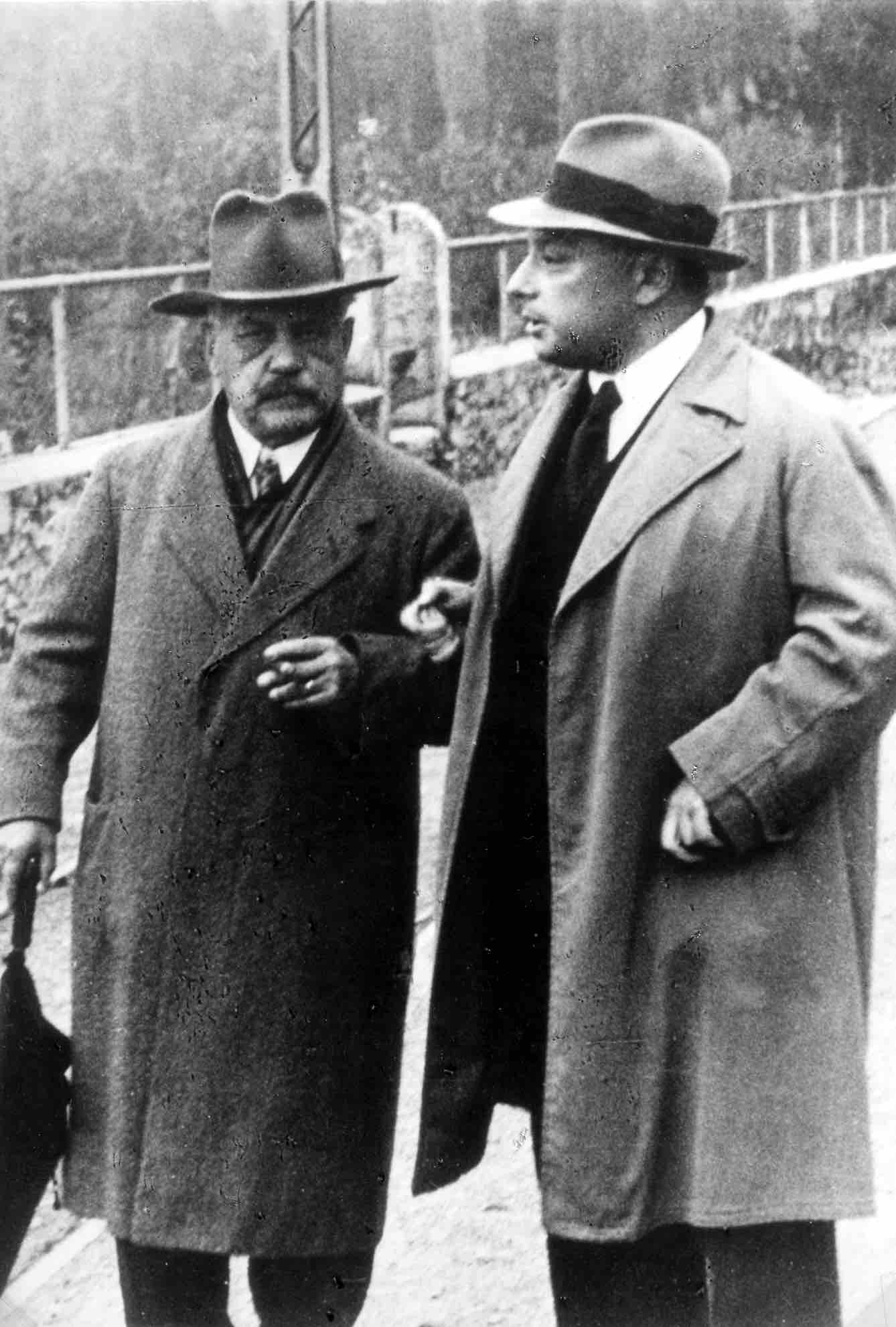}
  \includegraphics[scale=0.21]{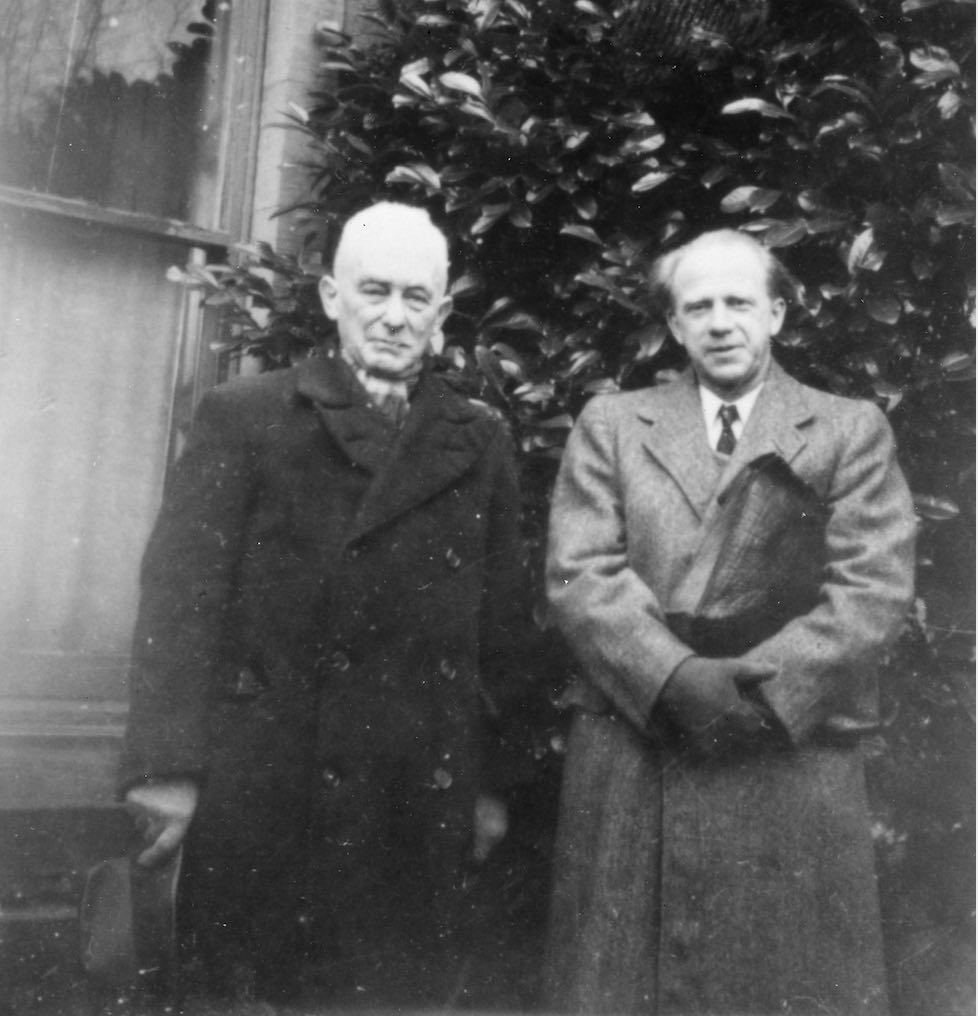}
  \includegraphics[scale=0.103]{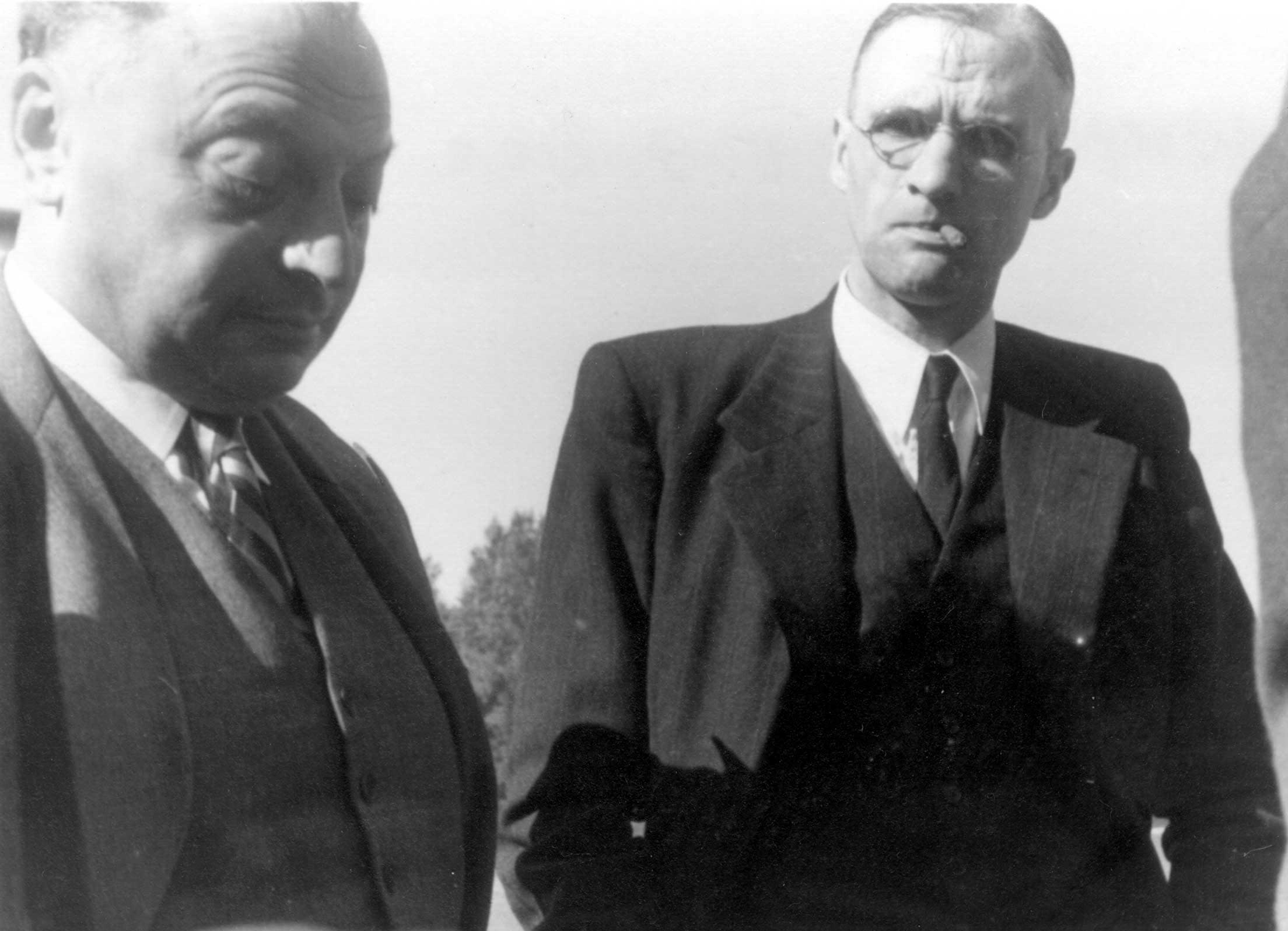}
 \includegraphics[scale=0.16]{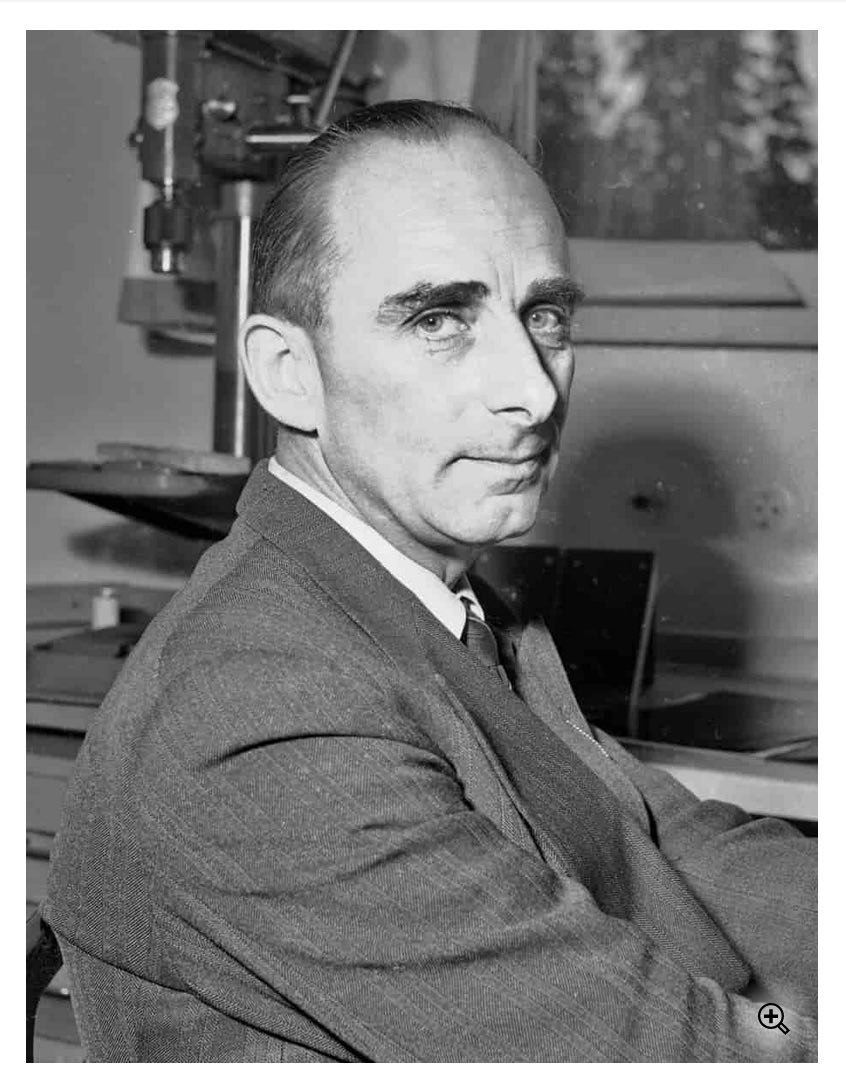}
  \caption{Bruno's mentors: clockwise from top left, October 1934 – Arnold Sommerfeld and Wolfgang Pauli,  in Geneva,   CERN Archives,  Pauli collection;  \href{https://repository.aip.org/islandora/object/nbla\%3A296208}{Max Born and Werner Heisenberg} outdoors at Edinburgh, November 1947, Max-Planck-Institute f\"ur Physik, AIP Emilio Segrè Visual Archives, Born Collection; 
  \href{https://snl.no/Rolf_Wider\%C3\%B8e}{ \RW  \ Oslo 3. mars 1953} 
  for Verdens, from \href{https://snl.lo}{Store norske leksikon}, Wolfgang Pauli and Philip I. Dee at  the \href{https://repository.aip.org/islandora/object/nbla:310379}{1948 Solvay Conference in Brussels}  CERN Archives, Pauli Collection.} 
 \label{fig:mentors-new}
 \end{figure}
After the war, Bruno's formal physics education started. After  obtaining   his Physics Diploma in \Gott\ with a thesis on the  theory of the betatron, he worked  for six months as Werner Heinsenberg's assistant, Fig~\ref{fig:mentors-new}. During this period, he went back to the problem of double beta decay which he had also been working on  during the war, as he says in a letter to his father: ``I've been struggling with neutrinos again for about 2 weeks - this time even Heisenberg is interested in the matter - but I'm getting nowhere. I must have known most of what's involved 2 years ago - but  the papers,  on which I wrote it all down,  are long gone, of course".\footnote{Touschek's letter to his father on 19th July 1946  from \Gott: ``Seit ungefähr 2 Wochen schlage ich mich wieder mit Neutrinos herum - diesmal ist sogar Heisenberg an der Sache interessiert - aber Komm und Komm nicht weiter. Das blode ist, dass ich das meiste was damit zusammenhängt vor 2 Jahren schon einmal gekonnt haben muss - die papierln  auf die ich das alles geschrieben habe sind natürlich schon langst nicht mehr da - [...].}
Bruno  was then  taken  to Glasgow, where Philip I. Dee, Fig~\ref{fig:mentors-new},  was planning
to build  a 350 MeV synchrotron. and where Bruno started his doctoral studies on April 1st, 1947.

Bruno's PhD in 1949  was  followed by  a three year position as Nuffield Lecturer. During the years in Glasgow,   Bruno came in contact with Max Born, Fig~\ref{fig:mentors-new}, attending his Colloquium in Edinburgh, where Born  was Tait Professor since {1936}, after having been suspended from his professorship at the University of \Gott.  At the time, Born  was  publishing    the new (fifth) edition of his famous book {\it Atomic Physics},  and  Bruno, who was making frequent visits to Edinburgh,  was involved in proof reading and checks. During this work, Bruno's studies of weak interactions brought him to understand their  universality and  he prepared an appendix about the analogies in  the decay of nuclei,  $\mu$ and charged $\pi$ mesons.\footnote{In a 13th February 1950 letter to his father, Bruno 
wrote:``In January, I worked with M. Born in Edimburgh and wrote  a chapter and an appendix for him. That was quite
entertaining", {\it Im J\"anner hab' ich mit M. Born in Edimburgh zusammengearbeitet und ihm ein Kapitel und einen Anhang f\'ur sein geschrieben.Das war ganz unterhaltsam.}}
 As Bruno writes in his CV, ``When the proofs arrived, he realized that this coincidence had also  been noted by the Italian physicist Giampiero Puppi, who was in fact cited in the final text."\footnote{ From Bruno's CV, in \RSUPD: 
 {\it In questo periodo visit\`o spesso Born a Edinburgo, e cur\`o la nuova edizione dell'{\it Atomphysik}.  Nel corso di questo lavoro si accorse dell'universalit\`a dell'interazione beta e prepar\`o un'appendice del libro di Born in cui sono messe in evidenza le strane analogie fra il decadimenfo beta dei nuclei, dei mesoni mu e dei pigreco carichi. Quando arrivarono le bozze, si accorse che le stesse osservazioni erano state fatte da Puppi, il quale \`e appunto citato nella stesura finale del libro.}} 
  It should be noticed that  the first paper with the proposal of the universality of
 the weak interactions is due to Bruno Pontecorvo \cite{Pontecorvo:1947}, who had been 
 inspired by the Conversi, Pancini and Piccioni
 experiment \cite{Conversi:1947aa}, which showed that the negative
 component of the heavier cosmic radiation was absorbed in
 the iron, but not in the carbon, before decaying.  In his paper, Pontecorvo discussed the possibility of a fundamental analogy between nuclear
beta-decay processes and absorption of negative muons.
 
While in Glasgow Bruno also   formulated  the classic  problem of infrared radiation \cite{Bloch:1937pw}  in  the covariant formalism of  relativistic Quantum ElectroDynamics (QED), together with   Walter Thirring \cite{Thirring:1951cz}. 

 In 1951 Bruno started searching for a position outside the UK, and when an  opportunity arose for a position in Rome  he was ready to accept it. This opportunity came through  Bruno Ferretti,\footnote{For more details about Touschek's  prewar visits to Rome and the post-war years  in Glasgow, where he obtained his PhD, and    came to know of Ferretti's work,  see  \cite[Chs. 2 and 7]{Pancheri:2022}.} who had visited Manchester and Birmingham in 1947, 
and   worked    on radiation damping \cite{FerrettiPeierls:1947} with Rudolph Peierls, later to be Bruno's PhD external advisor.
Bruno's   interest in Ferretti's work on radiation problems 
  was a reason for him to accept Edoardo Amaldi's offer of a research position at the newly established National Institute for Nuclear Physics (INFN) in Rome \cite{Amaldi:1981}
.\footnote{This fact was also mentioned by Luigi Radicati in an interview recorded by L. B. in Pisa on June 16, 1997.} 
Such a position entailed 
 research in  theoretical physics, and ``assist experimenters  in  [\dots] accelerator physics", a  field which was rapidly superseding the traditional cosmic rays experiments as a major investigation tool in particle physics.

When Bruno arrived in Rome in 1952, he had become what he dreamed of, and more.  He was  the heir of the great European scientists
of the first half of last century, Fig.~\ref{fig:mentors-new}, and  was able to talk and discuss on equal grounds with any other theoretical physicist in the world.
 He had learnt    mathematical physics from Hans Thirring as a student in Vienna,  studied  electrodynamics and atomic physics  from Arnold Sommerfeld's book and through personal exchanges and early correspondence, understood Quantum Mechanics directly from Werner Heisenberg and  Max Born.\footnote{For details of correspondence with Sommerfeld, Heisenberg and Max Born  see \cite[Chs. 2, 7]{Pancheri:2022}. }
 He had also acquired hands-on-experience of how to build an electron accelerator from scratch,  working with  \RW,
 and completed his knowledge on particle accelerators through  his participation in the UK   post-war effort  to build novel types of accelerators such as the 
 Glasgow synchrotron,
  with Philip I. Dee. 
  It was on the basis of such experience, that he had been hired by  Edoardo Amaldi, who envisioned to make Italy a protagonist of post-war scientific renaissance.

The physics institute in Rome was in those days a crossroads of ideas and projects, 
 such as  
 the creation of a National Institute for Nuclear Physics (INFN) \cite{Battimelli:2001aa}, and the definition of plans for the construction of a National Laboratory, where a modern type  accelerator should be built.  Many distinguished visitors from abroad enriched the atmosphere,  among them  Wolfgang Pauli from Zurich, and Matthew Sands from Caltech, while a new generation of students  was graduating, ready to be the protagonists of the reconstruction.    Touschek immediately entered into the spirit of the place and within three months of his arrival in Rome, together with  Matthew Sands wrote an article about {\it Alignment errors in the strong-focusing synchrotron} \cite{Sands:1953aa}, which immediately attracted the attention of the physicists working on a proposal 
 to build a proton synchrotron (PS) at CERN.\footnote{In a letter to his parents on February 24 1953, Bruno  mentions ``\dots  in connection with a new project (strong focussing synchrotron) I covered myself with fame, at least in small circles, and I believe that even if I got stuck in Rome I wouldn't have too much worry." Sands, in a later interview  is more explicit about the interest of their work to people at CERN working on the proposed PS, see   \url{https://www.aip.org/history-programs/niels-bohr-library/oral-histories/5052},  M. Sands, interviewed by F. Aaserud, 4-5 May 1987, Niels Bohr Library \& Archives, AIP.}

    \begin{figure}
\centering
\includegraphics[scale=2]{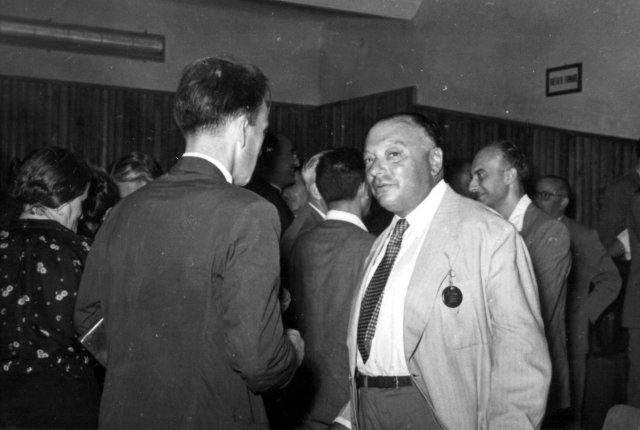}
\caption{Bruno Touschek (left) with Wolfgang Pauli at the 1953 National Congress of the Italian Physical Society, {Pauli Collection, CERN Archives.}}
\label{fig:Pauli-Touschek-1953-Cagliari-lowres096}
\end{figure}
   
 Arriving in  Rome Bruno also  forged a scientific and personal bond with   the last of his mentors, Wolfgang Pauli, Figs.~\ref{fig:mentors-new}, \ref{fig:Pauli-Touschek-1953-Cagliari-lowres096}.  They  had probably met before, but now 
 Bruno  could easily discuss physics on the same level with  a Nobel laureate and their friendship lasted until Pauli's death.     In Appendix \ref{app-BTML}, Gatto, nine years Touschek's junior,  remembers his wonder  attending a conference in Cagliari in September 1953 (in fact the annual congress of the Italian Physical Society),    and  being invited by Touschek  to have coffee with Pauli,  something that was quite natural for Bruno, 
  but much less so for the shy Raoul. These memories  also show  how close    Raoul and Bruno became  to each other shortly after they  met in Rome,   and bring interesting details about the birth of AdA and how Raoul   developed  an  active interest in electron-positron physics, as shall be seen in the next sections.

 \section{ The particle physics scenario of the 1950s}
   \label{sec:particlephysics1950}

In 1953, the reconstruction of European science entered in its operational phase. With  the expected approval of the establishment  of CERN by the national governments,  plans to build     powerful particle accelerators in Europe  moved forward.  These plans would lead to the construction of  the  Proton Synchrotron (PS) at CERN \cite{Blewett:1953ogy},  the  electron synchrotron at the Frascati National Laboratories (LNF) \cite{Salvini:1962aa,Bonolis:2021} and a linear 
accelerator at the newly built Laboratory  in Orsay \cite{Marin:2009}. 

The considerations which led to the construction of the CERN PS were the subject of a Conference on  the Alternating-Gradient  Proton Synchrotron,  held at the Institute of  Physics of the University of Geneva on October 26-27 and 28, 1953, to summarize the work of a study group set up the previous year.
The Conference Proceedings  \cite{Blewett:1953ogy} give  a fascinating description  of the work of the group, and the development of the project, from its beginning in Bergen (Oslo)  \cite{Dahl:1953cxg} to the final comments by Werner Heisenberg, who wrote: ``For research in elementary particles, the decisive quantity is not the energy in the laboratory system but the energy available for particle creation in the center of mass system of the two colliding nucleons". Having stated this fundamental physics goal, Heisenberg  went on to recommend that a proton beam  be accelerated to a final energy of 30 GeV, because  it would allow the creation of nucleon-antinucleon pairs, a process which could not be adequately studied with a 20 GeV machine.

Heisenberg's 1953   observation highlights the  strong interest in anti-matter  in relation  to matter, { which was present in the particle physics community, together  with} 
 the use of symmetry properties in Quantum Field Theory in the early and mid 1950s.
 With the discoveries of more and more elementary particles and their anti-matter counterparts, symmetry properties became a central theoretical issue, and the formulation of a  theorem of invariance of particle processes under Charge, Parity and Time  (CPT) transformations was proposed and developed. 
 The genesis of the theorem  is  complex, with roots in  the work of Julian Schwinger 
  \cite{Schwinger:1951xk},  Wolfgang Pauli \cite{Pauli:1955}, and Gerhart L\"uders \cite{Luders:1954zz}, 
as  detailed in a recent reconstruction  \cite{Blum:2022eol}. Generally called the CPT theorem, it  would play a crucial role in the development of electron-positron colliders.

How these developments were perceived by Gatto can be seen in   
a 2004 letter to one of the authors, L.B., translated and reproduced in  App.~\ref{app-LB}.  Gatto mentions how he had become aware of the CPT theorem \cite{Schwinger:1951xk,Luders:1954zz,Luders:1955fh,Pauli:1955}  through Bruno Zumino and Gerhard L\"uders.\footnote{In  a later paper by Zumino with Gerhart L\"uders  entitled {\it Some Consequences of TCP-Invariance},  a direct reference is given of Zumino's early attention to particle symmetries  \cite{Luders:1957zz}: ``\dots One of us (G.L.) wants to emphasize here again the importance of the role of the other (B.Z.) during all stages of the work that led to the theorem, both through personal discussions and through correspondence. In particular, the original formulation of the theorem [TCP-invariance], for parity-conserving interactions,  was suggested by B.Z in early 1953." }  Zumino had   graduated with Ferretti,  but soon left for the United States, returning occasionally to Rome.\footnote{Before moving permanently to Berkeley, Zumino  also spent  many years at CERN, where  he was  theory group leader in the 1970s, when he
proposed Supersymmetry as an important property of the world of elementary particles.}

As Gatto describes, Touschek was  his mentor during the first period he  spent  in Rome. Then, in 1956 Ferretti moved from Rome to the University of Bologna, his {\it alma mater}, where he held the chair of Theoretical Physics until he retired in 1988, and  Gatto left Rome  
 for a  stay   abroad,  following the traditional pattern of an educational trip, 
 a {\it Bildungsreise}, to gain a broader 
  international  research experience,  
  and, we can add, to acquire the  knowledge and the track record needed to become a university professor. The place  
  where novel discoveries were being made was  the United States.  Gatto had been awarded a Fulbright fellowship for studies abroad and, perhaps inspired by Edoardo Amaldi,  went to Berkeley, where the Lawrence Berkeley National
Laboratory, hosted a new powerful particle accelerator, that had been operating since 1954. It was
called the Bevatron, a weak-focusing proton synchrotron that accelerated protons
to energies of billions of electron Volts, BeV, a unit now most commonly called GeV. Such high energies could be used to discover hitherto unknown
 particles and, or, to confirm the existence of predicted antiparticles. 
Among them there was the antiproton, whose production and observation at the Bevatron was announced by Emilio Segr\`e and collaborators in the fall of 1955  \cite{Chamberlain:1955ns}.

{The other question of major interest in the '50s was the  understanding of weak interactions, in which the discovery of strange particles played the main role.  While the brilliant papers
on the oscillation of neutral kaons \cite{Gell-Mann:1955ipe} \cite{Pais:1955sm} were inspiration to Bruno Pontecorvo \cite{Ponte:1957}  to later propose neutrino oscillations
 \cite{Gribov:1968kq,BP:1978},\footnote{This proposal  has received experimental confirmation and has put an important constraint on a theory of elementary particles beyond the Standard Model.}
 the $\theta - \tau$  puzzle for the decays of charged kaons led Lee and Yang to propose that parity conservation was violated in weak interactions \cite {LeeYang:1956aa},    an explanation confirmed by  Madame Wu experiment on the beta  decay of polarized cobalt \cite{Wu:1957my}.}

\subsection{Antiprotons and new discoveries}
\label{ssec:antiprotons-etc}
The anti-electron, i.e, the positron, had  been discovered through cosmic ray searches before the war, but for the anti-proton  one had to  wait until after the war,
when it first appeared in emulsion stacks exposed at high altitudes, {such as the two events observed by Rossi's group at
MIT \cite{PhysRev.95.1101} and by Edoardo Amaldi's group in Rome. The latter event, named {\it Faustina}, had been interpreted as associated with the annihilation of an
antiproton} 
 \cite{Amaldi:1955}.\footnote{A letter exchange between L\"uders and Amaldi about the photograph of this event can be found in  Archivio Amaldi, Physics Department Archives, box 22. Such exchange {highlights}  that the Rome group discovery had submitted their results to the {\it Nuovo Cimento} on 18 February 1955, before the Berkeley group did so to {\it The Physical Review} on  24 October 1955, as discussed 
 by  G. Battimelli, D. Falciai, ``Dai raggi cosmici agli acceleratori: il caso
dell'antiprotone", in Atti del XIV e XV Congresso Nazionale di Storia della
Fisica (Udine 1993 - Lecce 1994), a cura di A. Rossi, Ed. Conte, Lecce 1995,
pp. 375-386, {and} G. Battimelli personal communication.} 


{In 1956, a collaboration was established between the  Berkeley and Rome Laboratories in which Amaldi and Segrè's goal was to expose a stack of emulsions  to the flux of antiprotons produced by the Bevatron, in order} to observe the annihilation process undergone by an antiproton inside the emulsion, and to compare it with the cosmic-ray event possibly due to such an antiparticle. The emulsions were analyzed by both groups and what they found corroborated the interpretation given in \cite{Chamberlain:1955ns}  that the particles observed in the Bevatron were antiprotons and also supported the hypothesis that the  {\it Faustina} event was indeed due to an antiproton 
\cite{PhysRev.101.909}.

 A second paper was published by the Berkeley/Rome groups \cite{Chamberlain:1956yy}
 in the  {\it  Nuovo Cimento}
 in parallel with Raul Gatto's theoretical paper, in which he was  discussing the capture and annihilation of antiprotons, as well as processes involving positive and negative kaons \cite{Gatto:1956}. Thus,  when Gatto had  left Italy for Berkeley in late 1956, he was
already deeply involved in the study of antimatter, and he continued
to work on consolidating the concept of antiparticles at the level of
the fundamental constituents of matter, as in his work on  the
annihilation of a nucleon-antinucleon system into a $K-{\bar K} $ pair \cite{Gatto:1957}. 

The discovery of the antiproton was presented by Emilio Segr\`e 
in New York in April 1956 at  the International High Energy Physics Conference, also known as the {\it Rochester Conference}, after  the place in New York where it  had first been held in  1950.\footnote{For a history of the Rochester Conference see R. Marshak's article in  \href{https://books.google.it/books?id=KAcAAAAAMBAJ\&lpg=PA92\&hl=it\&pg=PA92\#v=onepage\&q\&f=false}{The Rochester Conference},   Bulletin of the Atomic Scientists, June 1970, pp. 91-98.} At this Conference, Amaldi also presented a paper reporting both  what they had observed in their cosmic ray experiments and  what they had found in emulsions exposed at the Bevatron.\footnote{Segr\`e and Chamberlain received the 1959
 Nobel Prize in Physics  for their discovery. For  Edoardo Amaldi's  contribution, see \cite{Battimelli:1995}.} 
 
   The  year  1956  represents  a {\it watershed } for particle physics. At the Rochester conference  the discovery of the antiproton 
made its public debut, the  whole {\it zoo} of new particles took center stage and,  for the first time, parity conservation in weak interactions was questioned.

Touschek  had  travelled with Amaldi to the Rochester Conference, presenting  his work on  strange particle decays \cite{TouschekEtAl:1954}.
 and   was profoundly impressed by the discussion about possible non-conservation of parity in weak interactions, which had come up in the theoretical session,  {chaired by C.N. Yang}, following a question by Feynman \cite{Oppenheimer:1956hfa,doi:10.1146/annurev.nucl.52.050102.090730}. 
  Soon after the 
Conference, Lee and Yang thoroughly analyzed the question and in June
submitted a paper discussing possible experimental tests of parity
conservation in beta decay \cite{LeeYang:1956aa}.
 Such a question was discussed by Gatto in different contexts and from
different perspectives in papers published between 1957 \cite{Gatto:1957nn} and 1958 \cite{Gatto:1958aa}. 
Back to Rome, Touschek, who had until then mostly worked on  field theoretical problems \cite{CiniMorpurgoTouschek:1954,MorpurgoTouschek:1955,MorpurgoRadicatiTouschek:1955aa,Morpurgo:1956aa} and  phenomenological  studies with the Rome experimentalists \cite{FabriTouschek:1954}, turned his attention to weak interactions and neutrino physics, finding a relationship between the neutrino mass and non-conservation of the leptonic number \cite{Touschek:1957ab}, based on his previous work in \Gott \ on double beta-decay \cite{Touschek:1948ab}. 

\subsection{Proposals for Center-of-mass collisions }
\label{ssec:1956cmcollisions}
The year 1956 was to be remembered also for accelerator  physics. A few months after the Rochester Conference, a revolution 
started   at the CERN Symposium on High Energy Accelerators \cite{Proceedings:1956ipa}, when  
 proposals for building new accelerators  were presented by Donald Kerst \cite{Kerst:1956gxa} and Gerald O'Neill \cite{ONeill:1956cvf}, promoting    tangential rings for particle-particle collisions, in which center-of-mass collisions would allow higher and higher energies to be achieved. 
 
 When Kerst presented his idea of how to realize c.m. collisions  of two protons in a laboratory, through two tangential accelerators,
 he  was already famous for having built the first betatron \cite{Kerst:1940zz,Kerst:1941zz,PhysRev.60.53}, in 1941 at the University of Illinois. His  ``induction accelerator"  was based on \W's  1928 article \cite{Wideroe:1928aa} which proposed a magnetic ring, which would  accelerate electrons and, at the same time,  bend them in a circular path. 
 {The  {\it Strahlentransformator}, a  ``ray transformer", as \W \  called his invention, became known as a  betatron,  after Kerst built the first one.}
 
The idea of using the well known kinematic principle that  center of mass collisions can reach much higher  energies when using two oppositely traveling particle beams than just a single beam hitting a stationary target, was not new. In 1943 a patent had been filed with the German Patent office by Rolf Wider\o e, following  a discussion  with Bruno Touschek, when they were working together in Hamburg, constructing \W's 15 MeV betatron, 
The patent was ratified only after the war, in 1953. At the time \W\ was working in Switzerland constructing betatrons for medical applications and  was ready to discuss its merits at accelerator conferences.
In that same year, 1953, at the Geneva conference, Heisenberg
invited his colleagues to plan particle accelerators whose beam energies
were such that they could reach the nucleon-antinucleon creation threshold,
{\it de facto} opening up a new perspective for observing interesting processes
and achieving new results in the laboratory. The way was paved, and in 1956
 the first stone was laid 
 \cite{Kerst:1956whq,ONeill:1956iga}.

\begin{figure}
  \includegraphics[scale=0.5]{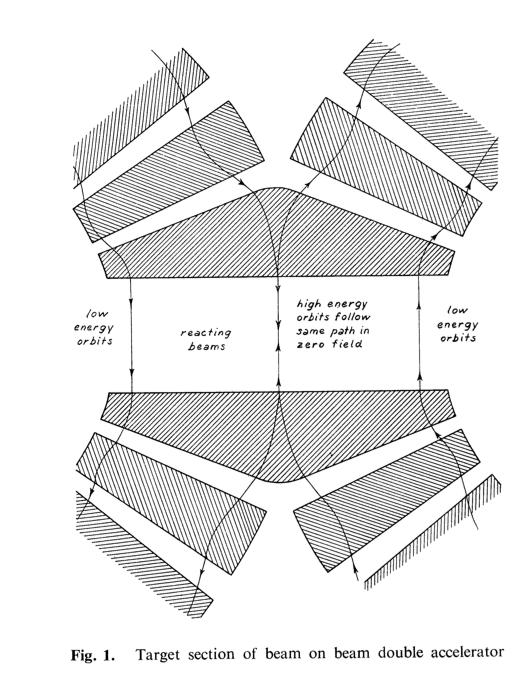}
  \includegraphics[scale=0.5]{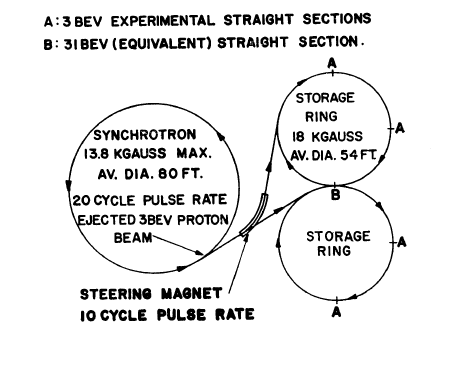}
  \caption{At left, Kerst's proposal for realization of c.m. collision obtained with two separate proton accelerators \cite{Kerst:1956gxa}, at right O'Neill's proposal for two storage rings fed by a separate synchrotron \cite{ONeill:1956cvf}, from \cite{Proceedings:1956ipa}.}
\label{fig:kerstOneill1956}
\end{figure}

 The two American proposals differed as   Kerst proposed two accelerators placed side by side, whereas
O'Neill idea was to realize  center of mass collisions by separately storing
two particle beams fed by a single synchrotron, as shown  in Fig.~\ref{fig:kerstOneill1956}. The other difference was that,  in addition to propose protons to collide against protons, O'Neil also envisaged electrons against electron collisions. Indeed this was the project he promoted in the following years  \cite{Oneill:1959ab,Barber:1959vg}.

\begin{figure}
 \centering
\includegraphics[scale=0.3]{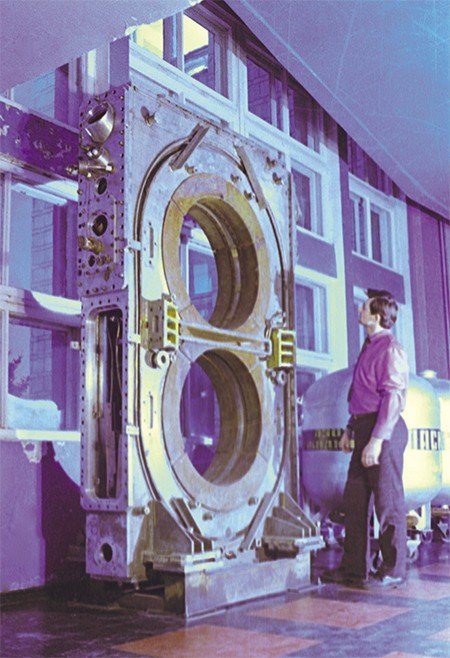}
 \includegraphics[scale=0.3]{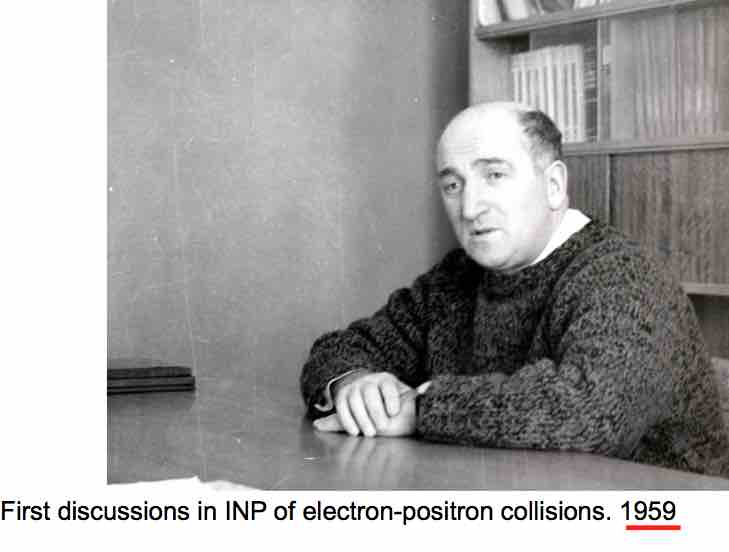}
\caption{ At left,  a  photograph of VEPP-1, the Russian electron-electron collider built by G. Budker, right panel;  slides 
from  A. Skrinsky's presentation  at the  2011 BTML Lectures in  Frascati, \copyright LNF-INFN, all rights reserved.}
 \label{fig:Budkertouschek-catania}
 \end{figure}

By the time Gatto was back in Rome   in 1958, 
the scenario had changed.   In the next two years,
  three projects for the  construction of accelerators with   center-of-mass collisions would arise. 
     Unknown to the West until 1963 \cite{Marin:2009},  Russian projects for both electrons against electrons and electrons  against positrons  were  discussed in the USSR \cite{Baier:2006ye} and set in motion at  the  Institute of Nuclear Physics (INP), the Syberian Branch of the Soviet Academy of Sciences in Novosibirsk,   by Gershon  Budker, Fig.~\ref{fig:Budkertouschek-catania}.
   \footnote{The INP 
   was renamed the Budker Institute of Nuclear Physics in honor of G. Budker, after his death in 1977.} 
   An American project  with electrons against electrons to test the limits of Quantum ElectroDynamics (QED)  \cite{Barber:1959vg} was started 
    between Princeton and Stanford, and, in March  1960  in  Frascati, Touschek's  proposal for electron-positron collisions was approved, following discussions 
     in Rome in Fall 1959.  His idea was 
completely revolutionary, and was based on the  symmetry properties of particles vs. antiparticles and
   the deeper concept of probing the quantum vacuum through  particle-antiparticle annihilation in c.m. collisions   
   \cite{Bernardini:1960osh}, \cite[p.76]{Valente:2007}.\footnote{Touschek himself wrote: ``Nothing can be more elementary than the vacuum, whose properties determine all of physics. . . The system
   $e^+e^-$ is very close to the vacuum, but not identical, of course." His emphasis was on the need to start from as anonymous an initial state as possible:
    ``$e^+e^-$ against $e^-e^-$: existence of the annihilation channel” [B. Touschek, ``Adone and the polarization of vacuum” (manuscript, B. T. A., Box 11, Folder 3.90, pp.  1-2)].} 
  
\section{AdA} 
\label{sec:AdA}
In this section we will describe the events which led to the  development of AdA between Rome and Frascati. AdA's success has sometimes been described as a {\it curiosity} {\cite{Richter:1997aa}} or a lucky idea, but it was in fact the convergence of many roads, among them advancements in accelerator science, a  national laboratory in Frascati searching for  new ideas after having built its first modern accelerator, a cadre of exceptional theoretical and experimental  physicists in Rome, picking up Fermi's legacy, the existence of CERN, where ideas would circulate between laboratories in a cross fertilization process.     
A single laboratory in fact may not have had the capacity to make AdA work, especially since the American scientists from Stanford doubted the idea could  be realized,as it appears from Gatto's memories in App.~\ref{app-BTML}.\footnote{As Gatto writes in App.~\ref{app-BTML}: ``Electron-electron collisions would allow to test the photon propagator. I remember 
a conference by Professor Panofsky, at the end of 1959, reporting on the pioneering 
work of Barber, Gittelman, O'Neil, Panofsky, and Richter \cite{Barber:1959vg}, on electron rings. 
Answering to a question, Panofsky mentioned that, to test the electron (rather than the 
photon) propagator, electron-positron collisions would have been suitable, through 
observation of 2-photon annihilation, but that such a development could present 
additional technical difficulties and that for  the moment had been postponed." A similar cautionary comment can be found  in  B. Richter's talk  at the 1961 CERN International Conference on Theoretical Aspects of High Energy Phenomena \cite{Richter:862684}, where he said: ``until we know what can we do in storing a beam, we cannot say anything about the positron experiment".  He was right in his caution, and Frascati alone could have taken longer to prove the feasibility of an electron-positron collider. But history was changed after  Touschek and Gatto's presentations  about AdA and ADONE, which followed Richter's at the same conference. Their presentation caught the interest of two French physicists from Orsay , and  what did not fully work in Frascati, worked when AdA was brought to France, at the \LAL, 
\cite[p. 335]{Pancheri:2022} \cite{Bonolis:2018gpn}.} 

Here we  shall be  focusing in particular on the theoretical physics developments which were crucial for the idea to rise and be feasible. This point of view shows  that the success of AdA is due to  three fundamental concepts: center of mass collisions to increase the available energy, the CPT theorem which ensures that particles and antiparticles accelerated in the same magnetic ring will move along the same circular orbit with equal and opposite velocity and   will annihilate, and relativistic QED. 
 \begin{figure}[h]
\includegraphics[scale=0.9]{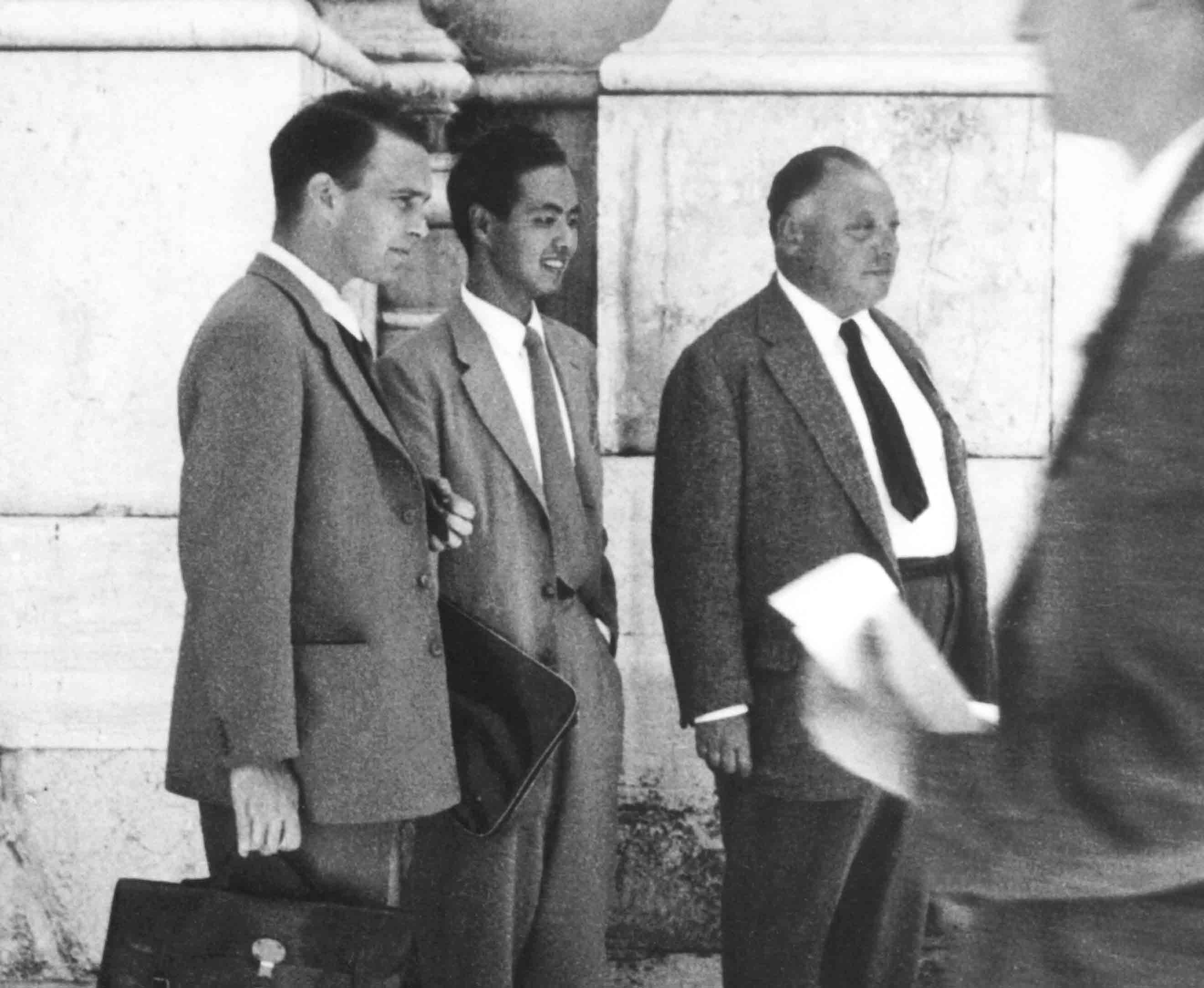}
 \hspace{+0.5cm}
\includegraphics[scale=0.1]{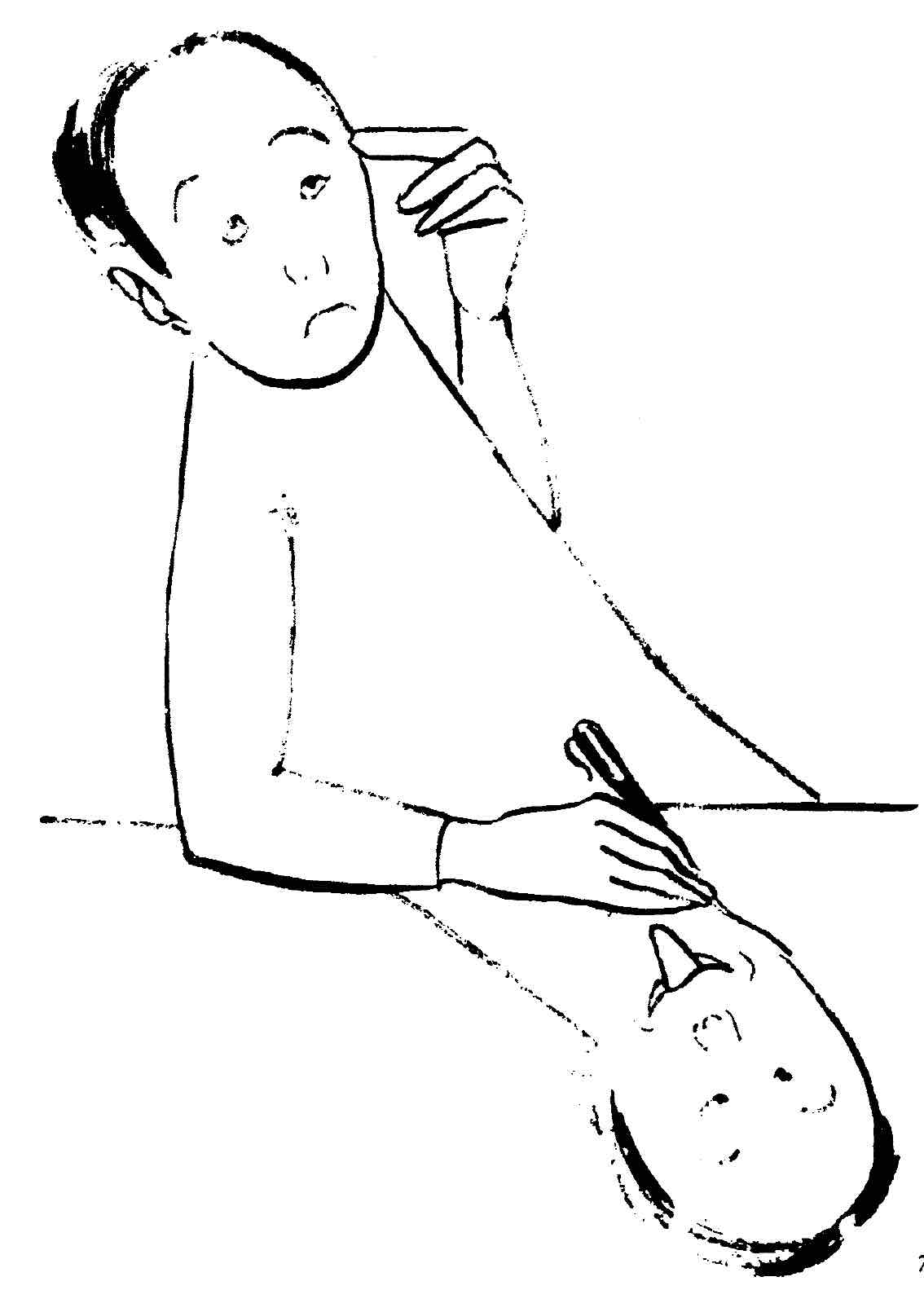} 
\caption{ From left to right: Bruno Touschek, T.D. Lee, Wolfgang Pauli and Robert Marshak at the  Padua-Venice Conference on mesons and newly discovered particles, September
1957, courtesy by M. Baldo-Ceolin to  L.B.; at right Touschek's caricature of T.D. Lee, with  hints of parity non-conservation, \copyright Touschek;'s family, all rights reserved.}
\label{fig:Venice1957}
\end{figure}
 Of crucial relevance to Gatto and Touschek's joint belief   in the importance and feasibility of  electron-positron collisions, was that 
 Touschek  was close to 
   Wolfgang Pauli, Fig.~\ref{fig:Venice1957},   during the years Pauli was working on the CPT theorem,  and Gatto had met  L\"uders and worked with him \cite{Gatto:1958aa}, namely both of them heard  and followed  the  development  of the CPT theorem in the early 1950s. It is then easy to 
argue that when their two paths crossed roads with the availability in Frascati of a modern type electron  synchrotron built by  a carefully  selected scientific and technical staff, all the conditions were met for an extraordinary 
breakthrough, such as the conception and realization of AdA, the first ever electron-positron collider.

\subsection{Different roads  to AdA}
\label{ssec:AdAroad}
Contrary to Bruno, with  his  experience and  knowledge in different fields of physics, Raoul, on his arrival in Rome, was relatively inexperienced. But in this post-war period, everything moved   along  accelerated tracks, discoveries were following discoveries, ideas piled up, fostered by the international exchanges, which had been interrupted by the war.  At first,
the scientific community had to focus more on post-war reconstruction rather than finding new ways of doing  science. Then, with the  new political asset in Europe, the creation of CERN, and American funded transatlantic travel,   fundamental research rapidly moved forwards. 

Immersed in the lively international atmosphere of the Rome Physics Institute,  Gatto had soon developed into one of the most prominent young  particle physicists of his generation.  
During his first years in Rome Gatto had written  phenomenological
articles   on nuclear and particle physics 
and a more formal
article  on a topic suggested by Ferretti. 
His publication record in those early years is remarkable for a young physicist of his age.\footnote{See his publication list in \url{https://ui.adsabs.harvard.edu/.}} He was strongly influenced by Touschek's presence and friendship, as one can see  from the letter reproduced in App.~\ref{app-LB}, which testifies that his
interaction with Bruno Touschek was very important during this period. 

Then, Gatto’s own trajectory towards particle physics had started, with his one year stay in the United States, a  trip which took place at a time  when very unique discoveries were made. He first spent some days at Columbia University,  where, 
C. N. Yang and T.D. Lee, seen in Fig.~\ref{fig:Venice1957} with Pauli and Touschek, were working, and,  then, a period in Berkeley. 

After several days at Columbia,
Gatto arrived in California, where he was influenced by the exciting atmosphere related to the experiment of the Alvarez group on the strange particles. He understood the relevance of the strange particles 
and of the weak interactions, and  wrote two important papers on the decays of the kaons \cite{PhysRev.106.168}  \cite{Gatto1958PossibleET},  before and after
the proposal of the V - A theory of weak interactions
\cite{Marshak:1957xk,Sudarshan:1958vf} \cite{Feynman:1958ty}.

His  brillant research 
at Berkeley had as a  consequence offers of work at  Chicago
and Columbia, but he  preferred to accept an 
invitation by Amaldi to come back to Rome.
His versatility led him  to contribute to different topics in
elementary particle physics, and in 1958   joined  Marcello Cini and Ezio Ferrari,   authoring 
 a paper, which used a dispersion relation approach   and  showed 
 that the neutron proton  mass difference
cannot  be  explained by their electromagnetic interactions \cite{PhysRevLett.2.7}.\footnote{Information about Ezio Ferrari's contribution to academic life at University of Rome, can be found  at \url{https://archivisapienzasmfn.archiui.com/entita/613-ferrari-ezio}.} 
In Rome, he also  gave lectures on weak interactions
\begin{figure}
\centering
\includegraphics[scale=0.53]{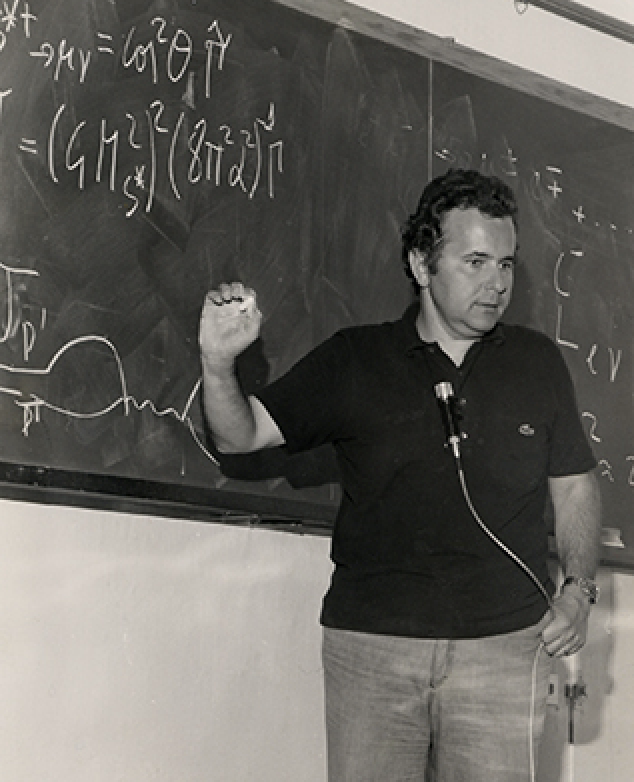}
\includegraphics[scale=0.056]{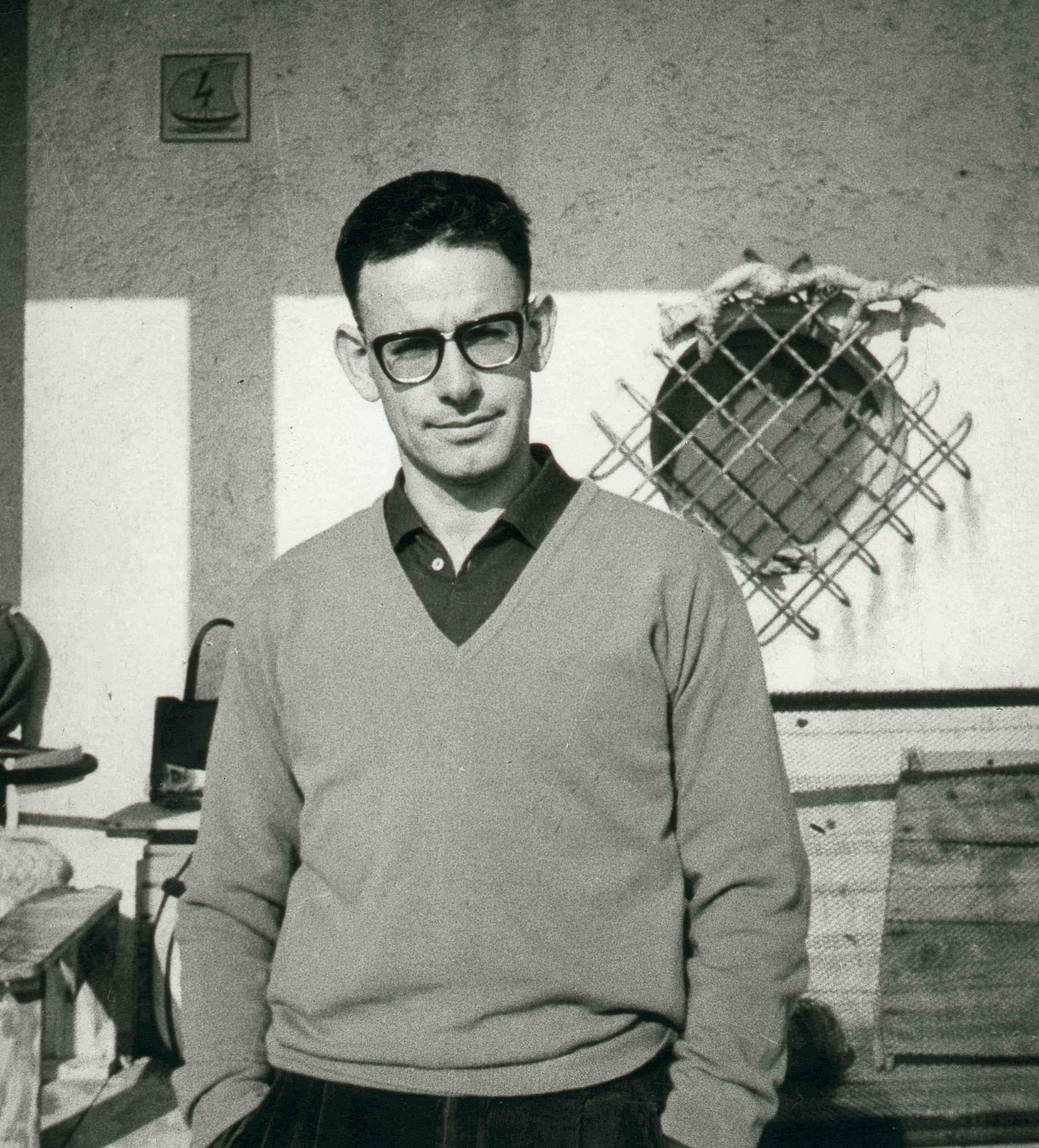}
\caption{At left: Nicola Cabibbo, around 1963, courtesy Cabibbo Family, all rights reserved; At right, Francesco Calogero in 1957, personal  courtesy to LB, all  rights reserved.}
\label{fig:cabibbocalogero}
\end{figure}
 and wrote a paper related to his lectures  \cite{Gatto1959PresentPI}, which shows his good knowledge of the status of weak interactions. 

Gatto's lecture notes   had  been prepared  by two  young physicists, who had graduated with  Touschek, Nicola Cabibbo and Francesco Calogero, Fig.~\ref{fig:cabibbocalogero}.
 As soon the eightfold way was proposed, he wrote an important paper on the weak currents in SU(3) \cite{Cabibbo:1959zz}, which is quoted in the famous work by Cabibbo, where he introduced the angle, which took his name
\cite{Cabibbo:1963yz}. In 1958 Gatto  was ready for a University professorship.  Together with Bruno Zumino and  Sergio Fubini, he was   one of the winners in the national  competition for  three   Chairs in  Theoretical Physics,  and  was called to teach at   the University of Cagliari, which he joined in 1959.

At this time,  Bruno's theoretical interests  had been focused   on neutrino physics, an old acquaintance  of his and  the object of Bruno's intense studies, with  their symmetry property transformations \cite{Touschek:1957ab}. 
 Since 1953 Touschek discussed with Pauli issues connected with time reversal in field theory. He  
   wrote more than one paper on the subject  \cite{Morpurgo:1954aa,MorpurgoTouschek:1955,MorpurgoRadicatiTouschek:1955aa,Morpurgo:1956aa}, 
  and discussed with Morpurgo the extension of the procedure to Parity and Charge Conjugation \cite[p. 85]{Morpurgo:2004aa}.
  In 1957/1958 Touschek exchanged several letters with Lüders, Pauli and Zumino discussing items
connected with symmetry properties of physical theories.\footnote{See correspondence in Bruno Touschek papers from  \RSUPD,   and \cite{Touschek:1958aa}.}
Touschek was led to  propose  chiral transformations \cite{Touschek:1957aa}, and in 1958 had started a work with Pauli.  Then, suddenly and tragically, on December 14th 1958, Pauli died. The work with Touschek was published posthumously  \cite{Pauli:1959kmw}. Bruno was deeply moved and thought of leaving theoretical physics.\footnote{Years later, he commented on his disenchantment with theoretical physics, as he wrote: {\it At the time I felt rather exhausted from an overdose of work which I had been trying to perform in the most abstract field of theoretical research: the discussion of symmetries which had been opened up by the discovery of the breakdown of one of them, parity, by Lee and Yang. I therefore wanted to get my feet out of the clouds and onto the ground again, touch things (provided there was no high tension on them) and take them apart and get back to what I thought I really understood: elementary physics. B. Touschek , "Ada and Adone are storage rings" (B. T. A., Box 11, Folder 3.92.4, p. 7).  } } 
This loss and the  death of his  aunt Ada, just a few months later, gave  him the  emotional and intellectual jolt which changed his life:  when  he went to Kiev in July 1959,  the opportunity for a change came, and,  a few months later, he argued to look for  unexplored grounds, proposing an experiment on electron-positron collisions, 
  a turning point for particle physics.

In 1959,  on the verge of joining the University of Cagliari, Gatto was also ~
open to  a change and,  despite his strong interest in weak interactions,  was ready to be conquered by Touschek's idea of  electron-positron rings. As we shall see next, after hearing Panofsky and Hofstadter's seminars in Kiev, and later again Panofsky's seminar in Frascati, top panel in Fig.~\ref{fig:panofskySeminar},
 he understood  the importance of the idea, and joined his friend and former mentor in Rome, participating  not only to the birth of electron-positron physics, but its further developments as well.

 \subsection{The birth of AdA between October 1959 and March 1960} 
 \label{ssec:AdAbirth}
  \begin{figure}
\centering
 \includegraphics[scale=0.1]{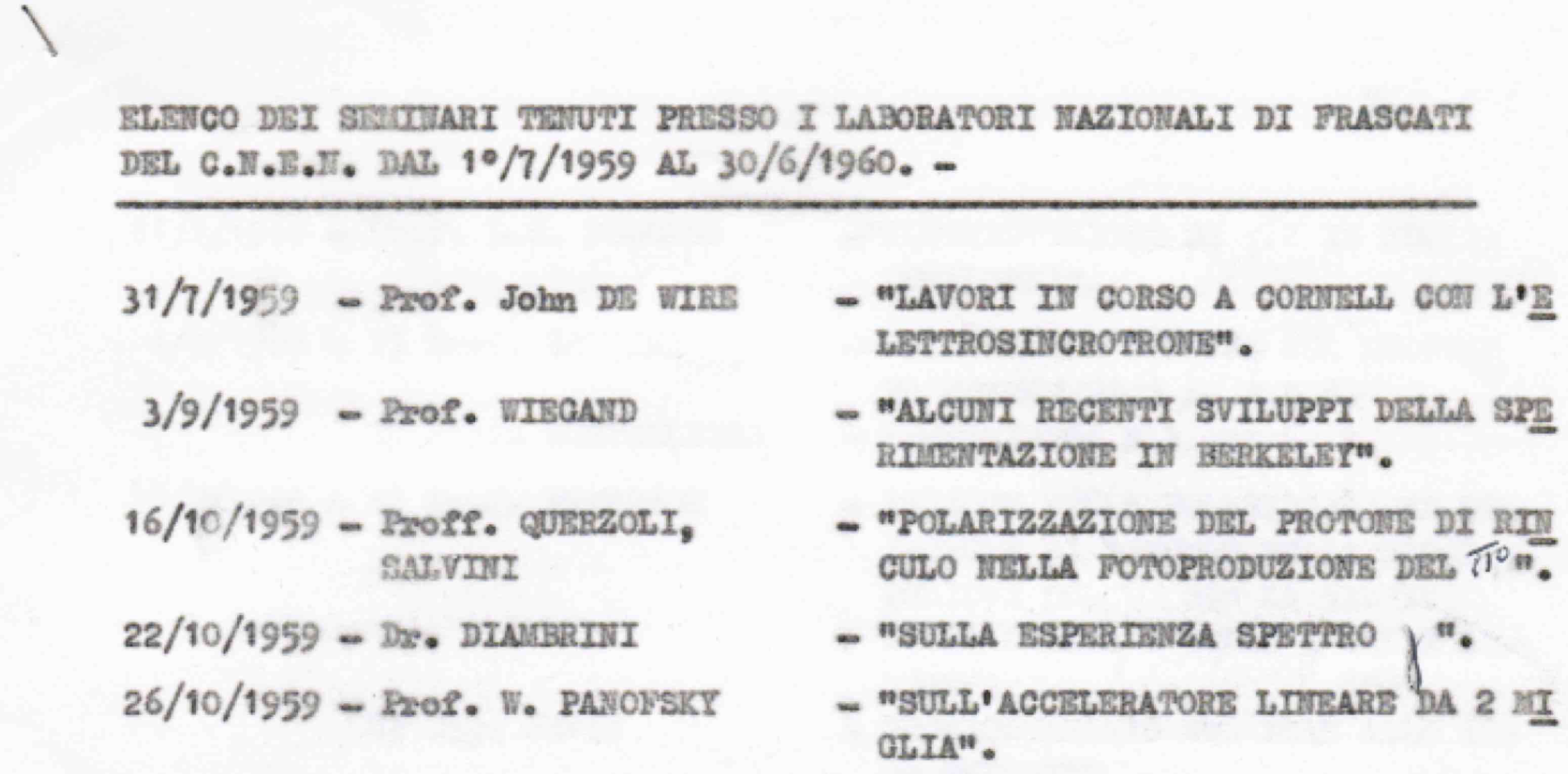}
\hspace{0.2cm}\includegraphics[scale=0.2]{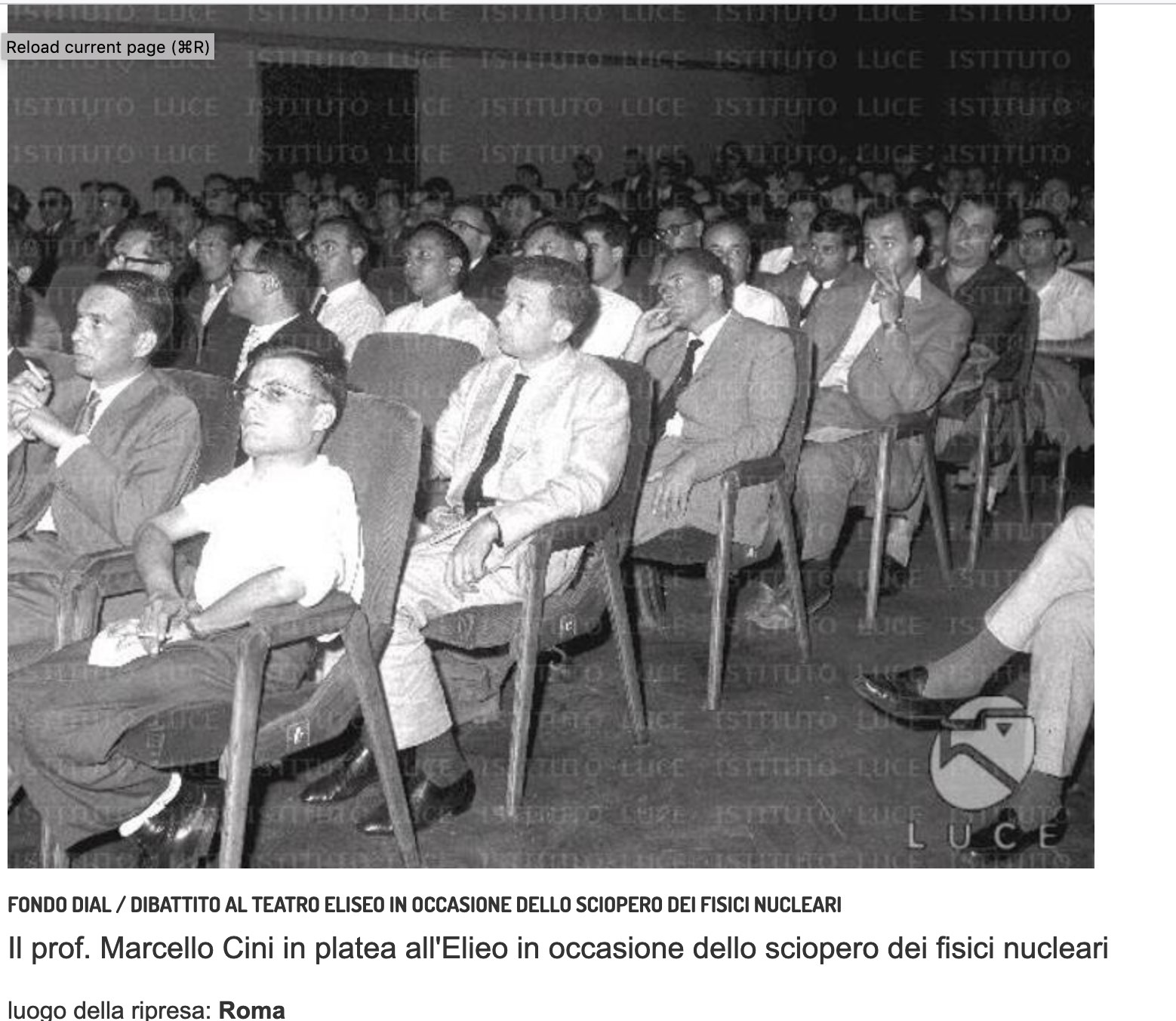}
\caption{On top, an excerpt from the  list of seminars held in Frascati during the year 1959-60, courtesy Vincenzo Valente to G.P., all rights reserved; below,  Bruno Touschek, extreme left,  and Marcello Cini, first in second row, 
 during  a 1960  debate at the Eliseo Theatre in Rome, on occasion of a \
 \href{https://patrimonio.archivioluce.com/luce-web/detail/IL0000008857/9/il-prof-marcello-cini-platea-all-elieo-occasione-dello-sciopero-fisici-nucleari-1.html?startPage=0&jsonVal={\%22jsonVal\%22:{\%22query\%22:[\%22marcello\%20cini\%22,\%22*:*\%22],\%22fieldDate\%22:\%22dataNormal\%22,\%22_perPage\%22:20,\%22archiveType_string\%22:[\%22xDamsPhotoLuce\%22]}}}{strike by nuclear scientists}, in favour of  more adequate salaries and better prospects for younger scientists,  Credits Archivio  Luce, Cinecitt\`a,   [patrimonio.archivioluce.com] [D128-24.JPG].} 
\label{fig:panofskySeminar}
\end{figure}
As it will be  shown, 
Gatto's recollections, together with those from Cabibbo in \cite{Cabibbo:1997aa,Cabibbo:2003aa}, 
 confirm how the idea of AdA developed \cite{Pancheri:2022}, namely that the path leading to electron-positron colliders began  on or around  October 26, 1959, after  {seminars} by Wolfgang Panofsky in  Frascati  and, probably, in Rome as well  \cite{Cabibbo:1997aa}.\footnote{It is not clear whether Panofsky held two seminars: Nicola Cabibbo in 1997 mentions a seminar in Rome, while Frascati documents show a seminar by Panofsky on October 26th, 1960. Both could have taken place.}
 A confirmation of the impression from such a seminar is found in \cite{Cabibbo:1960zza}, where the authors wrote: ``We are indebted to Professor Panofsky for a stimulating seminar  on the possibility of colliding beam experiments." Or perhaps, it started a few months before, when both Touschek, Gatto and Marcello Cini, seen with Touschek in 1960 in Fig.~\ref{fig:panofskySeminar},  participated in the 1959 Rochester  Conference 
  in Kiev, at the end of July \cite{Kiev:1960}.\footnote{The 1959 Conference  was originally scheduled to be held in Moscow, July 15-25, but the location  was later changed to Kiev, perhaps because of the contemporary International Cosmic Ray Conference, ICRC 1959, being held in Moscow on the same dates. The scientific competition of the Cold war was at its peak with the US organizing  an extraordinary National Exhibition at Sokolniki Park in Moscow, a showcase of American  life in the 1950s,  \url{https://www.rferl.org/a/Fifty_Years_Ago_American_Exhibition_Stunned_Soviets_in_Cold_War/1783913.html}.}
 
 Here is  the sequence of events, as it can 
be gathered from personal documents,  Touschek's letters to his father, conference proceedings, and bibliographic references.
\begin{itemize}
\item Kiev, July 1959: Among the rather large  number of Italian scientists attending  the conference, there were  three theoretical physicists from the University of Rome: Marcello Cini, Raoul Gatto and Bruno Touschek, who listened to Hofstadter's talk on his  measurements of the proton electromagnetic form factor, as well as to  Panofsky's talk on the  tangential  electron-electron collision project  at Stanford, and  on  the results from an experiment to measure electron-positron collisions in flight at the linear accelerator \cite{Panofsky:1959kz}. 
\item 
In  
September, Panofsky was in Geneva to attend   the 2nd Accelerator Conference  \cite{Panofsky:1959nua,Kowarski:1959lua}
and then came  to Rome, visiting  Frascati and  giving  a  seminar, which  
is known to have taken place in the Laboratories on October 26th, 
Fig.~\ref{fig:panofskySeminar} \cite{Pancheri:2022}.

\item {In October} Gatto,  Cabibbo and Touschek  all attended Panofsky's seminar (either the one in Frascati or another one in Rome)  at the end of which Touschek  pointed  out the advantage of using electron-positron collisions for the inherently superior physics discovery potential and  for   practical  reasons -- one ring instead of two.  
Nicola Cabibbo's recollections about the seminar clearly show how CPT had a major role in Touschek's idea about the importance -- and feasibility -- of a matter-antimatter experiment: ``It was after the seminar that Bruno Touschek came up with the remark
that an $e^+e^-$   machine could be realized in a single ring, `because of the CTP theorem’ ”
[Cabibbo 1997, p. 219]. Raul Gatto, too, well remembered that, ``Bruno kept insisting
on CPT invariance, which would grant the same orbit for electrons and positrons inside
the ring” [R. Gatto, personal communication to L.B., January 15, 2004.]
\item From November 1959 until February 1960, four theorists in Rome began to study the physics processes that might be observed  in electron-positron collisions: Gatto, two of Touschek's former students, Nicola Cabibbo and Francesco Calogero,  and Laurie Brown,  a distinguished American physicist visiting from  Northwestern University in the US, started  and finished work on two separate aspects of the possibility of studying strong interaction contributions in electron-positron collisions. The first article which was finished and received by {\it The Physical Review Letters} (PRL)  office on February 8th, 1960, is on the contribution of  pion-pion interactions to the time-like photon propagator, by Brown and Calogero \cite{Brown:1960}. The second one was  by Cabibbo and Gatto  on the time-like observation of the pion form factor in electron-positron  collisions \cite{Cabibbo:1960zza},  and was submitted on February 17th, the same day that Touschek {in Frascati} aired the idea of making an experiment to transform the newly built   electron synchrotron into an electron-positron collider. This proposal was not  approved by the Scientific Council of the Laboratories, but Touschek and Giorgio Ghigo were encouraged to prepare  plans to   build a small machine, 
to test Touschek's idea with an eye  toward future accelerators, where it could be applied 
to become 
a new tool for discovery. That same afternoon, a meeting   was held to prepare  a detailed proposal. 
Touschek and Ghigo were  joined by Carlo Bernardini, at the time a  theorist on the LNF staff, and Gianfranco Corazza, an expert on how to create an  extreme vacuum in an accelerator chamber. 
About two weeks later, on March 7th,   a  proposal  was submitted  to the Scientific Council of the Laboratories, which 
agreed that it should be  sent to INFN for funding. 
\item The two theoretical physics papers from Rome were published on March 15th, 1960 \cite{Brown:1960,Cabibbo:1960zza}, and, shortly afterwards, AdA's construction was approved by the INFN with an initial budget of 8 Million Lire.
\end{itemize}

\noindent Neither of the two theoretical papers mentions Bruno Touschek, or Frascati, which is obvious since Bruno's proposal came up only on February 17, the day both papers were already laying at the PRL editorial office. Only Marcello Cini is mentioned in Brown and Calogero's paper, where  he is thanked for his encouragement. {Marcello Cini (1923-2012), shown with Touschek in the bottom panel  of Fig.~\ref{fig:panofskySeminar}, was a theoretical physicist who joined the Physics Institute in Rome becoming Professor of  
Theoretical Physics Institutions 
in 1957, and is also 
 acknowledged in one of  De Tollis'  works about the influence of pion-pion interactions in 
 photoproduction of charged pions \cite{DeTollis:1960}}.\footnote{See \url{https://www.pg.infn.it/lintitolazione-dellaula-e-al-professor-benedetto-de-tollis}
 for a  Symposium, in memory of Benedetto De Tollis, and  contribution by Giorgio Parisi.}  This points    to the frequent exchanges and discussions about pion-pion interactions 
 {that took} 
 place in Rome in 1959 and to the possibility that discussions about electron-positron collisions and/or Hofstadter type experiments were started in Rome among the senior theorists, such as Cini, Gatto and Touschek, soon after 
 their return from Kiev,  
and continued 
after Panofsky's seminar in Frascati on October 26, 1959.\footnote{
 Early interest in Frascati  about  storing electrons in a magnetic ring, appears in some of Carlo Bernardini's  personal papers. 
Carlo   Bernardini,   a friend and  a chronicler of Touschek's life, had participated to the construction of the Frascati synchrotron and later became 
 a close collaborator of Bruno Touschek in the AdA project. 
 He 
 recalls that in his notes, dated February 24, 1958, (see Bernardini's personal papers at the Archive of the Physics Department in Rome, Sapienza University) he  had a sketch of a circular magnetic bottle which he had called the ``Storion'', a device for accumulating electron beams to be circulated in a ring, which points to ongoing 
 discussions about 
 the storage problem at least since that time.
He also recalls how towards the end of the 1950s, the accelerator community was particularly interested in high-energy proton collisions, to be achieved, for example, using the Fixed-Field Alternating-Gradient (FFAG) concept, promoted in particular by Donald W. Kerst, technical director of the Midwest Universities Research Association (MURA). However, as Bernardini pointed out, ``attention was apparently focused more on the kinematic advantage of colliding beams, than on the new physics to be learned from them,'' while Touschek ``considered the kinematics as rather obvious; to him the possible physics to be learned from colliding particles was far more significant.'' In Touschek's view, a physical system could be adequately characterized by studying its `geometry' and its `dynamics'. Its geometry is observable by employing space-like photons as in electron-proton scattering experiments, exactly what Robert Hofstadter was doing at Stanford. Like Touschek, he was not interested in the breakdown of QED. He had visited Frascati, and his experiments had a strong influence on Touschek, who, on the other hand, remarked that ``no one had as yet observed the dynamics; for that, one needed to produce time-like photons at sufficiently large energy [\dots].'' In fact, in contrast to those who sought to create electron-electron (tests of QED), or proton-proton collisions, Touschek saw electron-positron collisions  as a means of exciting the quantum vacuum to a much cleaner initial state \cite[pp. 161-162]{Bernardini:2004aa}.}

  \begin{figure}
  \vspace{-3cm}
\includegraphics[scale=0.27]{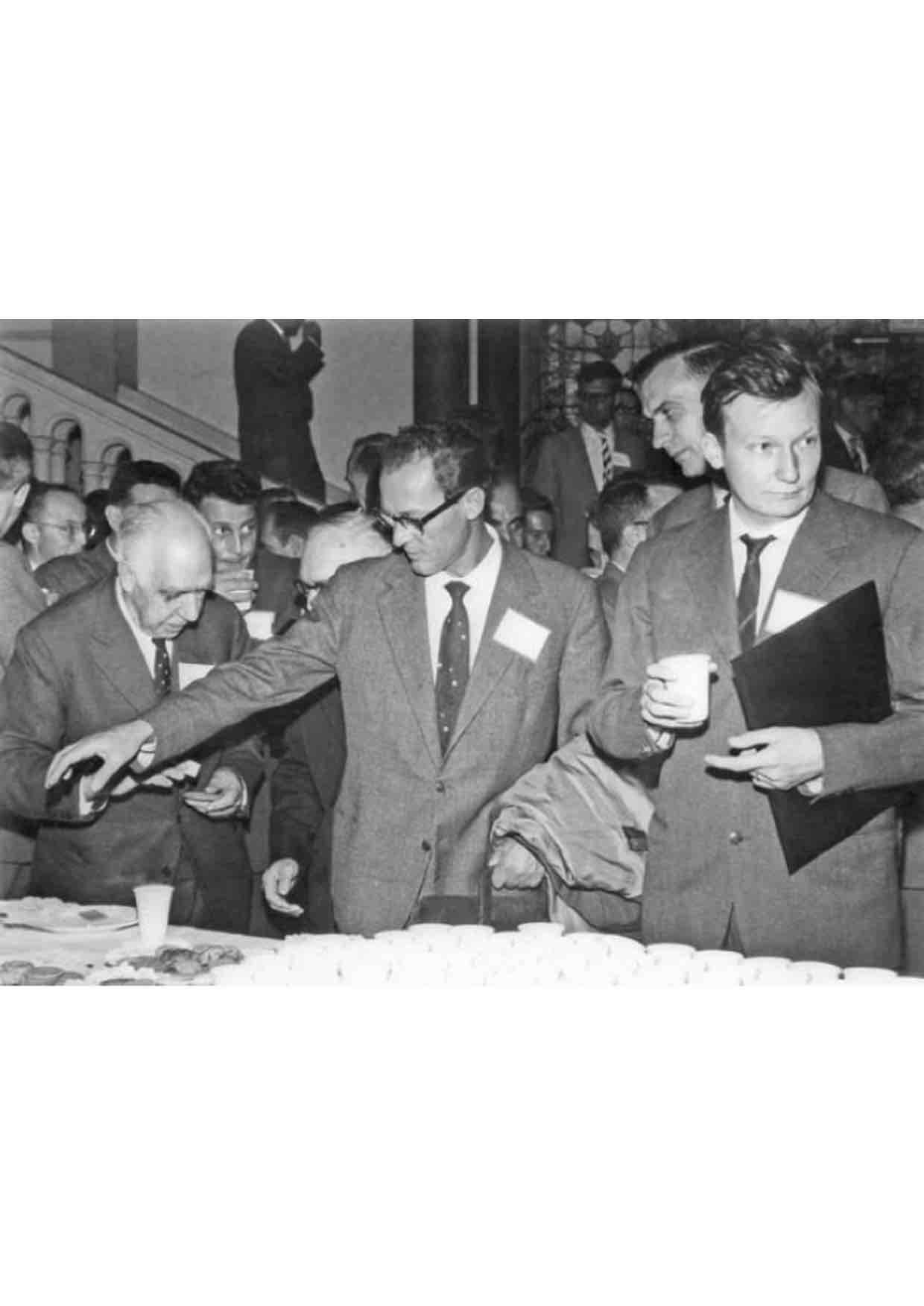}
\vspace{-4cm}
\caption{From left: Niels Bohr, Marcello Conversi and Carlo Rubbia, 1957,  \copyright Cern Distribution Services, all rights reserved.} 
\label{fig:BohrConversiRubbia}
\end{figure}

In Rome, theoretical physics work started immediately after Panofsky's seminar and  a fully realistic project   was approved less than 5 months after Touschek advanced the idea at Panofsky' seminar. Still,  the sequence of events, as outlined before, leaves  a question: which came first, the theoretical ideas or the experimental proposal.  And what was Touschek's involved in,   while   Cabibbo, Calogero, Gatto  and Brown were doing and checking  the calculations, writing the  papers, having them  typed and sent to PRL?  Cabibbo, in an interview with Luisa Bonolis, remembers that feverish period, saying ``\dots calcolavamo e calcolavamo \dots". 
Touschek's letters do not reveal much about this period, except that Bruno, in a  letter to his father, {dated January 16, 1960},  acknowledges 
a proposal 
  to create a theoretical physics group to support the newly operating electron synchrotron in Frascati.  It is noteworthy  that on October 29, 1959, he had already received  official authorization  for regular access to the Frascati Laboratories \cite{Valente:2007}.  He had  also started to teach a new course on {\it Statistical Mechanics} at the University of Rome, which is still remembered by many of the students who attended it {\cite{Margaritondo:2021,Dicastro:2023}}, and which can be considered the initiator of the intense and successful line of research later developed in  the Physics Institute of Rome. Was that all he did? From the fact that, on March 7 he presented a detailed proposal for the building of AdA,  with many details specified in the Storage Ring notebook, which he had  started  on February 18, it can be argued that while  his former students, Cabibbo and Calogero,  and his friend Gatto were doing the calculations on the physics, he was absorbed in the task of devising if the idea would be feasible, and developing  the ways  to make it work.  
  To do this, as he says, he had to bring together ``all he  had thought about it and much which others had suggested to me".\footnote{These words appear in a  typewritten   manuscript entitled ``On the storage ring", undated but probably prepared after  February 17th, 1960, from    Bruno Touschek papers in  \RSUPD. In this document, Touschek also remembers to have first heard the suggestion to use crossed beams from   \W, during the war. }

 The premises were clear to Touschek, who was `attracted by the perfection and the beauty of a machine capable of producing ``an excited vacuum" \cite{Bernardini:2004aa} and had an absolute faith in the CPT theorem, as Carlo Rubbia, seen in  Fig.~\ref{fig:BohrConversiRubbia} with Marcello Conversi, 
 later remembered him saying  in \cite[pp. 57-58]{Rubbia:2004aa}: ``I have met for the first time Bruno when I was a student at the Scuola Normale di Pisa [where he was teaching.] . . . Then I spent a few years in the United States. On my return to Italy, I moved to the University of Rome, where in the meanwhile Marcello  [Conversi], had become professor.  I met then often again Bruno in the wide and relatively dark corridors of the Physics Department . . . I still remember him saying with a very loud voice resonating in the corridors ‘the positron and the electron must collide because of the CPT theorem!’ ”.\footnote{Carlo Rubbia  had  graduated in physics from the SNS in Pisa, with  Marcello Conversi as thesis advisor.}

 This is how the extraordinary 
  achievement of AdA was born in those few months,  from October 1959 to March 1960, between  the Institute of Physics of the University of Rome 
  and the Frascati National Laboratories
  which were to  become the birthplace of electron-positron storage rings. 

\begin{figure}
    \includegraphics[scale=0.1]{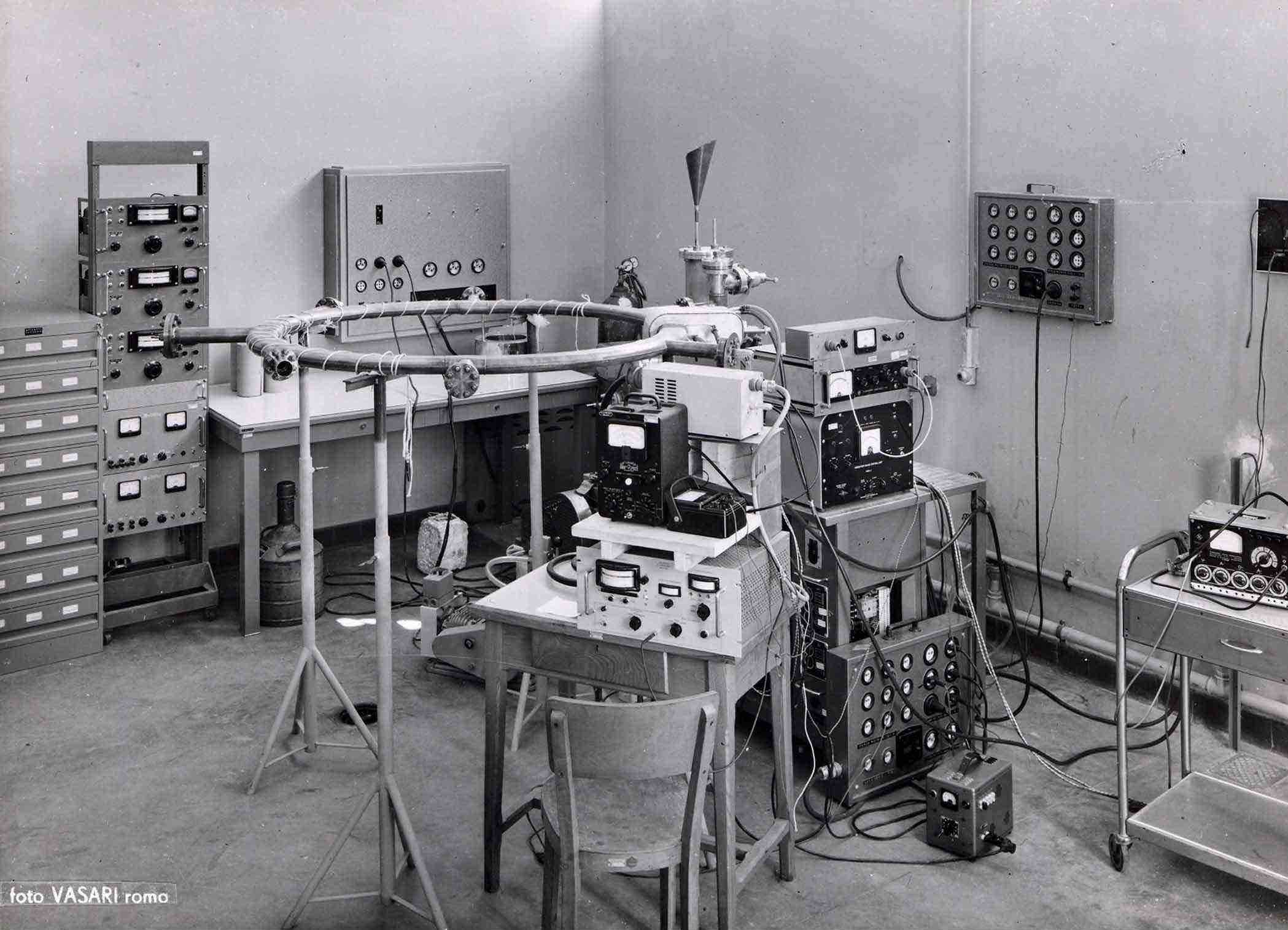}
    \hspace{0.5cm}
       \includegraphics[scale=0.18]{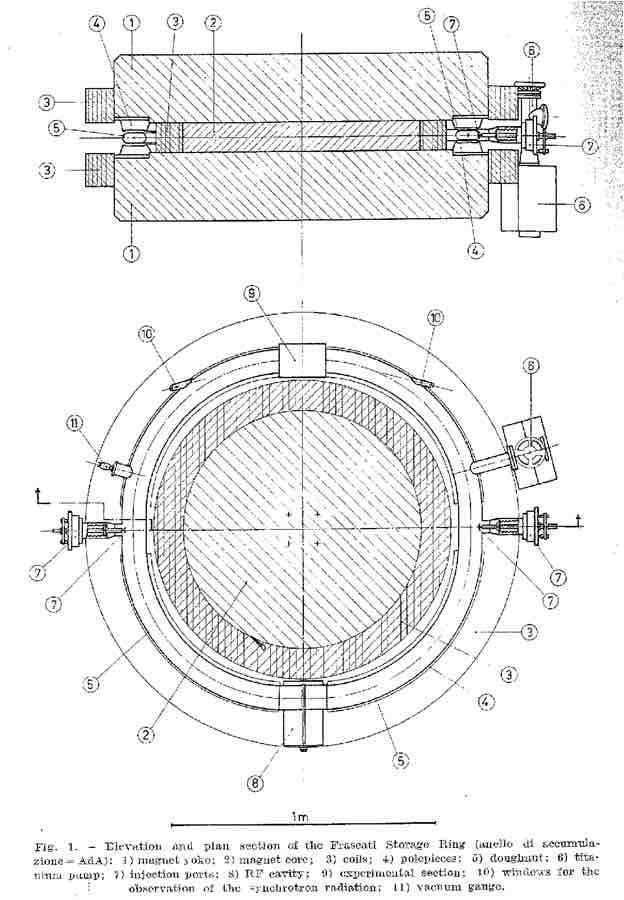}
    \caption{(Left panel) AdA's doughnut under construction at Frascati National Laboratories; (right panel) side and top view of AdA's scheme; 1960 \copyright \ INFN-LNF images}
    \label{fig:AdAscheme}
    \end{figure}

AdA's first Act was over, and  Act II began 
when the INFN Board of Directors 
approved funding for AdA's construction at  the end of March,
 {placing} 
  Touschek in charge of  the project. 
Bruno followed the day-by-day work in Frascati, in the machine shop, and through  weekly trips from Frascati  to the {city} of Terni, in central Italy, where  AdA's  magnet was being manufactured under the specifications  of the Laboratories, namely  of Giancarlo Sacerdoti and Giorgio Ghigo, see  Fig.~\ref{fig:AdAscheme} for AdA's scheme and the doughnut.
    At the same time, Gatto started a long paper with Cabibbo on all the processes that they could think  worthy to be measured in electron-positron collisions. This article was completed in a year and was published in {\it The Physical Review}. It became known among the Frascati physicists as {\it la Bibbia}, (the Bible,  in English),   since it contained all and more of the physics that could be done with electron-positron rings \cite{Cabibbo:1961sz}.  Less than 8 months later, AdA's magnet was delivered, and on  February 27, 1961, electrons began circulating in the ring:  AdA's first Act was over, and  Act II could begin. 

There was great enthusiasm in Frascati 
when AdA started to work, 
and the proposal to build a much higher   energy  collider, ADONE, was prepared \cite{Amman:1961} and presented to INFN \cite{Amman:1961relazione}. This proposal had been partly anticipated in a handwritten note by Touschek dated November 9, 1960, and
an internal LNF report,  completed by Gatto's theoretical introduction and practical details on ADONE's technical specification  and projected cost (1.5 Billion Italian Lire).

The final part of AdA's story developed between Rome, Frascati and the \LAL, in Orsay, where AdA was taken  in July  1961 \cite{Bonolis:2018gpn,Pancheri:2018xdl,Haissinski:2023ene}, to improve its luminosity performance by means of the powerful linear accelerator built in the Orsay Laboratory in France,  by \PM \ \cite{Marin:2009} and his collaborators.  

The final success \cite{Bernardini:1964lqa}  was achieved by  the observation of the process $e^+e^- \rightarrow  e^+e^- +\gamma$,  whose theoretical calculation \cite{Altarelli:1964aa} brought forward a new generation 
 of physicists, the class that  had entered the University of Rome in the fall of 1959 and started to graduate in November 1963.

\subsection{Making AdA work}
\label{ssec:AdAmaking}
Here we  outline the   contribution of Gatto and his students to the theoretical physics background needed to prove that AdA had observed collisions between electrons and positrons, although its luminosity was and remained too low to observe annihilation with the creation of new particles. 

After the burst of enthusiasm over the first electrons or positrons that circulated in the ring and emitted visible radiation, the task was to prove that collisions and subsequent annihilation  into new particles  had really occurred. This turned out  to be rather difficult.

\begin{figure}
\centering
 \includegraphics[scale=0.3285]{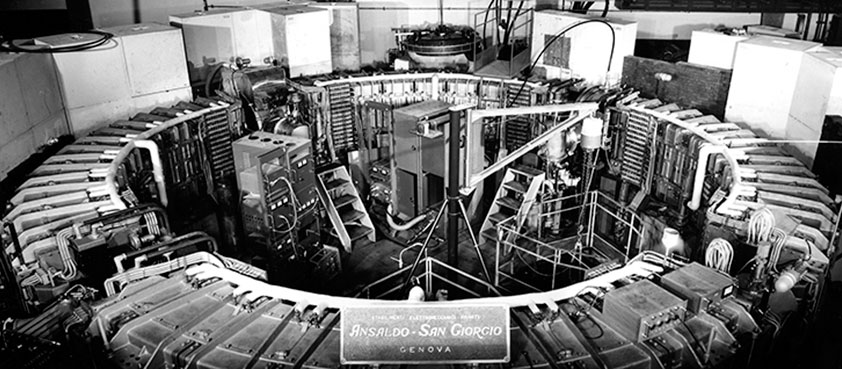}
\includegraphics[scale=0.36]{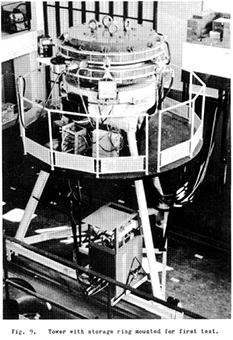}
\caption{At left, AdA, seen in the back, at center,  positioned next to the Frascati synchrotron, which acted as injector, \copyright \ INFN-LNF; at right,  photograph of AdA   presented by Touschek at the CERN conference, June 1961 from \cite{Bell:1961gi}.}
\label{fig:Ada-synchro}
\end{figure}

 AdA was fed with electrons and positrons from  the conversion in the doughnut of a photon produced by the synchrotron, next to which AdA was positioned, as shown in Fig,~\ref{fig:Ada-synchro}. But  no new particles were  seen to be produced  and it soon became clear  from Gatto  and Cabibbo's calculations that the cross section for the production of new particles required a much {higher}  {\it luminosity} than what AdA 
 could attain. Still the machine had been built and electrons or positrons were circulating in it, and this result was important even if the electron synchrotron could not feed  a sufficiently large number  of photons
   for  physics to be observed. The solution came from abroad, after Bruno and Raoul had presented their work at the CERN conference in Geneva in June 1961 \cite{Bell:1961gi}. 
The conference featured three successive talks on colliding beams, by Richter, Touschek and Gatto. Richter presented the ongoing electron-electron collider project at Stanford, Touschek presented AdA and ADONE, whose proposal had also been signed by Gatto and which was about to be approved by the Italian government agencies,  and  Gatto presented  his extensive work with Cabibbo on electron-positron colliding beam experiments,  {\it la Bibbia}.
  More than anything else, this sequence of the three speakers and the content of their talks show the rise of the idea of electron-positron colliders  and the  collaboration between Gatto and Touschek
 in making electron-positron physics  a mainstream subject for both experimenters and theorists.
\subsubsection{AdA and the \LAL \ in  Orsay}
 A crucial consequence of Touschek and Gatto's talks  at CERN, was the interest they arose in the French physicists who had attended the conference, and who mentioned to their colleagues at Orsay that  in Frascati  {\it \dots  ils se passaient des choses qui  intriguaient les esprits} \cite{Marin:2009}, namely very  {intriguing}   things were 
 {happening}. This interest materialized a year later in the transfer of AdA to the  \LAL, where it remained for the next three years and where the feasibility of electron-positron colliders was proved in a Franco-Italian collaboration.\footnote{For  interviews to the protagonists of   the Franco-Italian collaboration, see the 2013  docu-film 
\href{https://www.lnf.infn.it/edu/materiale/video/AdA_in_Orsay.mp4}
{\it Touschek with AdA in Orsay}. }
 
  AdA's move to the \LAL \ in Orsay had started with a visit by Pierre Marin and George Charpak in July 1961, and negotiations began   in September during a conference in Aix-on-Provence \cite{Cremieu-Alcan:1962oky}. Gatto contributed to the success of this first step by giving a talk on  the experimental possibilities with colliding beams of  electrons  and positrons  \cite{Gatto:1962fka}.
 
{AdA's transfer  to Orsay was an almost incredible adventure, as it had to  cross the Alps and the Franco-Italian customs,  with  the vacuum system continuously working, to maintain  the exceptional vacuum conditions in the doughnut, which would take months to produce anew. After AdA's arrival in Orsay, } 
 in  July  1962, it was placed next to the linear accelerator, the LINAC. 
 
\begin{figure}
\includegraphics[scale=0.1]{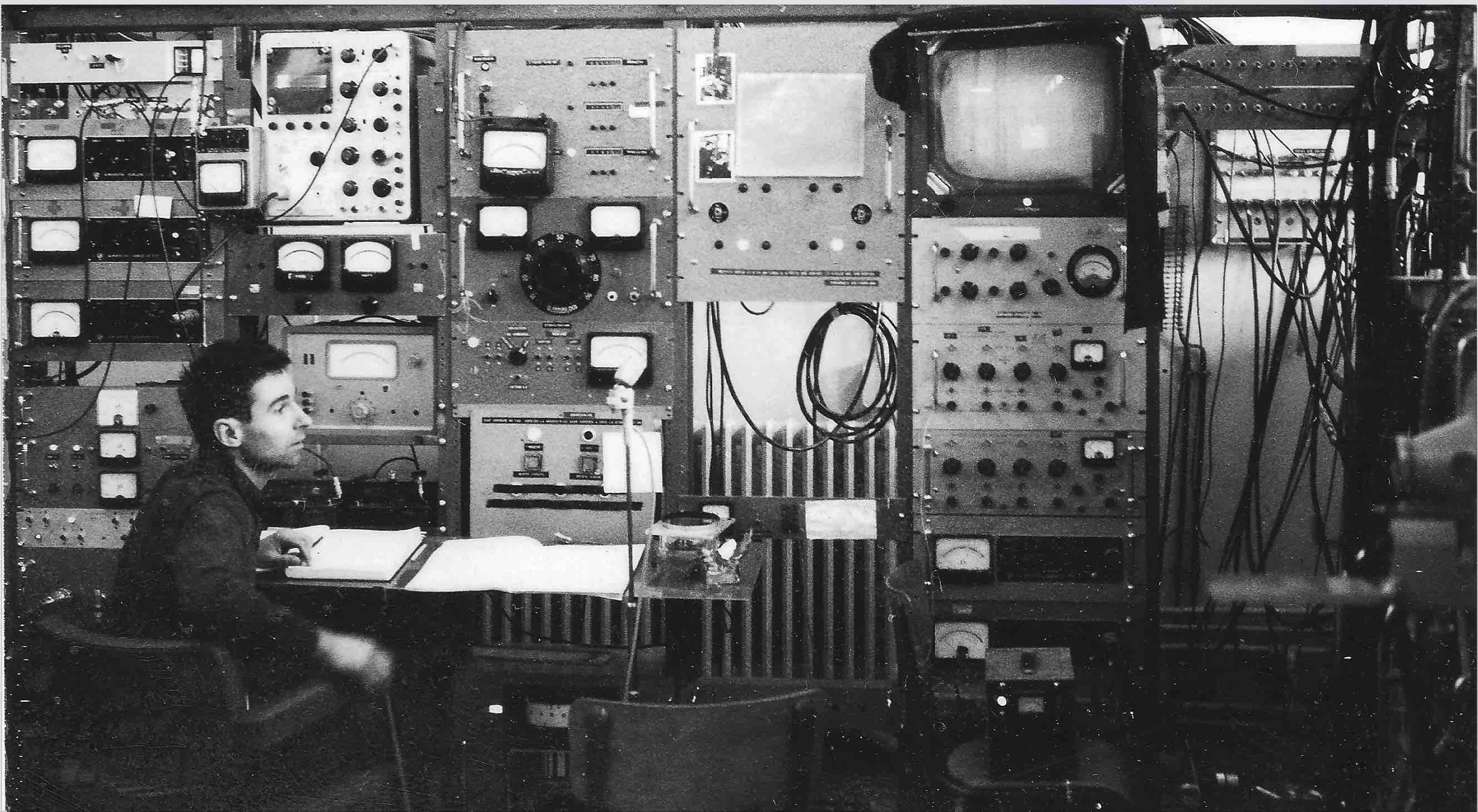}
 \includegraphics[scale=0.225]{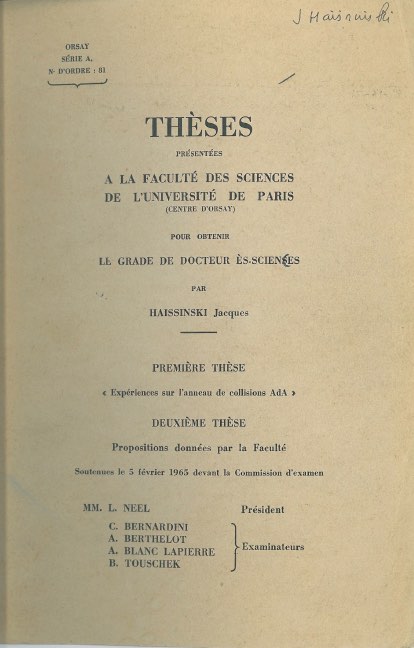}
  \caption{Jacques Ha\"issinski in 1963, in AdA control room in  Orsay, and the front page of his 1965 Doctoral Thesis,  courtesy of Mme. Y. Ha\"issinski, all rights reserved.}
  \label{fig:JH-these}
 \end{figure}

 In the fall, the actual experimentation started, with \JH \ joining the French team, Fig.~\ref{fig:JH-these},\footnote{\JH's Doctoral Thesis is a complete and  unique reference document  for the history of how AdA was made to work during the 3 years it remained at 
\LAL. \JH\ (1941-2024) was also instrumental in documenting the exchanges between Frascati and Orsay, with letters and other historical material used in \cite{Bonolis:2018gpn,Pancheri:2018xdl}.}    after  \FL \ left to work in a different field.\footnote{\FL \  had  participated to the preparation for AdA's transfer and to the initial installment in Orsay.} 
 Thus started three months of intense work, that culminated in February  with the discovery of a crucial phenomenon, 
 later to be known as the  {\it Touschek effect}. The effect established the existence of intrabeam
  M\o eller scattering 
  between electrons
   (or positrons in the positron beam), an effect which {\it de facto} limited the lifetime of the beams, but could be calculated and the energy dependence clarified \cite{Bernardini:1963sc}. The effect diminished with increasing energy, but at AdA's energies, Gatto and Cabibbo's calculations \cite{Cabibbo:1961sz} showed that no annihilation into new particles could be observed at AdA,  contrary to what  Touschek had hoped to see, and indicated 
in the Storage  Ring notebook,  as  can be seen in Fig.~\ref{fig:SRpag2}.  Thus, a different process had to be envisioned, which could prove that collisions had taken place.  Touschek did some {fast}
thinking, and order of magnitude calculations, figuring  out that single  {\it bremsstrahlung}, photon emission, could be the one to work, as shown  in some of his notes.   The month after  the publication of the {\it Touschek effect}, he  presented  this idea at a Brookhaven conference \cite{Touschek:1963zz}. 
But an  order-of-magnitude calculation was not sufficient to actually prove the collision. 
 To proceed further, Gatto's help was needed. 

 \subsubsection{The Rome thesis calculation which opened the way to electron-positron colliders}
 \label{sssec:thesis}
 
In  the spring of 1963, Gatto  started to divide  his time between Rome,  Florence where he had been called to the Chair of Theoretical Physics \cite{Casalbuoni:2021}, 
and the Frascati Laboratories, 
{where he  had an office} 
and from where he followed his Rome students.  
\begin{figure}
 \includegraphics[scale=0.195]{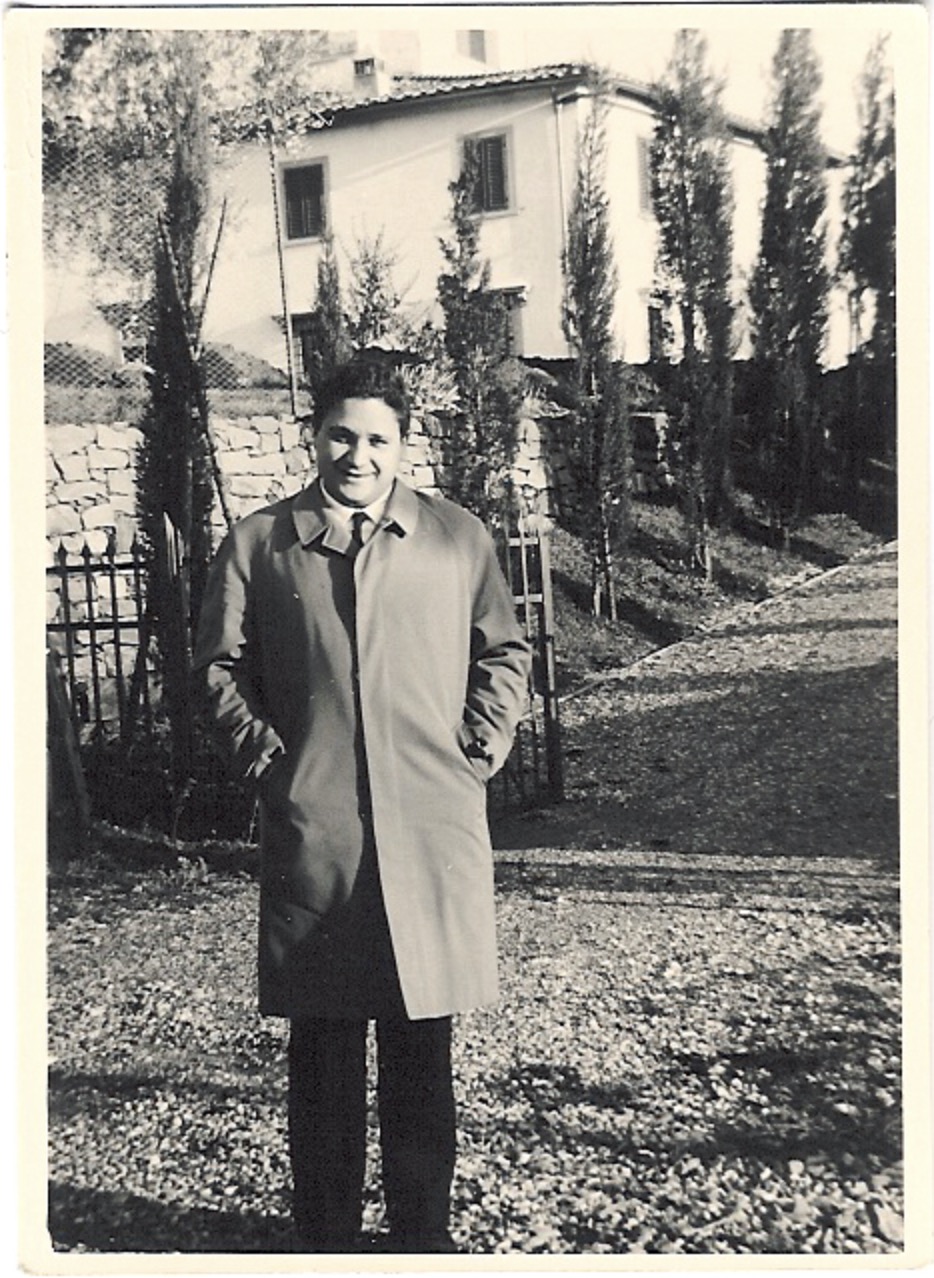}
\includegraphics[scale=0.385]{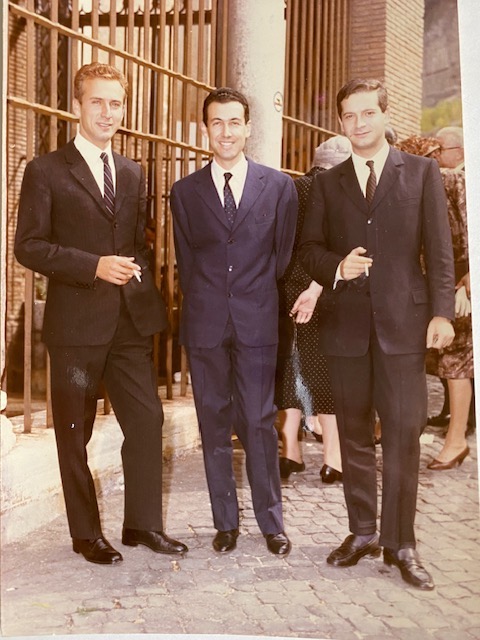}
\caption{Raoul Gatto in Arcetri, on  the Fiesole hills near Florence, when he started gathering around him an extraordinary  school in theoretical physics, courtesy of Gatto's family; in the right panel,  Luciano Maiani, first from left, and Giuliano Preparata, at right, who both joined Gatto's group in Florence after graduation,  together with their friend and university colleague Giorgio Capon, who  would 
 become  one of the protagonists of electron-positron physics at  the \LNF, 
  from ADONE to DAPHNE, photograph courtesy of Emilia Campochiaro Preparata from \cite{Preparata:2020}.}
\label{fig:maianicaponpreparata}
\end{figure}
An  affectionate and  vivid description  during this period in Frascati is given by Giuliano Preparata, 
 who would  graduate in physics  with Gatto, 
in November 1964, from the University of Rome.
Giuliano writes \cite{Preparata:2020}: \\
 {\it 
 On  a clear day in May 1963, 
I went up,  from the University of Rome towards the hills of the Castelli Romani, 
to  Frascati and  to the Laboratories, where  Professor Gatto worked, and with whom 
 I had an appointment. Although quite young -- he must   have been 
just over thirty years old -- Gatto was considered one of the leading theoretical physicists
in the world. The meeting took place in a very informal way. Gatto was a big boy
a little overweight, with an open and friendly face, perhaps a little shy;
I immediately liked him very much. [\dots] 
Once our conversation was over, he offered to take me back to Rome, where
he had to go for the evening.}\footnote{Translated from the original Italian version.}
 
 In Fig.~\ref{fig:maianicaponpreparata} we see Gatto  in Florence, around 1963, and, at right, three of the  brilliant students of those years, Luciano Maiani, Giorgio Capon,  later an experimental physicist at LNF, and Gatto's  student and future collaborator Giuliano Preparata.
 
 The Touschek effect had been  observed in  Orsay during AdA runs in January and February 1963.   During the following months, Touschek mulled  about  the  possibility of  using  the bremsstrahlung process as a proof-of-principle that electron-positron collisions were feasible, and, when he was in Rome,  he
 discussed with  Gatto the need to do the exact calculation, which nobody had yet done. 
 
 The process of photon emission in electron-positron annihilation is  a radiative effect,  belonging to the general topic of radiative corrections to a given process,   extensively discussed in  a  text book, by Jauch and Rohrlich \cite{Jauch:1955}. When Franco Buccella 
  approached Gatto to graduate under his supervision, and asked for a thesis subject, Gatto told him to read the  book by Jauch and Rohrlich in preparation for a thesis,  and then come back to him.

 Some time later,  another brilliant
  student of that year, Guido Altarelli,  went to see Bruno Touschek for possible supervision  and  a thesis subject. 
 The interview between Guido and Touschek did not  work out \cite{Greco:2008,Pancheri:2022}, and Altarelli approached Gatto, as a possible tutor. Thus
 Guido and  Franco joined  forces under Gatto's supervision and Touschek's encouragement,  and through summer 1963 calculated the  cross-section for ``Single photon emission in high-energy 
 $e^+e^-$ collisions", i.e,  the  process
  \begin{equation}
 e^+e^- \rightarrow e^+e^- \gamma \label{eq:1}
 \end{equation}
This calculation has both numerical and theoretical complexity, and 
  their work encountered some  difficulties. 
  
\begin{figure}
 \includegraphics[scale=0.0775]{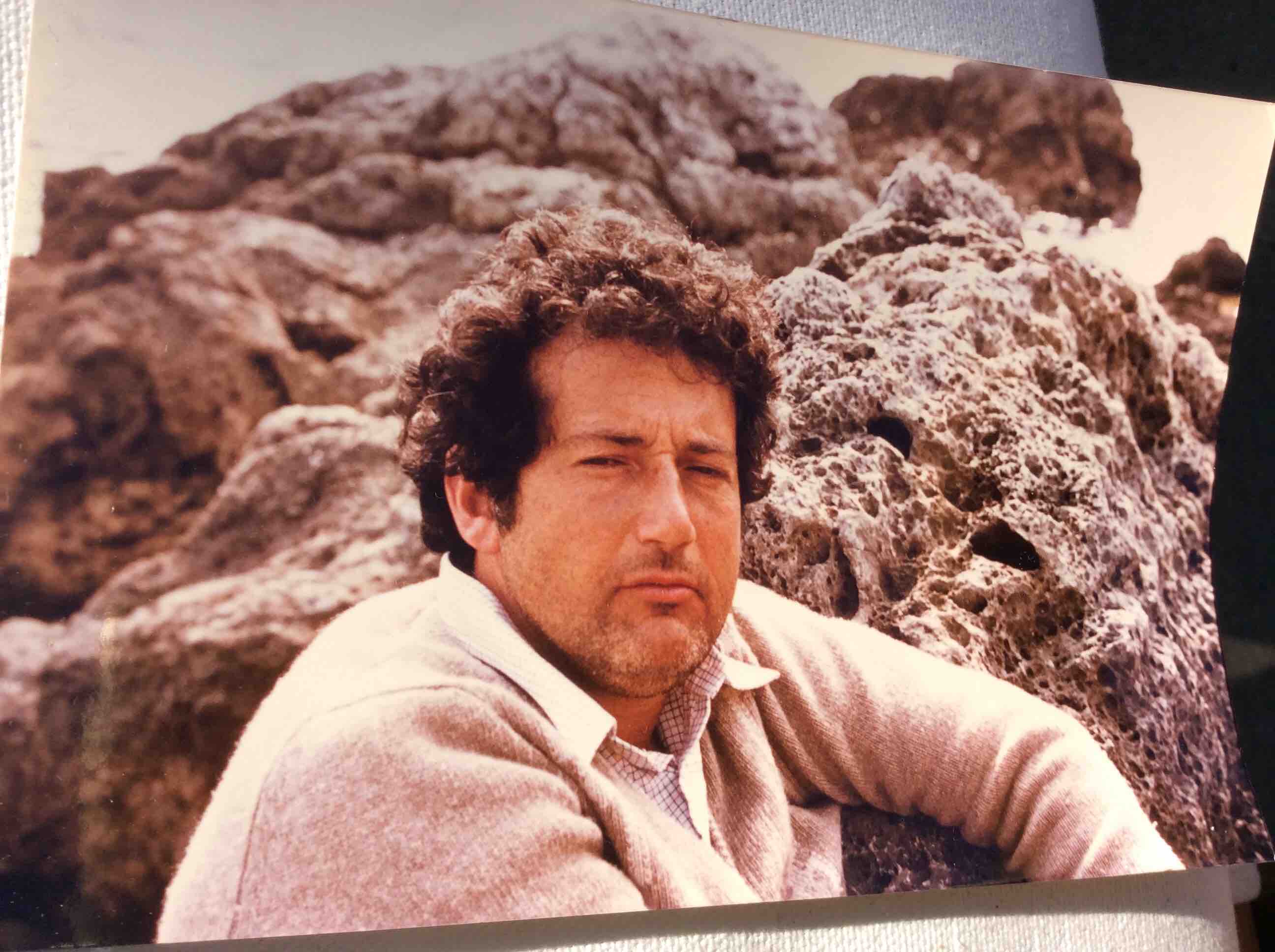}
 \includegraphics[scale=0.135]{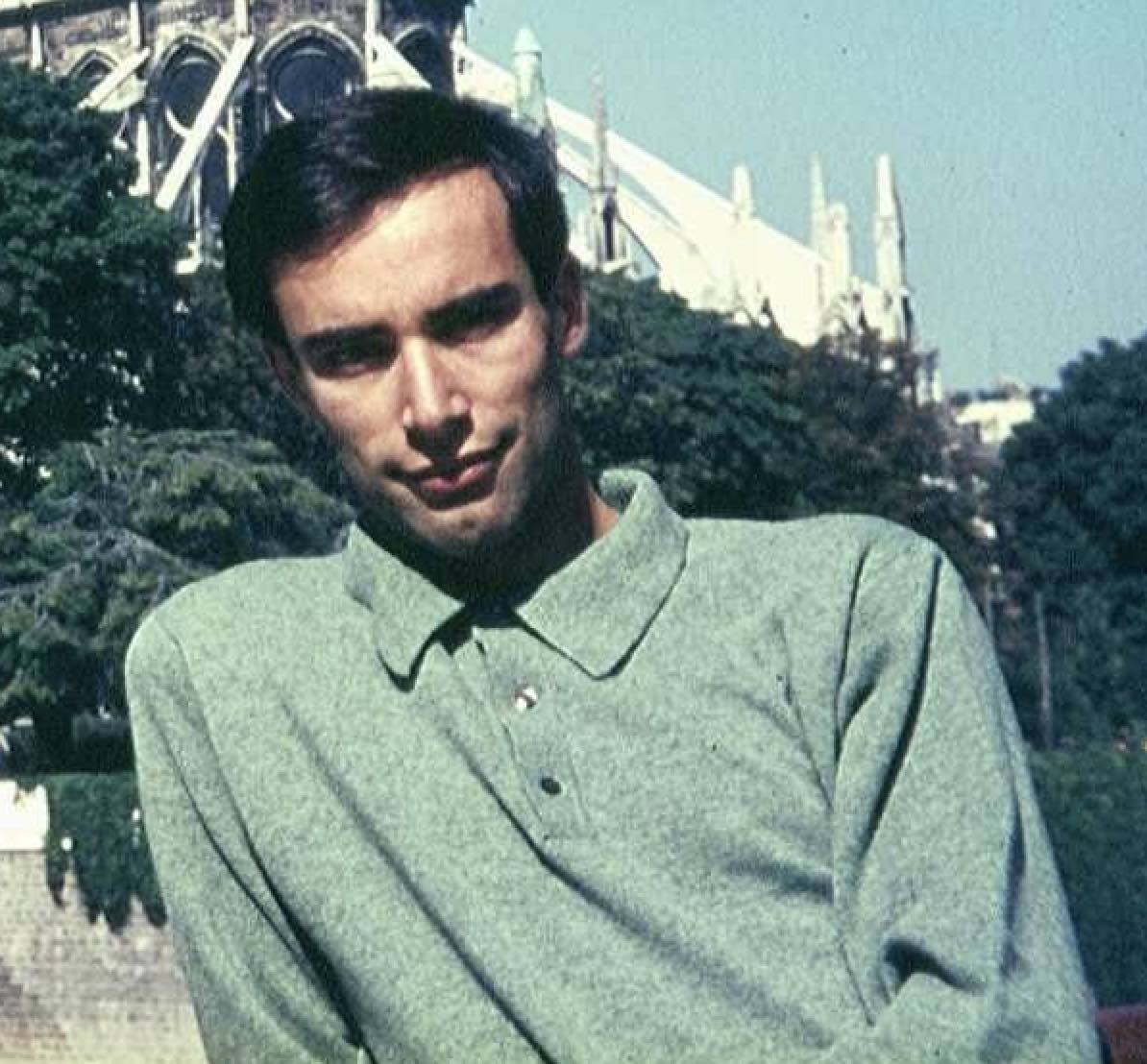}
\caption{Franco Buccella at left, in the 70s,  and Guido Altarelli in Paris around 1966, personal photographs, all rights reserved.}
\label{fig:buccella}
\end{figure}

   After an initial phase in Rome, the two students,  Fig.~\ref{fig:buccella}, decided to use the Frascati computer for the required numerical integration. At the Laboratories,  
  they  oscillated between   the computer room and the young researchers' office,   discussing  their work with Gatto and his assistant Umberto Mosco, or with Touschek and Carlo Bernardini.

In his contribution to the BTML, Gatto describes the special atmosphere pervading  Frascati in those days. 
    A spirit of cooperation and camaraderie prevailed during the ten years of construction of the Synchroton and  AdA, from 1957 to 1964,  and until 1967, when ADONE was under construction. Short   satirical poems about the director were circulated to everybody's merriment, letting off  the pressure the intense work entailed.  Carlo Bernardini, in particular, was famous for his prolific production of poems of this kind. 
    Elaborate pranks were also prepared and realized  by both the researchers and the technicians.
      Gatto and Touschek's voice were easy to imitate and their  students  could  be the occasional  targets of practical jokes.
 \vskip 0.2cm
 \noindent\fbox{ \parbox{\textwidth}{\footnotesize
Physically,  Touschek and Gatto were very different, one lean and nervous, the other quiet and soft. They also had extremely unique and recognisable ways of speaking. In Touschek’s case,  his very good spoken Italian was  accompanied by a strong Austrian accent characterised by a rounded pronunciation of the letter ``r”, which made his talking easily recognizable and often imitated when he was mentioned. Unforgettable for one of the authors of this article is his sentence ``Signorrina, dobbiamo guadagnarrci il pane e  il burrro”, {\it Miss, we must earn our bread and butter},  all pronounced with many ``r’s”. This was how he meant that work for the paper on the infrared radiative corrections to electron-positron experiments with ADONE had to be started.

Touschek’s  Austrian accent and loud voice, were  a strong contrast to Gatto’s Sicilian accent and soft carrying.   Gatto’s demeanour was cat-like,  true to his name. His voice was also easy to  recognize and imitate. This characteristic way of speaking was at the heart of a memorable prank, which took place in Frascati in the summer of 1963,  and was still remembered many years later  \cite{Greco:2008}. 

That was the summer, when Guido Altarelli and Franco  Buccella  were struggling to calculate the cross-section for  single {\it bremsstrahlung} in $e^+e^-$ scattering. The calculation was lagging behind, due to its computational difficulties. The calculation of the bremsstrahlung in electron-positron
can not be analytically completely performed and the computer
gave contradictory answers, very large values or even negative
values for the cross-section: the reason is that the subtraction
of very large numbers to give a very small result was beyond 
the precision of the computer.  Their difficulties and the resulting frustration  were shared with the young members of the Frascati theory group, Giovanni, nicknamed {\it Gian}, De Franceschi, who had graduated with Marcello Cini and was  able to imitate Gatto, and Giuseppe, {\it Beppe}, Da Prato, who had graduated in Rome with Ezio Ferrari. 
While the two students were struggling with their computational difficulties, 
De Franceschi  called Altarelli on the phone, pretending to be Gatto, 
and warned him to check what Buccella was doing. Guido, loyally
defending his colleague, anyway told him about the call: luckily, shortly after,  Da Prato disclosed the prank’s author, telling them 
that  Gatto had nothing to do with the call.} }
\vskip 0.5cm

As {mentioned in \cite{Pancheri:2022} and \cite{Buccella:2023},
 after the very inspiring 
discussion with Touschek the two students, Fig.~\ref{fig:buccella}, found the relativistic
approximation, that allowed them to complete in Rome  the analytical
computation, which gave rise to the formula for the cross-section for the emission
of a photon. Their work \cite{Altarelli:1964aa} is cited in the Landau-Lifshitz book
on  relativistic field theory  
 for electron-electron scattering,\footnote{The work is cited in V. B. Berestestki, E.M. Lifshitz, L. P. Pitaevskii, {\it Quantum Electrodynamics. Landau and Lifshitz Course of Theoretical Physics}, Vol. 4, 2nd edition, chapter 10 (Interaction of electrons with photons), paragraph 97 ({\it Electron-electron bremsstrahlung in the ultra-relativistic case}), p. 427.} since
the approximation, which neglects the annihilation diagrams, gives
the same result for electron-positron and electron-electron scattering. The electron-electron scattering calculation was of particular interest to Budker's group at the Laboratory of New Methods of Acceleration of  the Institute of Nuclear Physics in Novosibirsk,  where, in 1963  an  electron-electron collider, VEPP-1,  had been built  and an electron-positron ring, VEPP-2, was in an advanced stage of construction \cite{Marin:2009}. 
 As Baier writes \cite{Baier:2006ye} when the Frascati results began to appear in print,  the Novosibirsk group felt encouraged to continue  their efforts, which remained basically unknown to their Western colleagues until the International Conference on High Energy Accelerators, which was held in  Dubna in August 1963 \cite{Kolomenskij:1963wlo}
  and the post-conference  trip to the Novosibirsk laboratory \cite{Marin:2009}. 
 

Altarelli and Buccella graduated in November 1963.  
Their  work
showed compatibility between AdA's output and the theoretical predictions, once other issues, such as volume of interaction in the collision,  had been clarified.
 The article they  wrote reporting the result of their calculations 
 was prepared and sent for publication to the {\it Nuovo Cimento}, where it was received on  June 17, 1964, and published in the December issue of that year \cite{Altarelli:1964aa}.   
  In AdA's final paper \cite{Bernardini:1964lqa} this work  is cited as   ``G. Altarelli and F. Buccella: {\it Thesis } (unpublished)", although it had been received at the Nuovo Cimento's office  
a month before  AdA submitted  its final results},\footnote{The submission date of AdA's fourth and final article is July 16, 1964,  with publication date December 16, 1964.} with publication in the same Vol. 34 of the journal. This article is a landmark and {\it per se} widely known, but the timing and importance of this work in connection with AdA's success does not always appear in the history of the first electron-positron colliders and Gatto's role in the last step towards  establishing the proof of principle of the feasibility of electron-positron storage rings is blurred.

  \section{The  development of the Frascati theory group}
  \label{sec:lnftheory}
Altarelli and  Buccella, {whose work was pivotal to demonstrate AdA's feasibility,}   belonged to a group of students who had enrolled in the 1959-60 academic year, a year which saw the number of physics students almost doubled compared to the previous one. 
This class had witnessed the launch of the Sputnik satellite, but had also listened to Giorgio Salvini's lectures on physics broadcast by the public television company, RAI-TV, {and read the  Italian newspapers chronicles about the electron  synchrotron built in Frascati, and put in operation in April 1959.}\footnote{For images of the  synchrotron inauguration see \url{https://w3.lnf.infn.it/multimedia/picture.php?/528/category/8}.}
 This class and the ones which followed as Gatto and Touschek's students or collaborators,   was to play an important role 
in the post-war  renaissance of theoretical physics in Italy. 

{Gatto and Touschek's theoretical physics legacy 
appears jointly 
in a 1984  article \cite{Altarelli:1984pt}, where  the calculation of the  $W-boson$ transverse momentum ($p_t$) calculation proceeds via   the Altarelli-Parisi equations \cite{Altarelli:1977zs}, proposed by Guido Altarelli\footnote{Guido Altarelli (1941-2015) \cite{Maiani:2016} was Professor of Theoretical Physics first at Sapienza University of Rome  and then  at {Roma Tre}, and theory division leader of the CERN Theory group.} and Giorgio Parisi.\footnote{Giorgio Parisi was awarded the 2021 Nobel Prize in Physics "for the discovery of the interplay of disorder and fluctuations in physical systems from atomic to planetary scales", sharing one half of the prize with  Syukuro Manabe and Klaus Hasselmann, who jointly received it ``for the physical modelling of Earth’s climate, quantifying variability and reliably predicting global warming".}  While the  $W-p_t$ paper does not  refer to Touschek's work, some of its  authors   come directly from  Gatto and Touschek's school of theoretical physics: Altarelli had graduated with Gatto,
Mario Greco, a member of the Frascati theory group since 1965 \cite{Greco:2018},    had   formulated  a coherent state approach  to  Touschek's resummation ideas   \cite{Greco:1967zza}, while Keith Ellis was  Altarelli's student, and   Guido Martinelli had graduated with Nicola Cabibbo, who had been one of Bruno Touschek's first students \cite{Martinelli:2023}. 
  Touschek's legacy is  particularly explicit in an earlier  work by  Giorgio Parisi and Roberto Petronzio,\footnote{See Giorgio Parisi's memories of Roberto Petronzio in \cite{Parisi:2017}. }
 both of whom had graduated under Nicola Cabibbo's supervision. Their well known \cite{Parisi:1979se}  paper  about transverse momentum in strong interaction processes acknowledges Touschek's resummation technique in QED  \cite{Etim:1966zz,Etim:1967}.}
 
We will now take a step back in the timeline of events, as we want to  
  focus    on one very relevant aspect of the story: the beginnings of the Frascati theory group and the interplay between Gatto and Touschek in shaping it, in parallel with their influence on theoretical physics developments in the University of Rome.

There is a golden thread linking the \LNF \  to Enrico Fermi, and in particular to the theory group, which passes through  Fermi's colleague and high school  friend Enrico Persico \cite{Amaldi:1979uq}.\footnote{The other important tie between the Frascati theory group and  Fermi is of course Bruno Ferretti, who had been Fermi's youngest assistant and taught  Fermi's theoretical physics course after Fermi   left Europe for the US  in December 1938.  Ferretti was never directly involved with Frascati, only indirectly  through Touschek, who came to Rome attracted by their common theoretical physics interests,  and through the students who graduated with him, among them Gatto and Carlo Bernardini.}
\begin{figure}
\centering
 \includegraphics[scale=0.198]{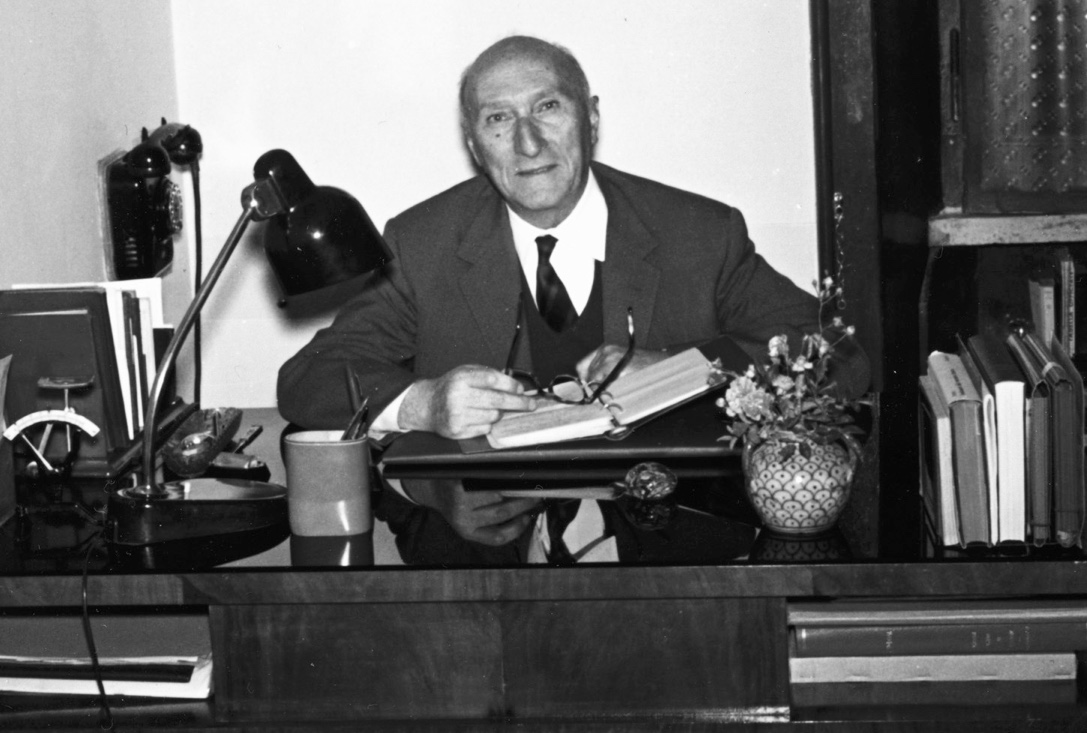}
\includegraphics[scale=0.44]{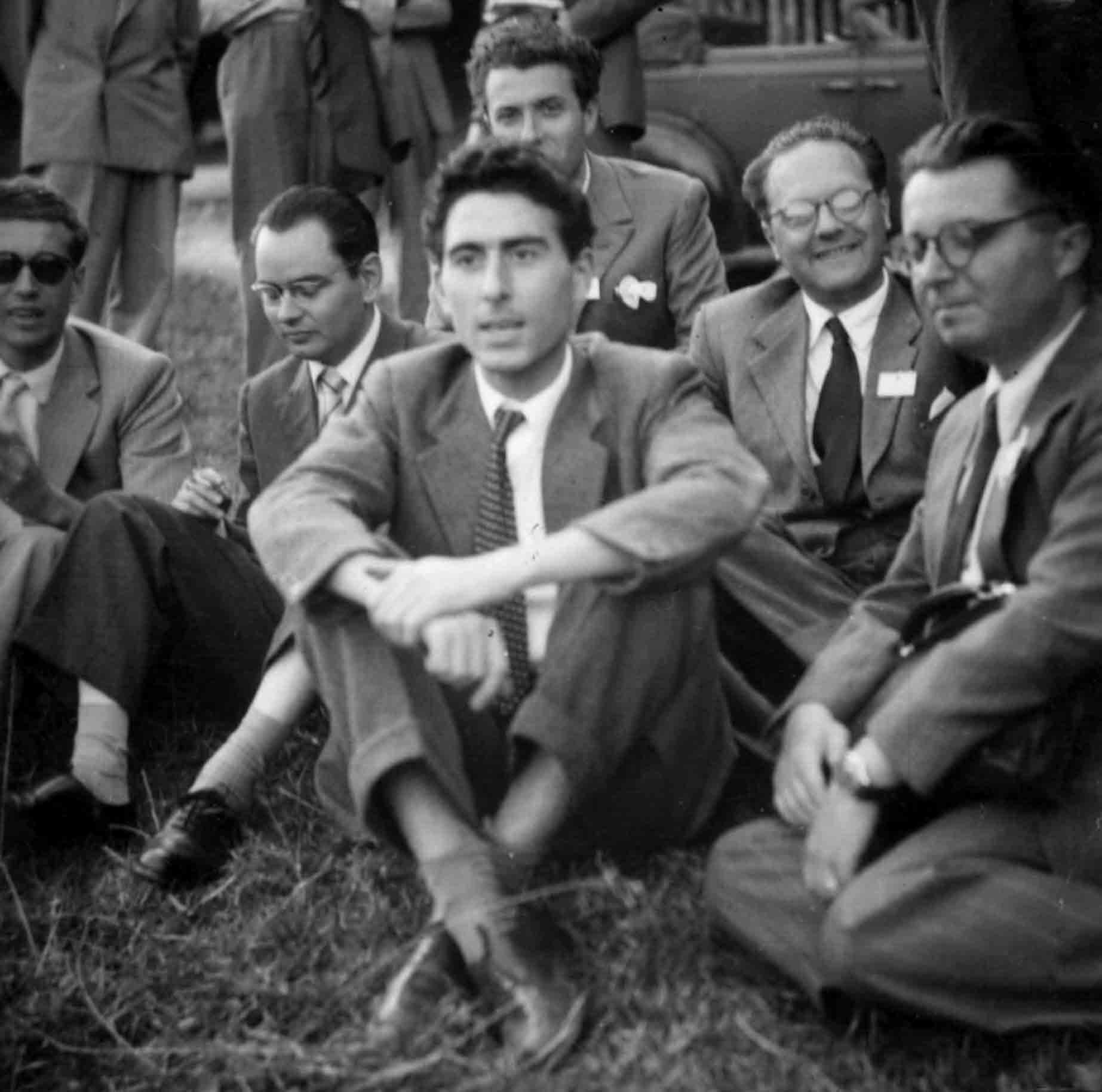}
 \caption{Left: Enrico Persico in the 1950s, courtesy Alessandra Raggi, all rights reserved; (right panel)  Giorgio Salvini (center),  Edoardo Amaldi and Bruno Ferretti (right), at the 1949 Como-Basel International conference on Cosmic Ray Physics, \RSUPD, all rights reserved.}
 \label{fig:persicosalviniamaldiferretti} 
\end{figure}

 Persico, Fig.~\ref{fig:persicosalviniamaldiferretti}, had left Italy in 1947 for Queb\`ec in Canada, returning to Italy in 1951, and was at the University of Rome when    
plans  began in 1953 to build a national laboratory to  host a modern type of particle accelerator, such as a  weak-focusing electron synchrotron, as it then became. The designated laboratory director was Giorgio Salvini \cite{Salvini:2010}, {shown  in Fig.~\ref{fig:persicosalviniamaldiferretti} together with Edoardo Amaldi and Bruno Ferretti.}\footnote{Amaldi and Gilberto Bernardini, heirs of Enrico Fermi and Bruno Rossi, the pioneers of modern physics in Italy, had chosen Giorgio Salvini to lead the project to build the Frascati National Laboratories and the electron synchrotron, after the project's approval by the newly instituted INFN \cite{Battimelli:2001aa}.}
{ Salvini  forged a first-class   team of   both senior and junior scientists,  scouting   all the Italian universities and technical institutes, and hiring the best Italian graduates, } including  theorists in the laboratory staff.
Aware of the need for advanced mathematical calculations and advice, Salvini  asked  Enrico Persico  
 to head the theoretical division of the synchrotron construction, as senior scientist. 
 Among the   younger staff, he hired  Carlo Bernardini, who had just graduated with Bruno Ferretti, 
and would later play an important part in both AdA and the ADONE projects \cite{Bernardini:2004aa,bernardini2006fisica}.
Persico, who  had just returned from Canada, accepted and, starting from 1953, was the author  of  many \href{http://www.lnf.infn.it/sis/preprint/search.php}{Laboratory Notes}, some of them with Carlo Bernardini as co-author.\footnote{Persico authored  the first Laboratory Note. His  last two 
appear in 1959 and in 1962.}

\begin{figure}
\centering
  \includegraphics[scale=0.26]{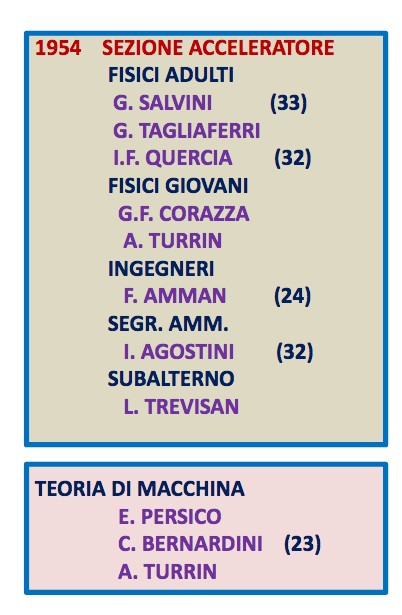}
\includegraphics[scale=0.35 ]{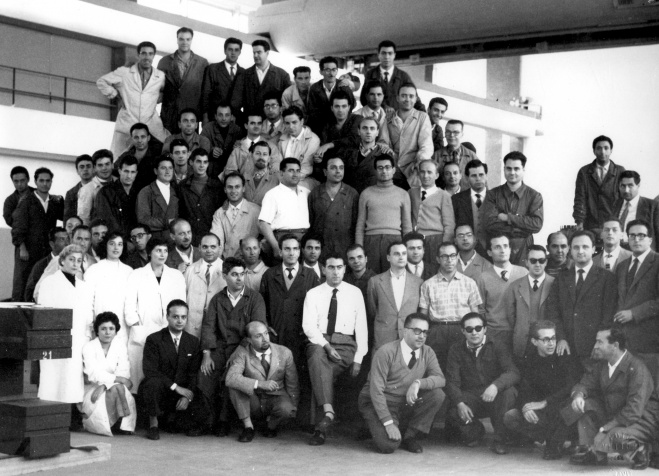}
\caption{At left, listing of the perspective Synchrotron staff  assembled by Salvini in 1954  \cite{Salvini:1962aa}, when the laboratories had not yet been built, with age between brackets,  in a cartoon from a talk by G. Mencuccini,   experimenter  at both the Frascati synchtrotron  and  ADONE;  at right  a photograph of LNF personnel in 1957, during the construction of the synchrotron. One distinguishes Giorgio Salvini, seated, in the middle, Carlo Bernardini and Giorgio Ghigo second and third from right in first row,  Gianfranco Sacerdoti,  third row, first at right. \copyright \ INFN-LNF, all rights reserved.}
\label{fig:synchrostaffAdA}
\end{figure}

The Laboratory grounds were ready in 1957, and the team of scientists, administrators and technicians started  in earnest to assembly  the synchrotron, Fig.~\ref{fig:synchrostaffAdA}.
In 1959, as soon as the synchrotron started operations, Persico asked to be released from his duties at Frascati to resume teaching his course 
on ``Instituzioni di Fisica Teorica",  and to give lectures on  accelerators at the Corso di
Perfezionamento in Fisica Teorica Applicata (1958-59).
The notes were prepared   by Ezio Ferrari and Sergio E.  Segre, with citation 
in  chapter VI of 
 the work by Sands and Touschek \cite{Sands:1953aa}.
  In the IX chapter, which is devoted to the
intersecting rings, there is mention of  O'Neill, Barber, Richter
and Panofsky, about ``A proposed experiment on the limits of Quantum Electrodynamics"  (High Energy Physics laboratory, Stanford University),
Stanford (California) 1958: unpublished report, \cite{Barber:1959vg}.

It was also at this  time that Persico   created with Edoardo Amaldi, Fig.~\ref{fig:persicosalviniamaldiferretti}, the {\it Laboratorio   Gas Ionizzati}   that was to pioneer plasma and fusion research in Italy \cite{Bonolis2022}.
Until then, the efforts in the Laboratories and at the Rome  Physics Institute had focused on commissioning  the accelerator and start the planned experiments. At the same time, it was necessary to   look forward to future projects, which would keep Italian physics  at the forefront, 
and the INFN management began looking for a scientist who could provide the Laboratories with advice and a vision of the future in particle physics. 
 As we have seen in the previous sections, there  were  two obvious choices, Raoul Gatto and Bruno Touschek.
  In 1959, Gatto and Touschek were both on the brink of a change in their lives. Gatto had become a Professor at the University of Cagliari, in Sardinia, and was going to start teaching in the 1960-61 academic year, commuting between Cagliari and Rome. Touschek had lost his mentor in theoretical physics, Wolfgang Pauli, in December 1958, and in May 1959, his maternal aunt Ada, the closest connection to his life before the war and to the mother he had lost.
  Both Raoul and Bruno were  ready for new directions. The search for a head theorist in Frascati  
  began  with the latter.

In the already mentioned  letter of January 16,  1960, to his father, Touschek mentions the possibility 
of taking on greater responsibility in the laboratory in the future. In this letter he even envisages  
 finding  a villa to live in Frascati, noting that, in such a case,  his wife   Elspeth, ne\'e Yonge, would have to learn driving, moving out of  the city at least until their son Francis (born in 1958) would begin school. But Bruno was suspicious of becoming a ``house theorist", from his wartime experience  with \W \ and his betatron,  or from the post-war period in the UK,  when he was involved  in the construction of the Glasgow synchrotron, while also following the construction of  other synchrotrons under  developments in  the UK.
 
  The situation reached a crisis point   in February 1960, when the Scientific Council of the Laboratories called a meeting  to discuss the creation of a 
 theoretical physics group or school in Frascati.\footnote{The establishment of such a group was mentioned as a priority for the short-term evolution of research activities in Frascati Laboratories in \cite{Ageno:1959aa}.} 
 
  During the meeting, 
 Touschek dismissed  such   a need   beyond what the University of Rome could already provide, and  instead suggested  what he called an electron-positron {\it experiment}, as the best way to attract people and generate new ideas. After rejecting the initial idea of transforming the just commissioned synchrotron into a storage ring, the proposal caught the imagination of Salvini and the scientists who were present,  and the  project  went ahead at a smaller scale, 
 with Touschek  appointed to be in charge. Such a commitment was clearly  going to take up most of Touschek's energies and time,  but  theoretical guidance for the Laboratories was still needed, especially since it was evident that more calculations had been and could be done to show the way to electron-positron physics. Gatto stepped easily in. Having  won his Professorship in Cagliari he  could be 
 expected to commute between Cagliari and Rome and follow students who could work on Frascati-related  physics problems. 

After it was decided 
 that a dedicated theoretical physicist was  needed,  Raoul Gatto
 began 
  his  official affiliation with Frascati   sometime in the spring of 1960.\footnote{Gatto's   affiliation with  Frascati  first appears 
 in July 1960 \cite{Cabibbo:1960zza}.  Since  in Gatto's February  paper  with Cabibbo \cite{Cabibbo:1960zza},  there is no Frascati affiliation,  one can 
  easily date Gatto's   appointment  in   Frascati  to have occurred sometime in spring 1960.} In this new capacity, he also
  followed  the developments  of AdA's construction and, in February 1961, co-authored the ADONE proposal with Touschek. Since they could literally   see the  first electrons (or positrons) circulating  in AdA, they understood that the road was open for building a larger and more powerful collider. Gatto then embarked in the longer paper about $e^+e^-$ collisions with Cabibbo, presenting the work at many international conferences, including the one in Aix-en-Provence in September 1961, where  
the possible transfer of AdA to France was  first realistically discussed  
\cite{Bonolis:2018gpn}.

In 1962, Gatto's  commuting between Rome and Cagliari came to an end. Problems with research funding in Sardinia  and relative isolation, led him to leave and move to the University of Florence. 
 Touschek, after the discovery of the {\it Touschek effect} in February 1963, decided  to prove AdA's feasibility through the process in Eq.~ (\ref{eq:1}), and  Raoul   assigned the  calculation  as  thesis to  Altarelli and Buccella. Gradually Gatto began to spend more and more time in Florence, and in the summer  one finds him installed  in Arcetri, as his student Giuliano Preparata remembers him, when he went to see Gatto with the results for the thesis that had been assigned to him \cite{Preparata:2020}:\footnote{Ref. \cite{Preparata:2020} is a revised and expanded edition of the  2002 volume,  posthumosly published by Boringhieri.}\\
 {\it  It was around eleven o'clock on a summer Saturday morning, when on the platform
at the Santa Maria Novella station [in Florence] I saw my teacher again, who welcomed me
with a broad smile. We got into the car towards the Arcetri hill, where
the Physics Institute of the University of Florence is located.
Once we reached our destination, Gatto led me into his large study, with a marvellous 
view of the Florentine hills, he pointed to the blackboard and sat down comfortably
at the desk [to listen to my presentation]. } 

 As Gatto's involvement in Florence grew, his presence in Frascati necessarily diminished, and his role was taken over by Touschek, who had returned from Orsay after  AdA's successful  runs.   It was now time for Touschek to start his active  engagement  in   ADONE and worry about the  extraction of  significant physics from possible experiments. ADONE was a much larger undertaking than AdA, both scientifically and financially,  and Touschek could now see the need for  the creation of a Frascati-based  theoretical physics group. 

\begin{figure}
\centering
\includegraphics[scale=0.45]{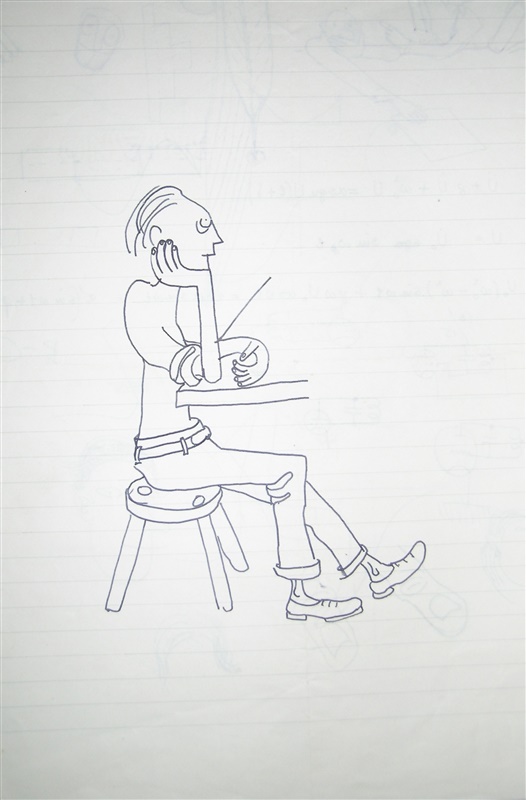}
 \includegraphics[scale=0.4]{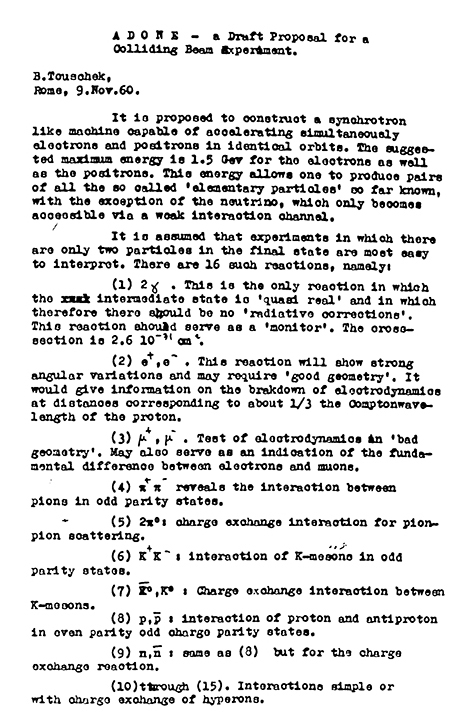}
\caption{(Left) A drawing by Bruno Touschek  \copyright \ Touschek family,  and his early  draft for a  proposal to build ADONE, November 1960, from  \RSUPD.
all rights reserved.}
\label{fig:adoneandproposal}
\end{figure}

Touschek followed the early stages of AdA's construction, in particular travelling to the town of Terni in central Italy, to follow the construction of AdA's magnet together with Giancarlo Sacerdoti, who had  designed the magnet with Giorgio Ghigo, and with whom Bruno shared memories about his past  and the dark years  in Germany \cite[p.93]{Greco:2004}.

As the fall approached,   Bruno felt more and more confident that electron-positron collisions were the road to future discoveries in particle physics, and on 9th  November 1960, in a type-written memo,  Fig.~\ref{fig:adoneandproposal},    outlined the main ideas about the physics which could be studied at  a 3 GeV c.m.  energy electron-positron collider, which could produce actual physics, not just a prototype such as AdA was. He  called it ADONE, {\it la bella macchina}, the beautiful machine, as Amman, its future director, would later call it.
{Then, soon  after electrons (or positrons) circulated in AdA on February 21st, 1961,  Gatto and Touschek  put forward  the  proposal for the construction of  the storage ring ADONE { \cite{Amman:1961,Valente:2007}}, together with Fernando Amman, Carlo  Bernardini 
 and Giorgio Ghigo.
  Amman, Bernardini and Ghigo had all participated in the construction of the Frascati electron synchrotron, which came into operation in April 1959, and 
had  encouraged Touschek in his  proposal to build an electron-positron collider. They were joined in the AdA's project by Gianfranco Corazza, whose great expertise in producing high vacuum in  the synchrotron ring would be indispensable to AdA's success.  Fernando Amman 
would later be in charge of the construction of   ADONE, {which was  to be an electron-positron collider much larger and more powerful than AdA, meant to  explore the creation and properties of all known elementary particles.}

Touschek's proposal for an electron-positron collider at   3 GeV c.m. energy 
was very courageous for the times, and echoes Heisenberg's 1953 suggestion to study nucleon-antinucleon production. Although  bitterly  regretted  in 1974 for being too low 
in energy to discover new states of matter just lying above the 3 GeV threshold, it was proportionate to the laboratories capacities and  the  national priorities.

\section{Physics at ADONE}
\label{sec:ADONE}
ADONE was undoubtedly a marvellous machine, but its accomplishments and contribution to particle physics were partly obfuscated by the political and  labor unrest which characterized its early years. Now, however, 55  years after it began operating  in 1969,  it is easier to  see its legacy in  accelerator,  experimental and theoretical  particle physics. 
A generation of accelerator physicists, who would later contribute much to
science, learned first-hand how to build a large, high-energy
 electron-positron collider such as  ADONE. Among them was Claudio Pellegrini,
 who collaborated with Bruno on one of  his last physics articles and  later moved to the United States \cite{Ferlenghi:1966}.\footnote{Claudio Pellegrini was the recipient of the US  Enrico Fermi Presidential Award, 2015, for 
``pioneering research to advance our understanding of relativistic electron beams and free-electron lasers " and ``contributing 
to the development of the world’s first hard x-ray free electron laser (XFEL), which has given researchers new resources to understand our natural world, enabling the study of new areas of ultrafast x-ray physics, and fields spanning atomic physics, plasma physics, chemistry, biology, and material science."}

When  two beams of electrons and positrons first circulated in ADONE in  1969, much  in  the world of particle physics was transformed. The geometry of the machine, where  the highest c.m. energy in the world was attained in the collision, required novel particle  detection techniques. It was a first, and required  the construction of 360-degrees detectors, a novel concept of experimental apparatus, as needed for colliding beam experiments. 
In this planning for ADONE's detectors, much is due to 
Marcello Conversi, both as a teacher and for his experience.\footnote{Conversi was a member of the so-called $\mu \pi$ experiment. For a complete list of ADONE's experiments and the researchers involved see Ref. ~\cite{Valente:2007}.} 
He had participated in the famous Conversi-Pancini-Piccioni experiment of the 1940s \cite{Conversi:1947aa}, which, according to Louis Alvarez, had inaugurated ``modern particle physics''.\footnote{See Alvarez's Nobel Lecture at \url{https://www.nobelprize.org/prizes/physics/1968/alvarez/lecture/}.}
  Conversi had also invented a novel type of detector, the hodoscope, in the 1950s \cite{Conversi:1955}. The hodoscope was the prototype of the scintillation chambers whose development led to  the experimental successes crucial  for the establishment of the  Standard Model of  Elementary Particles (SMEP).

 Touschek's guiding idea in proposing the experiment on electron-positron annihilation in February 1960 had been to look for the unknown, to find  ``the frequencies on which the vacuum oscillates," as his friend Carlo Bernardini used to say to summarize Bruno's thinking \cite[pp. 62-63]{bernardini2006fisica}.\footnote{Bernardini writes that Touschek's  mental representation could be summarized by asking what characterizes  the dynamic properties of a system, such  as the {\it characteristic frequencies} in acoustics or the colors for a light sources. In Bruno's vision,  if one deposited a certain amount of pure energy in the  simplest system in the world, the ``vacuum", then the vacuum could be excited and its virtual modes of existence would materialize.}
 But  the discovery of the unknown by a large  scientific and financial enterprise such as  ADONE also required planning
 for what  could be measured when the machine would start. Among the processes under consideration, there were precision  studies of
   vector mesons production, $\rho$ and $\omega$, with searches for  their possible recurrences. Attention was also focused on   rare Quantum ElectroDynamics (QED)  processes such  as  two-photon processes   \cite{Brodsky:2005wk}.\footnote{Brodsky's talk on past and future of photon-photon collisions  was presented at a Workshop on photon-photon  physics, organized by Maria Krawczyk (1952-2010) in Kazimirz, Poland.}  
   These studies were planned to be pursued  at  all the $\ep$ \ colliders under constructions after 1963, including ACO in France \cite{Marin:2009}  and VEPP-2 in the USSR \cite{Budker:1963aa}.
     In Paris,  studies of two photon processes were  promoted  at  Coll\`ege de France  by Paul Kessler \cite{Kessler:1962ffa}, following    Calogero and Zemach's  proposal  \cite{Calogero:1960zz}.%
 
In theoretical physics  both in Rome and in Frascati,  the generation of students who had  enrolled in physics since the years 1959-60 was reaching professional and scientific maturity. Among their teachers we reckon  Gatto, Touschek, Cabibbo, Cini, Calogero, De Tollis, Ferrari,   fully engaged in front-line research. In Fig.~\ref{fig:pacini-pancheri-ferrari}, contemporary photos show  De Tollis  and Ferrari, respectively in left and right panels.
 \begin{figure}
\centering
\includegraphics[scale=0.78]{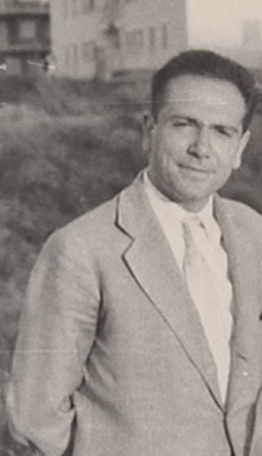}
\includegraphics[scale=0.39]{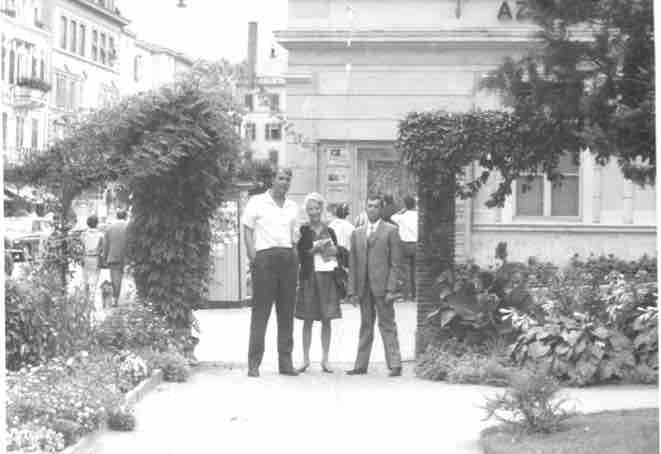}
\caption{Benedetto De Tollis, left panel, probably dated in the mid 1960s, when he supervised the thesis work of M. Greco and G.P.,  {photograph} courtesy of  S. Pacetti; at right Ezio  Ferrari, in grey suit, later to become Professor of theoretical physics in Rome, with G.P. and the astrophysicist Franco Pacini in Meran, summer  1966, personal photograph by G.P., with kind permission from Ferrari's family, {all rights reserved. }}
\label{fig:pacini-pancheri-ferrari}
\end{figure}

QED studies in Rome were pursued by   Enrico Persico's assistant, Benedetto De Tollis, 
who  started a {long series of studies on} processes such as $\gamma \gamma \rightarrow \gamma \gamma$ with both real and virtual photons in the initial or final state \cite{DeTollis:1964una},  
for  which De Tollis developed a powerful simplification technique based on dispersion relations. Such calculations required  complex QED manipulations,  and  provided    De Tollis' students   with the extensive  training 
  needed at Frascati to pursue and exploit electron-positron physics.\footnote{Among De Tollis' graduates, one counts Mario Greco, G.P. and Galileo Violini. The latter has recently received the 2024 Joseph A. Burton Forum Award of APS for establishing programs in physics education and research in Latin America, and for promoting international scientific cooperation.}
 
\subsection{Beyond perturbation theory: Touschek and resummation in Frascati}
\label{ssec:resummation}

While ADONE was 
being built on a plot of land opposite  the original site of the Synchrotron, in Via Enrico Fermi  in Frascati, Touschek  began to gather  a group of young researchers, to prepare a  future  theoretical physics group: Paolo Di Vecchia,  Giancarlo Rossi, Francesco Drago,  Etim Gabriel  Etim, G.P.,   Mario Greco, who had been supervised by 
 Benedetto De Tollis for his  1965 thesis on new vector mesons photo production and had then been hired  by Frascati in the accelerator division, and  
 Maria Grazia ({\it Pucci}) De Stefano,  who had graduated with Francesco Calogero with a thesis on the problem of scattering on singular potentials \cite{Calogero:1966zz}.

The importance of radiative processes in charged particle acceleration and in planning  the extraction of physics from experiments  was very clear in Touschek's mind. It had  been demonstrated in obtaining AdA's feasibility, and it started being his major preoccupation while ADONE, after its official approval, was under construction.  The single bremsstrahlung calculation had highlighted two problems, the need to go beyond a first-order QED  correction and the need to adopt strategies for dealing with very small momenta of the emitted photons. It was also clear that as the c.m. energy was pushed beyond existing  higher-order  corrections, two photon processes {would be detectable}.  When Paolo Di Vecchia went to see  Touschek and  ask for a thesis subject, Touschek assigned  him the  calculation of   the emission of two real photons in $\ep\rightarrow \ep$. 
  Touschek was still commuting between Rome and Orsay, and  the 
   task of    supervising the student became the responsibility of Mario Greco  \cite{Greco:1967zz}, Fig~\ref{fig:yogiMarioLia-EtimElspethFrancis}. Calculations involving  two-photon processes as in   Kessler's studies, were also discussed with Ugo Amaldi, Edoardo's son, who had a post-doc position at Istituto Superiore di Sanit\`a \cite{AmaldiU:2017a,AmaldiU:2004aa}. 

\begin{figure}
\centering
\includegraphics[scale=0.335]{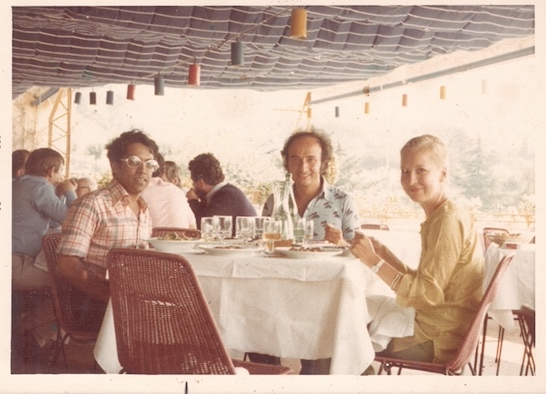}
\includegraphics[scale=0.203]{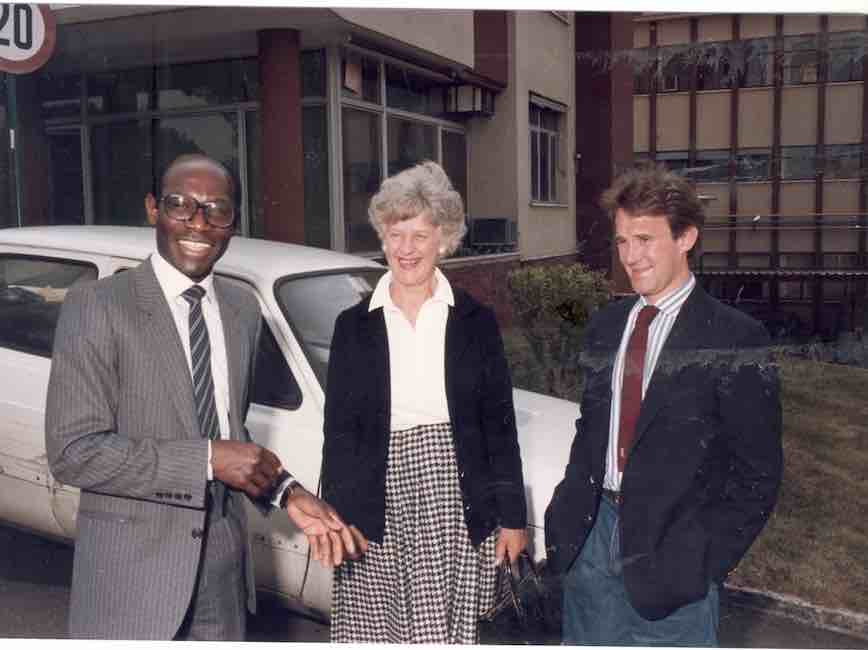}
\caption{At left: Yogendra Srivastava, Mario Greco and G.P. in the  1970s, G.P.'s personal photograph, unknown author, all rights reserved; at right, Etim G. Etim, Elspeth Yonge Touschek and Francis Touschek,  BTML 1987, \copyright LNF-INFN, all rights reserved.} 
\label{fig:yogiMarioLia-EtimElspethFrancis}
\end{figure}

 The year 1965 is probably  when Touschek started to  realistically consider  multiple {soft} photon corrections to ADONE's experiments. The  problem {of soft photon radiation}  had already been extensively discussed  in the literature, starting with Schwinger's 	{Ansatz} in 1949 \cite{Schwinger:1949}, {up to
the calculation of  its four-dimensional energy-momentum distribution  in 1961 \cite{Yennie:1961}.\footnote{A review of how this field of resummation developed after Schwinger's PRD article,  can be found  in \cite{Pancheri:2020abt}.}  {Touschek's interest in radiation problems has a long history, starting from the time he was reading Heitler's book  {\it The Quantum Theory of Radiation} during the war, \cite{Amaldi:1981}, and  working  with Walter Thirring in Glasgow. As ADONE entered the  construction phase, he began   considering  how these calculations could be applied to the realistic case of ADONE's very high energy, but  was also, mostly,   interested in  a general derivation of the theoretical problem.} 
The prompt for Touschek  to definitely go beyond the perturbative treatment 
may have finally  come  from an article by Steven Weinberg  entitled  {\it Infrared photons and gravitons} \cite{Weinberg:1965nx,Weinberg:1965rz}. Soon after the publication of Weinberg's paper, he  
 assigned a thesis  on the ``administration" of radiative corrections to  Etim,  a brilliant student from Nigeria \cite{Etim:1966zz}, who was studying at University of Rome with a fellowship from the government agency  ENI, seen in a later photograph from the BTML 1987, together with Francis and Elspeth Touschek, Fig.~\ref{fig:yogiMarioLia-EtimElspethFrancis}, right panel. 
 
 Bruno   then engaged  Etim and G.P. in  a more extensive general treatment of the problem \cite{Etim:1967}, and  suggested that  Mario Greco and  Giancarlo Rossi  extend his approach to a coherent state formalism \cite{Greco:1967zz}. 
 Giancarlo Rossi   had graduated under  Touschek's supervision  \cite{Rossi:2023}, and  would later write with him the  book on Statistical Mechanics \cite{Touschek:1970aa} and participate in  the  development of Lattice Gauge Theories studies at the University  of Rome, together with  Guido Martinelli. 

These  works opened the way to a novel approach   to infrared  quanta resummation in QED \cite[pp. 37-43]{Bonolis:2011}, including narrow resonance production \cite{Pancheri:1969yx,Greco:1975rm}, and  studies of zero momentum mode   in  gauge theories \cite{Palumbo:1983aa}.
Touschek's treatment had obtained the well known
 energy distribution  of the radiation, 
 $P(\Delta\omega)$,   through an  elegant derivation of the exponentiated form, 
\begin{equation}
P(\Delta \omega)=N(\beta)(\frac{\Delta\omega}{E})^\beta
\end{equation}
but he had  been equally interested in  the spatial momentum distribution, without actually been able to solve the problem in closed form. While not much relevant to electrodynamics, the spatial distribution, in particular the transverse momentum distribution, {was to later become} the road  to unravel some characteristics of hadronic processes.

Through the 1970s, the members of  theory group, together with a regular visitor from the USA, Yogendra Srivastava, Fig.~\ref{fig:yogiMarioLia-EtimElspethFrancis}, started considering how
to extend Touschek's technique  to  transverse momentum distributions in strong interactions. 
  Recasting Touschek's treatment with a large coupling constant,  
the resummation procedure was first applied
 with a constant coupling \cite{PancheriSrivastava:1976tm}, and then, through Greco and Rossi's   coherent state approach  \cite{Greco:1978te,Greco:2019} and  asymptotic freedom effects \cite{Curci:1978kj,Curci:1979sk,Curci:1979bg}.  
 The long life of transverse momentum resummation in QCD includes  a seminal paper such  as  the already mentioned work by Parisi and Petronzio on transverse momentum distributions in the Drell-Yan processes, 
and Greco's collaboration with Pierre Chiappetta \cite{Chiappetta:1981vv}. Resummation in hadronic processes  plays a role in studies of the total hadronic cross-sections \cite{Grau:1999em,Grau:2009qx}, and  remains of central interest to this day.

\subsection{The unexpected multiple hadron production}
\label{ssec:multihadron}
In the summer of 1969, Touschek gathered in Varenna some of the most
 important electron-positron accelerator scientists in the world, on the
 occasion of the 46th International School of Physics "Enrico Fermi",
 dedicated to physics with intersecting storage rings,  Fig.~\ref{fig:varenna1969}.
 \begin{figure}
 \centering
 \includegraphics[scale=0.21]{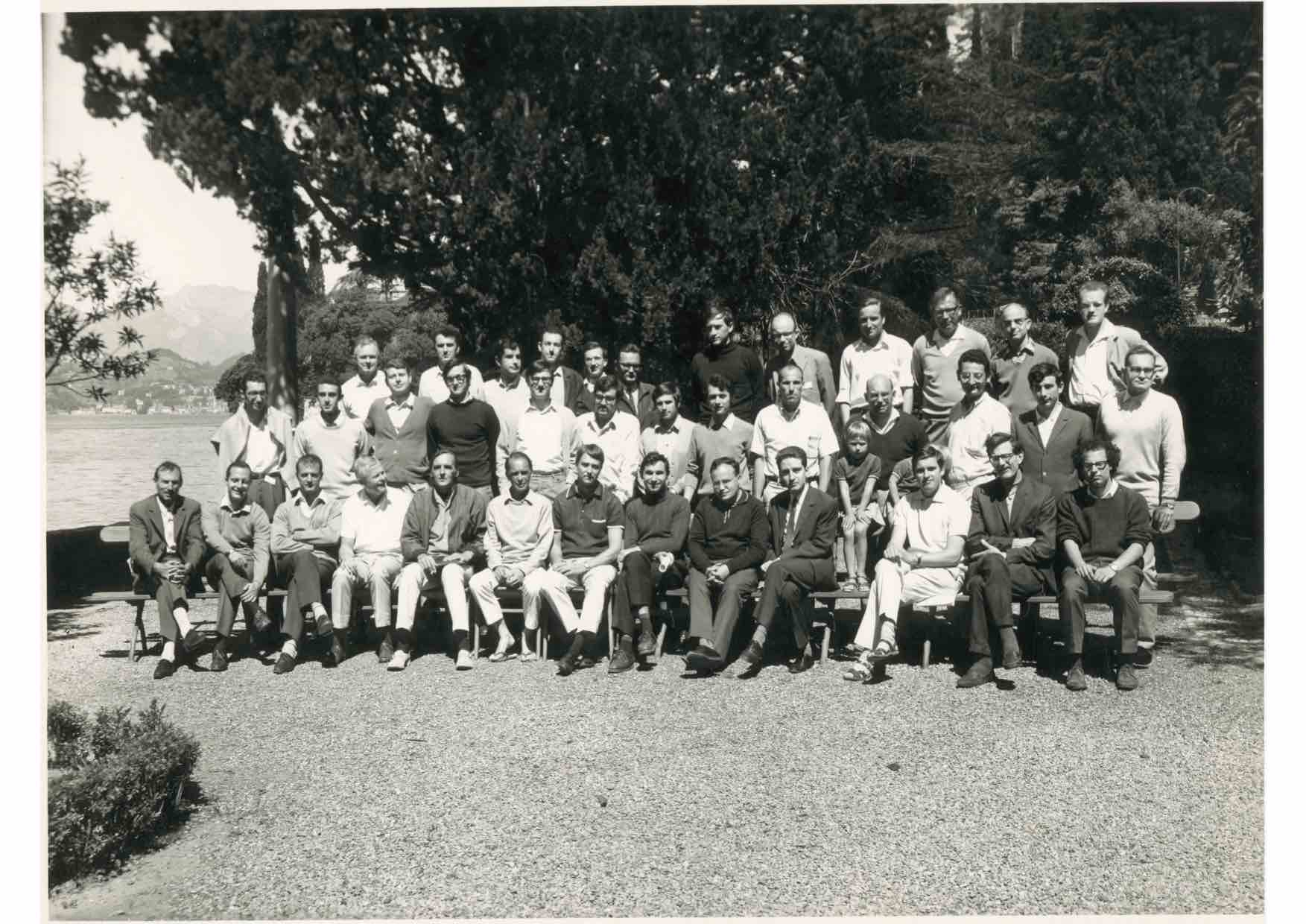}
 \caption{Varenna 1969: Touschek sixth from left in first row, between   Matthew Sands at his right, Gerald O'Neill, Jacques Ha\"issinski and  the Russian theoretical physicist V.N. Baier at his left. 
  Next to Baier, there is Enrico Celeghini, one of  Raoul Gatto's   Florence collaborators, and, in  second row, third from the right  there is   Claudio Pellegrini.  G. Capon is  fourth from left  in the last row, from \cite{Touschek:1971aa},  by kind permission from The Societ\`a Italiana di Fisica.}
 \label{fig:varenna1969}
 \end{figure}
 
 ADONE's first beam had  circulated  on December 8, 1967 and during 1968 the instabilities had been studied and cured. In April/May 1969 the first measurements of the luminosity of the two beams were made and then the straight sections for the experiments were installed {\cite{Amman:1969}}. But while they were closing the vacuum system, the  Laboratories activities were stopped on Friday 30 May, because of the starting of the protest movement, which 
 was part of the general unrest that enveloped   Italy,  from  universities to political life.\footnote{As recalled by Fernando Amman, who underlined the ``substantial unity of the AdA-ADONE enterprise [\dots] achieved through the contributions of Bruno, Carlo Bernardini, Gianfranco Corazza, Ruggero Querzoli'' (and himself), he also remarked that, while Touschek, the initiator and element of continuity, thanks to his scientific and human qualities, was ``decisive in maintaining the connections which were essential in achieving success [\dots] The epilogue of this adventure was not consistent with its beginning for a series of reasons among which the contestation in Frascati was the most apparent, but certainly not the only one [\dots]'' \cite[p. 39]{Amaldi:1981} }
 The machine was reopened in September 1969, with both electron and positron beams and   measurements of four leptons in the final state, $\ep\rightarrow \ep + 
  \mu^+\mu^-/\ep $.
   Correctly interpreted as coming from  two photons emitted from the final $\ep$ pair and converted into a $\mu^+\mu^-$ or another $\ep$-pair, this measurement came in second place with respect to a similar measurement by the Russian group in Novosibirsk  \cite{Balakin:1971hz,Bacci:1971nk,Balakin:1971ip}. 
   
  The bitter taste caused by this second place was coupled to the
fact that the planned precision measurements of the vector meson
properties had already been performed in ACO  \cite{Augustin:1968zz}. Such bitterness was
particularly felt by both Carlo Bernardini, who in 1965 had auspicated
``Vector Boson Hunting with ADONE " \cite{Bernardini:1965}, and Bruno Touschek, who was
also witnessing the disruptions and delays caused by the student movement
in their protest against the university management,     Fig.~\ref{fig:1968}.  
   
   \begin{figure}
\includegraphics[scale=0.18]{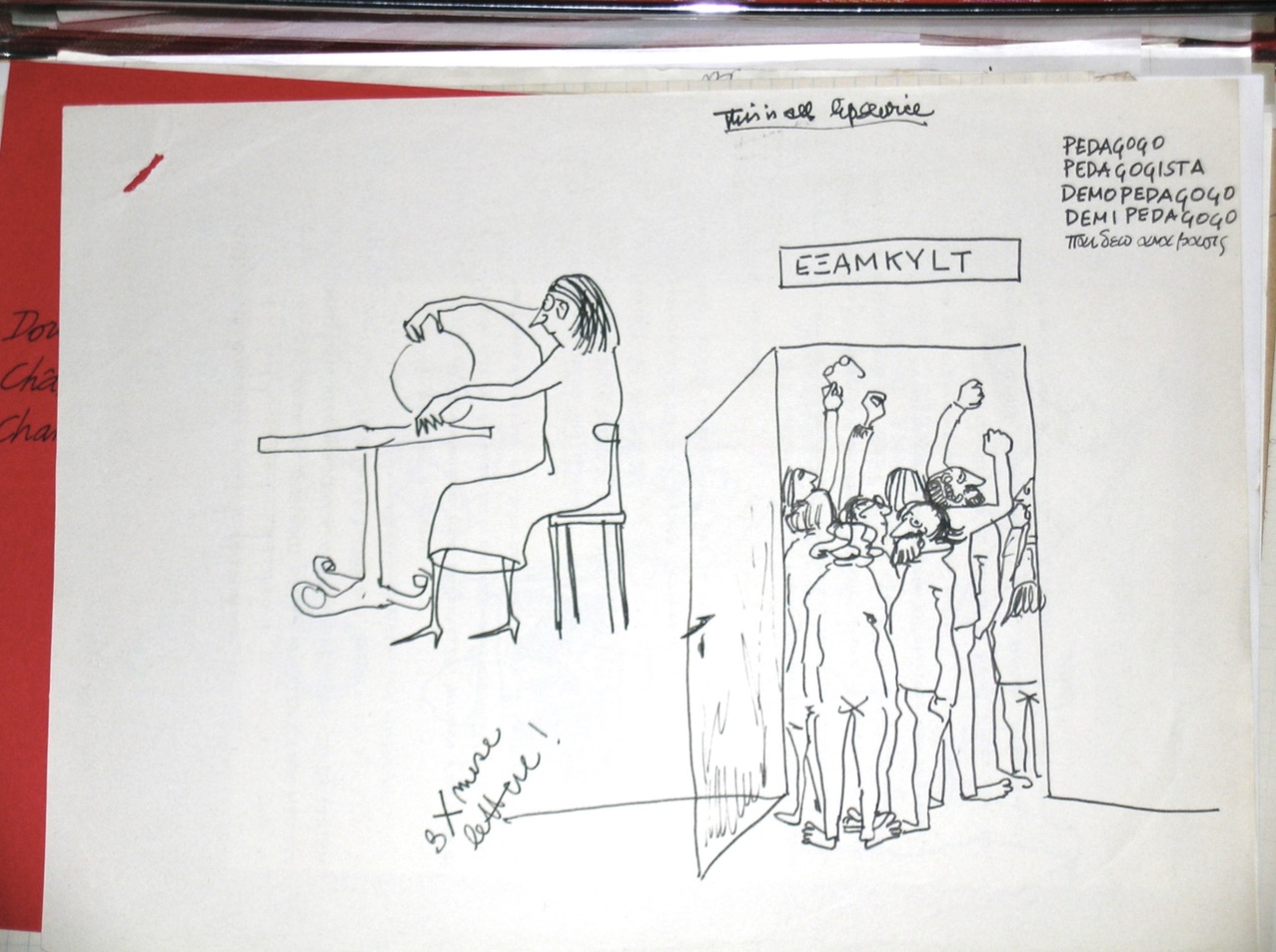}
\includegraphics[scale=0.21]{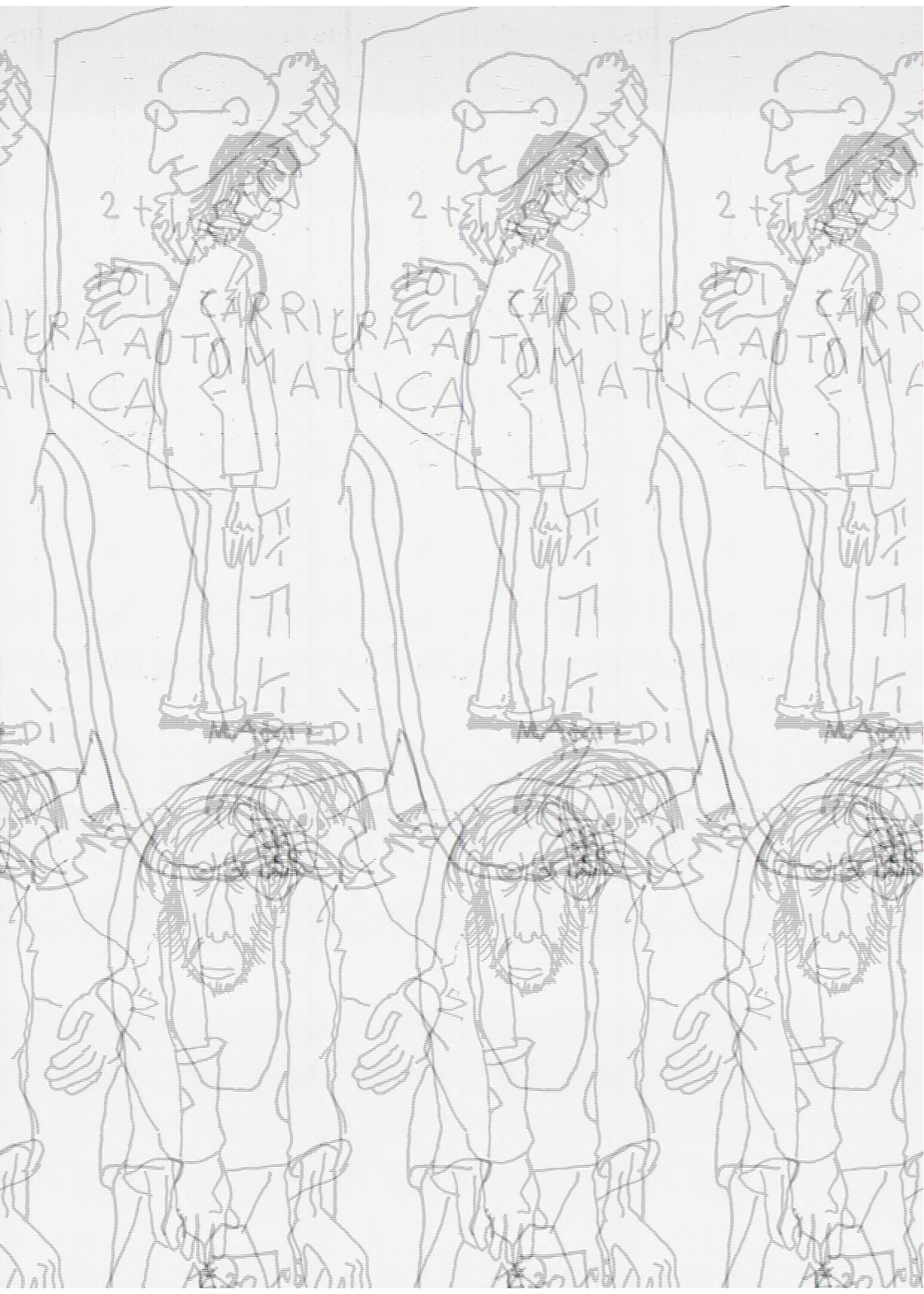}
   \caption{Two  of Bruno Touschek's drawings reflecting his views about the student protests in the Rome Physics Institute in  early 1970s, \copyright \ Touschek family, all rights reserved.} 
   \label{fig:1968}
   \end{figure}
 
{However, ADONE's higher beam energy did bring   new physics to light, as the real surprise came when the search {for new particles} included final states with only 
particles interacting strongly among them, called  ``hadrons" in short, and labelled as  $ h^{\pm,0}$. 
In early 1970, the detectors reported results in the 1.6–2 GeV $e^+e^-$ c.m
energy range, indicating a ``multiple particle production (most likely
hadrons)" in electron-positron interactions, in excess over the well known
QED background \cite[p.629]{Bartoli:1970df}. Both the “boson” and the ``$\mu/\pi$” group
 presented their data at the 15th International Conference on High Energy
Physics held in Kiev in August-September \cite{Bartoli:1970df,BarbielliniEtAl:1970}.\footnote{The
article of the boson group, received at the Nuovo Cimento in October 1970,
reported a total number of 239 events collected at at c.m. energies between
1.6 and 2 GeV.} 

Meanwhile, such preliminary measurements, indicating the possibility of an
abundant hadron production, had immediately attracted the attention of
theorists. On May 30, Cabibbo, Testa and Parisi submitted an article on the
hadron production in $e^+e^-$ collisions, based on private communications about
work in progress  at ADONE  \cite{CabibboParisiTesta:1970}, as Luciano Maiani, back in Italy after his
stay at Harvard, recalls  \cite[p. 635]{Maiani:2017hol}:  ``[\dots] at my arrival in June 1970, I found Nicola working with Giorgio Parisi and Massimo Testa on the preliminary results of Adone. They were very excited and wrote an important paper on the parton model for the  $\ep$ annihilation." 
Using data provided by Frascati groups, these authors studied the total cross-section of electron-positron annihilation into hadrons 
(``The typical high-energy event should consist in the production of a pair of virtual partons, each of which develops into a jet of physical hadrons'') 
and speculated on the possibility that the discussed parton mechanism was operative in the energy range explored by ADONE.\footnote{They extended the field-theoretical treatment of the parton model studied by Bjorken and {Paschos} and successively by Drell, Levy and Tung Mow Yan and were able ``to recover some of the experimentally observed properties of this process''.}  

The results of more than 5000 measurements of multiple particle production carried out at c.m. energies 1.4--2.4 GeV were reported in August 1971 \cite{Bartoli:1971zc}, and the existence of the reaction $e^+e^-\rightarrow h^+h^-$ was discussed by the BCF group in \cite{Alles-Borelli:1972iaf}. Since 1972, the phenomenon of multihadron production was
explicitely mentioned in titles \cite{BacciEtAl:1972}.\footnote{For a general overview of the results obtained by the ADONE groups named 
``$\gamma\gamma$, ``$\mu\pi$'', ``boson'', and ``BCF'' (Bologna-CERN-Frascati), see \cite[footnote 131]{Amaldi:1981}, for details about first and second generation experiments see 
\cite[101-106]{Valente:2007}. }

At first, there were  some inconsistencies  between the reports of the different groups working on the machine. Moreover, the errors were very large. This led to some skepticism about the validity of the surprising results, but they ``had a great impact'' on Burton Richter {in Stanford}. 
He was not skeptical, because ``the results at high energy disagreed much more with theory than they disagreed with each other'' \cite[p. 272]{Richter:1997aa}.\footnote{Since 1966, much time and effort was devoted by Touschek to the Committee for Experimentation with ADONE, an experience that altogether, was considered negative by Bruno ``in the sense that it led to foreseeing those difficulties in the use of the machine that actually occurred later.'' As recalled by Edoardo Amaldi, Touschek became particularly disappointed by the final decisions regarding the actual use of the machine: ``Between the tendency to assign all, or almost all, the available resources to a single group that thus could have disposed of high-performance equipment and the opposite tendency of dividing the same funds between various groups, each by necessity endowed with an apparatus of limited performance, it was certainly not easy to find the right compromise! The solution finally adopted involved an excessive fragmentation of the financial means, with consequences not completely favourable from the scientific stand-point'' \cite[p. 39]{Amaldi:1981}.}
And indeed, confirmation soon came from other colliders. 

The group at the Cambridge Electron Accelerator (CEA),  which had been developed into an electron-positron collider, presented their first results on $e^+e^-$ interactions at higher energies (4 GeV center-of-mass energy) at the XVI International Conference on High Energy Physics (ICHEP) held at the University of Chicago and at the National Accelerator Laboratory in Batavia (soon to be renamed Fermi National Laboratory),  held in September 1972 \cite{Jackson:1972gfg}.\footnote{See list of the contributed papers at the end of the Proceedings (Vol. 2) available at \url{https://digital.library.unt.edu/ark:/67531/metadc875096/m2/1/high_res_d/4202855.pdf}\label{batavia}.} The results from electron-positron rings were mentioned as ``among the most interesting'' presented at the 
 Conference 
 \cite[p. 75]{cern:1973}. And indeed, the ``multihadron production greater than expected" observed in experiments at Adone and CEA, became the title of a review article on the Conference in {\it Physics Today}: ``Although the error bars are large, the multihadron production cross section is, for example, much higher than a simple quark model would predict." \cite{Lubkin:1973}.
 ADONE  had also a key role  for the improvement of second-generation detectors, as Richter understood during a meeting in Frascati in 1968-1969 where he had the opportunity to see the ADONE data and talk with the experimenters: ``We were under a great pressure to reduce the cost of the project. One possibility was to eliminate the magnetic detector in favor of a much less
costly detector with no magnetic field. We were traveling home from am meeting at Frascati in 1968 or 1969, where I had my first opportunity to see
the Adone data and talk in detail with the experimenters. We spent much of the time talking about the results, and I came to the conclusion that the
4$\pi$ magnetic detector was essential to understanding the physics. Thanks to the early Frascati results, the project still had its 4$\pi$ magnetic
detector, which was so essential to the experimental program that led to the 'November Revolution'." \cite[p. 275]{Richter:1997aa}.
 
 Touschek was elated by Adone's unexpected results.  He understood  that his machine was indeed probing and reporting from unexplored regions.  As Vittorio Silvestrini emphasized already at ICHEP in Batavia, ``{\it [\dots] there was a garden where a desert was expected"} [our emphasis] \cite[p. 10]{Silvestrini:1972aa}.

 Meanwhile,    in 1971 Gatto had been called to the University of Rome from Padua. 
In a memory about \href{https://www.sif.it/riviste/sif/sag/ricordo/gatto}{Gatto}, Luciano Maiani remembers:
``It was the time of Bjorken scaling in deep inelastic electron-nucleon scattering, interpreted by Feynman as the scattering of electrons off essentially free, pointlike partons inside the nucleon". 
It was a very exciting period for particle physics, chronicled by Chris Llwellyn-Smith in  \cite{LlewellynSmith:2023avw}  with a personal perspective about those times. 

Back in Rome, Gatto set up a new collaboration   with a group of young investigators recently hired in Frascati, i.e. Aurelio Grillo, Sergio Ferrara and Giorgio Parisi \cite{Ferrara:1972kab}, and also started to look for correlations between deep-inelastic scattering and (what he called) deep-inelastic electron-positron annihilation, together with his former student Giuliano Preparata.
He attended the Batavia conference, together with the ADONE experimental groups, and discussed the above cited paper  as well as further work on the same problem with P. Menotti and I. Vendramin.\footnote{See list in the Proceedings mentioned in note \ref{batavia}.} After a few months, on the basis of the Frascati and the recent CEA results confirming the existence of large annihilation cross sections of $e^+e^-$ into hadrons, Gatto wrote with Preparata a new article entitled ``One-particle and two-particle inclusive deep-inelastic electron-positron annihilation in a massive quark model'' \cite{GattoPreparata1973}.

The excess, now revealed  in a wide range of energies, could be quantified by comparing  the cross-section 
for $\ep \rightarrow hadrons$ with 
the one for a typical point-like QED coupling 
to be measured in the reaction $\ep\rightarrow \mu^+\mu^-$.
The  ratio was  found to be larger than 1,   an occurrence not unlike   the wide-angle scattering results in the Rutherford experiment: namely, it  signalled  internal  structure, i.e.  that the  final-state hadronic particles were not point-like as electrons or muons. But something was not quite consistent  with the current picture of three  quarks constituents of the observed hadronic matter, with three known types, { the {\it up}} with  with charge 2/3 of that of the electron, and two, the {\it down} and the {\it strange},   with charge -1/3.
The ratio would have been 1 if hadrons were point-like,  or $2/3$ if they were composed of   fractionally charged  quarks with {no additional quantum numbers.


The observation of multihadron production 
spurred a large number of theoretical papers, as can be seen in  a review talk given by John Ellis in 1974, shortly before the discovery of the $J/\Psi$ \cite{Ellis:1974zra}. Among the many references and predictions about the ratio of cross-sections  $R=\sigma_{hadrons}/\sigma_{\mu^+\mu^-}$, the summary table  of  values of R also lists a value consistent with   ADONE's data (before the $J/\Psi$ discovery), as $R=2$. 
Such a number  had been mentioned by Gell-Mann in a seminar in Rome on April 4, 1972, as being based on the production of elementary constituents such as the {three}  known quarks, but carrying a new quantum number, called {\it color}, which 
could take three values \cite{LlewellynSmith:2023avw,Greenberg_2009}.This  increased by 3 the probability that the observed excess was due to the production of  {\it colored} quarks, which then morphed into the observed hadrons.  This value was close to what ADONE had observed and is reported in \cite{Bacci:1973fb}
  together with the date of the seminar. Conversi, who was in the audience, remarked that the value observed at ADONE was actually converging to a ratio $R\sim 2$.\footnote{Actually, the ratio $R$ is a function of the center of mass energy, thus, for higher energies, in particular after the $J/\Psi$ peak, $R$ further  increase, because of the additional production of charm quark pairs. See \cite[p. 77]{Maiani:2023} for a brief discussion of  the correlation with the expected further increase of $R$  beyond the threshold of the heavier quarks.}  

 In this seminar, Gell-Mann also stressed the relevance of the transformation
from constituent to current quarks. During the discussion
one of us, F. B., went to the blackboard to show that he had 
found {the generator, Z, of}  that
transformation  at CERN, in March  
1970 \cite{Buccella:1970av},
{by continuing} the research begun in Florence by Gatto, Maiani and Preparata
\cite{PhysRevLett.16.377} and {with an important contribution} by Cabibbo and Henry
Ruegg \cite{CR:1667my}.

Nearly in parallel with the Gatto-Preparata paper \cite{GattoPreparata1973}, the CEA group published a more extensive report on their studies on a  ``total cross section for electron-positron annihilation into three or more hadrons'' with measurement at  a center-of-mass energy of 4 GeV, where they had found a cross section of $4.7\pm 1.1$   times the theoretical cross section for $e^+e^-\rightarrow\mu^+\mu^-$ \cite{LitkeEtAl:1973}.\footnote{In Fig. 2, they report results from ACO, Novosibirsk, CEA, Frascati groups ($\gamma\gamma$, boson, $\mu\pi$), showing $R=\sigma(e^+e^-\rightarrow multihadrons)/\sigma(e^+e^-\rightarrow \mu^+\mu^-)$ versus the square of the center-of-mass energy $s$ in GeV$^2$. They specify that ``The elementary quark model gives $R=\frac{2}{3}$. Quarks with `color' give $R=2$.''} The cross section excess is also clearly shown   in   a 1973 review of ``$ e^+e^-$  colliding beam experiments"   by C. Mencuccini \cite{Mencuccini:1974}.
A final compilation of all  data on multihadron production, including a review of the quoted errors, was later  prepared by Salvini and Conversi, who  presented it  in  Pisa  in 1976 \cite{Conversi:1976io}.
\begin{figure}
\includegraphics[scale=0.07]{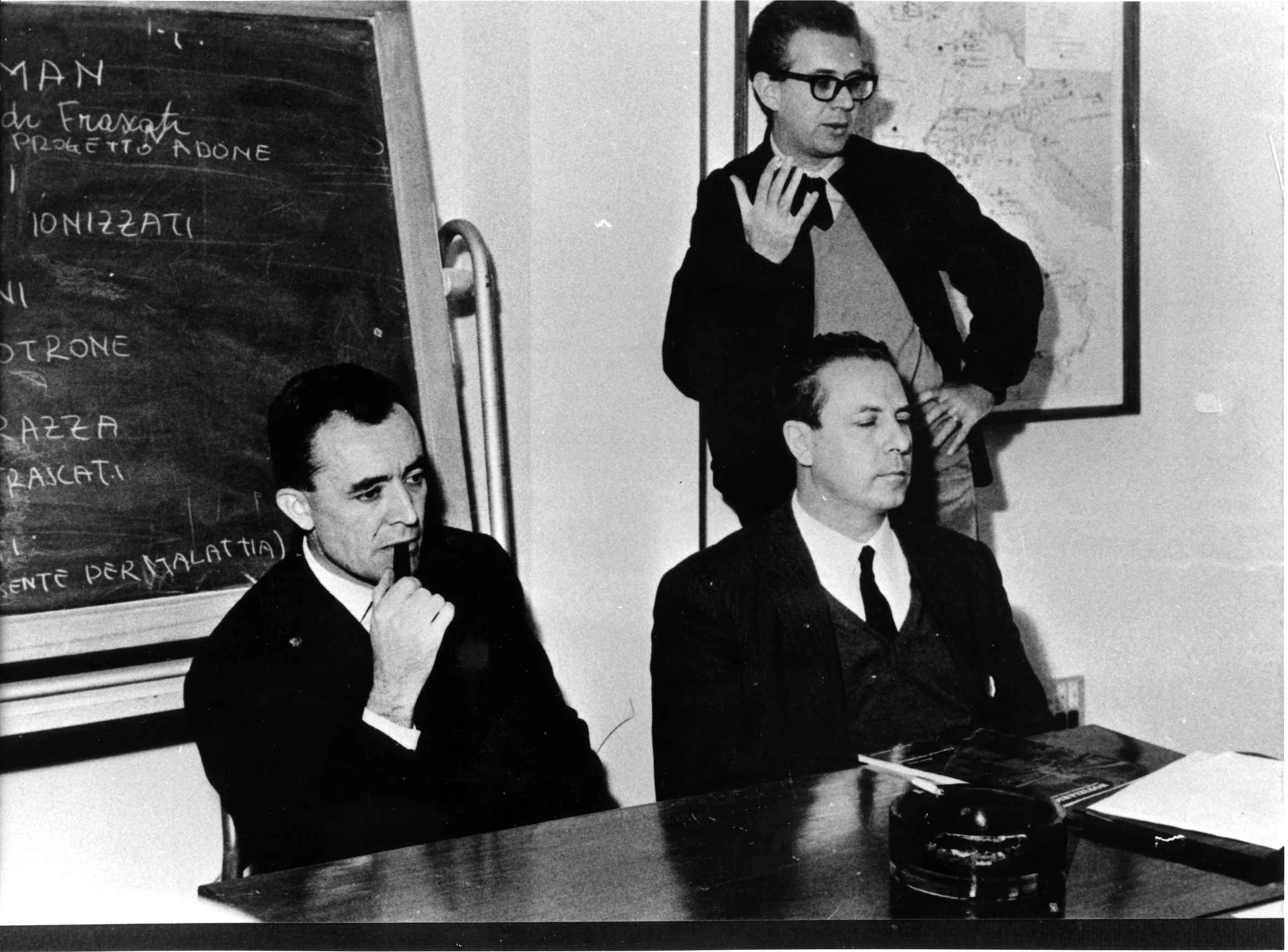}
 \includegraphics[scale=0.23]{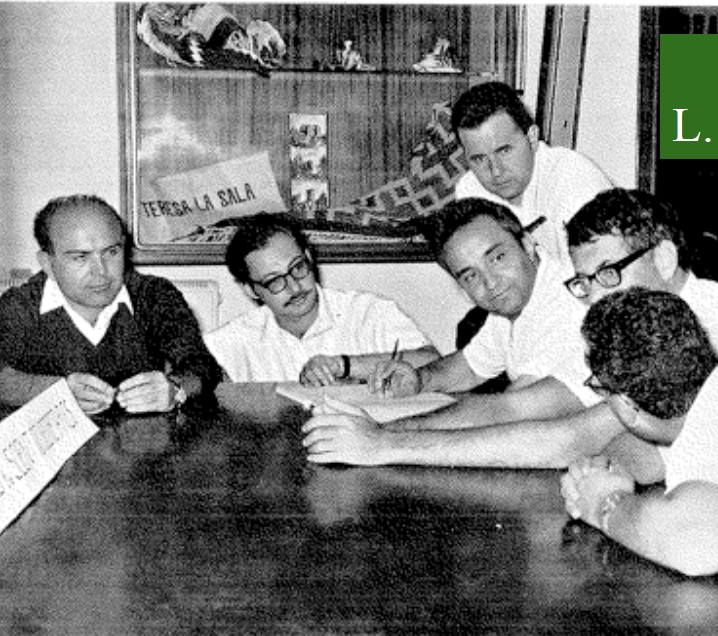}
\caption{Left: Fernando Amman, the director of the ADONE project, plasma physicist Bruno  Brunelli, and Carlo Bernardini, standing, during the construction of ADONE, around  1967, \copyright \ INFN-LNF, all rights reserved. Right panel: seated from left,  Bruno Zumino, Sidney Coleman, Antonino Zichichi, Sheldon Glashow and Murray Gell-Mann, with Nicola Cabibbo standing,  ERICE 1967, Gatto-CERN Courier Volume 7, Number 8, p.149,  August 1967. }  
\label{fig:erice1967}
\end{figure}

The observation of multihadron production is one of the early discoveries which led to the establishment of the Standard Model, which  has its fertile grounds  in a conjoint accelerator, experimental and theoretical physics work, spanning many countries and many decades. The central  decade  is the 1960s, starting with AdA and progressing through the many colliders to follow,  the development of new detectors and intense theoretical physics work,  some of whose  protagonists appear  in Fig.~\ref{fig:erice1967}. 

 When the  1970s  opened with the observation of multihadron production, all the known   particles could {still} be  described in terms of the  three fundamental constituents, the quarks {\it up, down}  and {\it strange}. {The discovered excess was attributed to  a new quantum number,  {\it color}, but} further discoveries were on the horizon, such as  the discovery of the {\it charm} quark, where electron-positron colliders played a central role in the interpretation of results from traditional fixed target accelerators, 
 {giving  a  major contribution to $\ep$ physics and the  establishment of particle colliders} 
 { as the  discovery tool for} the future of particle physics .}
 
 

\subsection{The discovery of charm}
\label{ssec:charm}

The excess in multihadron production revealed by  $e^+e^-$ 
at ADONE and confirmed by other colliders that had started operating at similar or higher energies, was the first striking demonstration of the potential of storage rings for particle research and the fundamental role of the electron-positron interactions as a unique tool for extending the study of hadron structure. Exactly as it was in Touschek’s initial motivations. It signalled that fixed target particle accelerators would soon be surpassed by colliders, and, with them, the detection techniques from which physics had been extracted so far would also change.

{In 1974, a}  revolution in particle physics, to which ADONE had opened the way, was at the door. This was understood and clearly stated by Sheldon Glashow in a talk he gave in Boston at the Conference on Experimental Meson Spectroscopy, held at Northeastern University on April 26-27, 1974 \cite{Wolfe:1974qcy}. At the end of his talk entitled {\it Charm: An invention awaits discovery} \cite{Glashow:1974ru}, he told his audience: ``I bet my hat that in one year from now charm will be discovered. And, if not, then you will eat your hat”. Which is exactly what happened at another conference held in Boston one year later, where Glashow had gummy sweets in the form of hats distributed to the attending colleagues.\footnote{Personal memories and communication by Y.N. Srivastava, then at Northeaster Universty.} 

This is what Glashow  wrote in the  {1974} Conference Proceedings:
\begin{quote}
\begin{tt}
I would bet on charm's existence and discovery but I am not so sure it will be the hadron spectroscopist who will find it. Not unless he puts away his fascination with such bumps, resonances and  Deck effects as have been discussed at length at this meeting.
\end{tt}
\end{quote}
As a matter of fact,  charm was discovered both by hadron spectroscopists and at the same time by the  more direct means of  electron-positron collisions. 

The first hint had come 
from  a fixed target accelerator, in an experiment by Leon Lederman, at Brookhaven, where a bump in the spectrum of $\mu$-pairs produced in the reaction $p+A\rightarrow \mu^+\mu^- + X$ had been observed \cite{Christenson:1970um}. The prediction of the existence of charm had been  made in 1964, but it was only after the 1970 article by Glashow, Iliopoulos and Maiani \cite{Glashow:1970gm}, that it was shown that its existence was theoretically necessary and search for it should start. 



Luciano Maiani was 
one of the ``gattini", the  young researchers that Gatto gathered in Florence between   1963 and 1969.   Maiani  had joined the Florence group in 1964, after graduating  with a thesis in experimental physics  {under the supervision of} Giorgio Cortellessa. After graduation, Maiani joined Mario Ageno's group at the  Istituto di Sanit\`a in Rome, but his real  interests were in theoretical physics and   was given leave to  go to  Florence and  work
with Gatto.  In 1970, as was usual for the best young researchers at the time, he was given the opportunity to spend a year  at Harvard University, where he  soon collaborated with Sheldon Glashow and John Iliopoulos on the  landmark paper on the theoretical reasons for the existence of a new quark, one with a higher mass than the three other quarks,  {\it up}, {\it down} and {\it strange}, 
already known to be the constituents of the so-called hadronic matter. 

In the docu-film {\it Touschek with AdA in Orsay},  Maiani remembers that time at Harvard and 
that at a certain point together with Glashow and Iliopoulos they decided to go and talk to Samuel Ting, down at MIT, towards the river:  ``We went to   see Ting, and told him to look for  it [the charmed quark]". 
 Recently, Giorgio Parisi in \cite{Parisi:2021}  remembers Maiani calling Cabibbo from Harvard in spring 1970, to tell him  about this extraordinary work they had done which would change the vision of particle physics. Maiani himself, in his interview with L.B., {recalled a lunch in Boston, where his wife Pucci  told Shelley ``how happy and excited'' Luciano was about the new results and the work they were doing. To which Glashow replied: ``He is right, this paper is going to be in all the textbooks'' \cite[p. 633]{Maiani:2017hol}.} 

As SPEAR entered its second year of operation, Glashow's call for the discovery of charm was not ignored. The search started. Its conclusion was  the discovery of a very narrow resonance  
now known as  the  $J/\Psi$, jointly announced by Burton Richter and Samuel Ting on November 11, 1974, in  California.
In Frascati the news arrived almost {simultaneously},  first through a call from one of Ting's collaborators, Sau-Lan Wu of  Columbia University, to the \LNF \ director Giorgio Bellettini,\footnote{See also video by Giorgio Bellettini  of a talk on ``Adone: from the Multihadron Production to the Observation of the $J/\Psi$, at  the 2014 BTML in Frascati, about both multihadron production and the $J/\Psi$ discovery, \url{https://www.youtube.com/watch?v=6XDRCeAekaQ&t=1458s}.}  and then a phone call from Mario Greco, {who happened to be} visiting Stanford in the days of the announcement. By pushing ADONE  beyond its design energy, the Frascati physicists were able to confirm the discovery within two days, preparing a paper which was received at the PRL office on November 18, and published in the same December issue of the journal as the two articles by the American groups  \cite{Aubert:1974js,Augustin:1974xw,Bacci:1974za}. Greco was on his way to the University of Mexico  where he prepared an article with his host Dominguez,  explaining the new state as what indeed it was,  a charm-anticharm bound state \cite{DeRujula:1974rkb,Dominguez:1974be}. Back in Frascati, Touschek immediately saw the need  to apply infrared radiative corrections to the production of the new state and urged Greco, G.P. and Yogendra Srivastava to  use  the resummation formalism especially  developed in Frascati under his guidance, for the extraction of the resonance width \cite{Pancheri:1969yx}. 
 While similar radiative correction calculations had also been {performed} at SPEAR \cite{Yennie:1974ga},  Touschek's inspired technique\cite{Greco:1975ke,Greco:1975rm} provided the most {accurate} calculation for quite some time.

Even after his affiliation to Frascati ended, Gatto's 
involvement  with the Frascati theory group continued  through the ADONE years, 
as he himself well describes  in a letter to G. P., that we reproduce in App.~\ref{app-GP}, translated from the original Italian version.\footnote{The letter  can be dated around 2010, in response to inquiries about Gatto and Touschek's role in the development of  the theory group in Frascati. Notice that in addition to what Gatto writes in this letter, there are other papers cited in  {\it inspirehep.net} with  affiliation to Frascati (or CNEN)  and a  number of LNF reports as well.}
 His  interest  in  $e^+ e^-$ experiments and their physics  
continued in the 1970s, when, together with Preparata, he studied the inclusive
production of  protons  and pions in a series of
works described in the review  \cite{Gatto:1974zz},
 and with Riccardo Barbieri, 
  when
he studied  positive parity states with the same quark content
of the $J/\Psi$ \cite{Barbieri:1975nb}.



Gatto's presence in Frascati reappeared thirty years later.
 In 1995, when the scientific personnel of the Frascati National Laboratories moved from one side of Via Enrico Fermi to a new building on the other side of the street, a box full of books showed up in the corridor of the first floor where the theory group had been relocated.\footnote{{The Frascati National Laboratories had started as a joint enterprise between the Comitato Nazionale dell'Energia Nucleare {CNEN}, 
   INFN -- both government agencies -- and a number of Italian  Universities, in particular  Rome. Funding for construction and infrastructures, including experimental equipment, were  responsibility  of CNEN,  which also employed technicians and technical staff, in addition to offer a few fellowships to new graduates, most of whom would later move to university positions. The arrangement was that  the
   academic institutions and INFN 
   would have full and sole responsibility for scientific planning and exploitation.    In time this partnership changed, as the scientific goals of CNEN and INFN evolved and diverged. In 1975 the two  components of the laboratories separated: CNEN kept the old grounds on one side of Via Enrico Fermi, INFN  moved equipments across the street, where ADONE had been built and the accelerator division created. Researchers and other staff split as well, some of them remaining with CNEN, others  becoming INFN employees. }} The books were mostly paperbacks, in English, and of a literary nature. Some of them  bore the name of their owner, Raoul Gatto,  having survived  in some corner of  the library in the old CNEN building, from the time when Gatto had  an office in the Laboratories,     when he  was  head of the theory group, before passing the {\it baton} to his friend and colleague  Bruno.


\section{Epilogue}
\label{sec:epilogue}

Raoul Gatto passed away in Geneva on  30 settembre 2017, 87 years old,  having had the opportunity to see the birth of the Large Hadron Collider and the discovery of the Higgs boson. His life spanned the  years of the development of the Standard Model and the experimental discoveries which confirmed it. He was blessed by 
 the lasting affection of students and his many  collaborators, becoming a member of many honorific institutions, including the national academy in Italy, the Accademia dei Lincei,  Fig.~\ref{fig:Lincei}. His influence on the course of particle physics was incisive and long lasting thanks to his work  as editor of the particle physics section of Physics Letters. 

\begin{figure}
\centering
\includegraphics[scale=0.26]{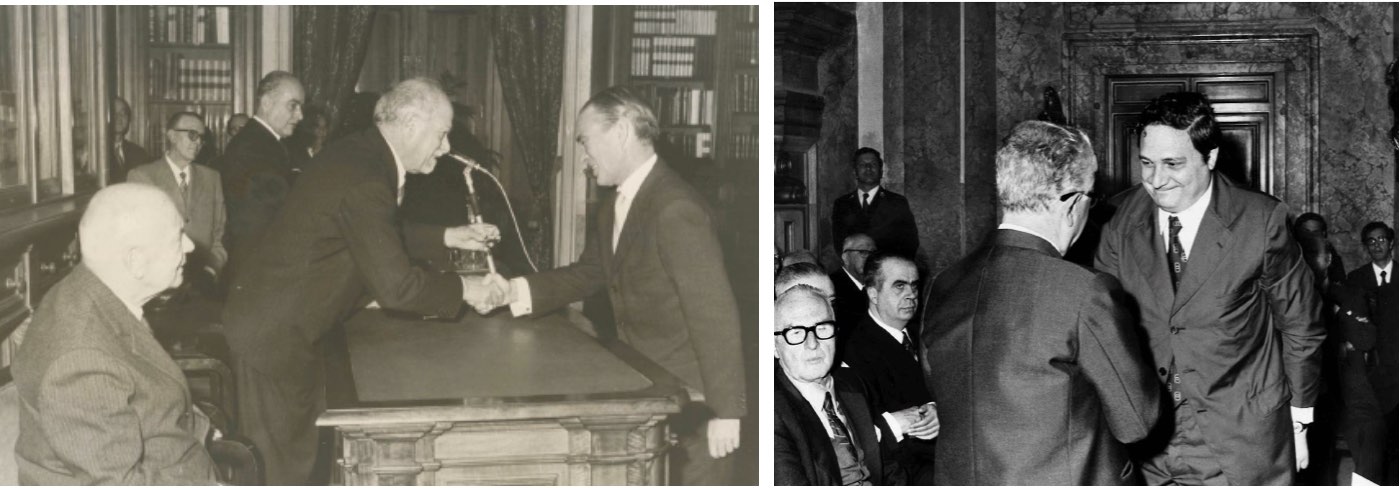}
\caption{Bruno Touschek (left) and Raoul Gatto (right ) being nominated  members of the Accademia dei Lincei,  Courtesy of  Touschek's  and Gatto's  family, all right reserved. }
\label{fig:Lincei}
\end{figure}

Unlike  his friend Raoul,  \BT\ died at a young age, on May 25th, 1978, after a series of hepatic comas. He had contributed to the early development of the Standard Model of particle physics, through his vision  of electron-positron colliders as the discovery tool of the future, but hardly saw its establishment and the experimental confirmations, most of which  came after his death. His  ground-breaking  work is of course in accelerator  physics, but his just as pioneering theoretical physics contributions, such as  the   chiral transformation \cite{Touschek:1957ab}, the work with Cini \cite{CiniTouschek:1958} and, later, in QED resummation,  are much less widely aknowledged.
\footnote{In or around 2009, G.P. gave  a seminar on Touschek' s life at University of Bern, attended by  a professor from the Mathematics Department who mentioned   currently   working  on the Cini-Touschek transformation, a theoretical physics approach, which has had a seminal following \cite{CiniTouschek:1958}. This work is also cited in  the book {\it Relativistic Wave Mechanics}, that is a collection of lectures by E. Corinaldesi, Edited by F. Strocchi, Dover Publications Inc. Mineola New York (2015), page 167, were it is said: ``Cini and Touschek have found a representation which is suitable for the treatment of the high-energy limit of the Dirac equation, in the same way as the Foldy-Wouthuysen transformation is for the non-relativistic limit", courtesy of Y.N. Srivastava.} 
In Italy his contributions to physics were confirmed by his becoming a foreign member of the Accademia dei Lincei in 1972,  Fig.~\ref{fig:Lincei}, but he could not become a university professor until he was almost at the end of his life because he did not want to renounce his Austrian citizenship  \cite{Amaldi:1981}.\footnote{Touschek  never relinquished his Austrian citizenship. Due to the  1928 laws about Italian universities concerning foreign nationals, he could not hold a  Professor Chair, until 1969, when the law changed and he became a {\it Professore aggregato}. A subsequent change in the law in 1973  allowed him to become {\it Professore Straordinario}, a position which would have allowed him to become {\it Professore Ordinario}, the top position in the university career -- with more prestige and better pay --  in three years' time. As Amaldi writes, this final step  took  place only in 1978,{the year of Touschek's death}, thanks to his close friends and  colleagues, who collated and presented all the needed paperwork, since he refused  to do it, considering it an ``unbearable obligation".}

Gatto and Touschek's friendship  had started in Rome, where they both arrived in  1952. Gatto, who  had graduated the year before,  entered the world of theoretical physics when he met Bruno Touschek, and Wolfgang Pauli through him. Joining the Physics Institute at a time of exciting experimental work and institutional building of facilities,  he  completed   his formation path  by travelling  abroad, acquiring  the experience and vision needed for the new science of particle physics. On the contrary, Bruno's road had been tragic and complex,  acquiring his knowledge of  both theoretical and experimental physics crossing     Europe in space and time,  from Austria, before and during the war, to Germany, during and after the war,   to the UK, where he completed his formal education, having  gathered  experience and learning from the  great pre-war masters of theoretical physics. 

Touschek and Gatto's coming together 
allowed the birth of electron-positron  physics in Italy, between Rome and Frascati in early 1960s. Almost at the same time, the study for a proton-proton collider began at CERN. 
While Touschek and his colleagues were planning AdA's move  from Frascati to Orsay, a working group gathered  at CERN  to plan the construction of the Intersecting Storage Rings accelerator (ISR),\footnote{In February 1962,   Touschek was invited to attend a meeting of  the Study Group on New Accelerators led by Kejll  Johnsen, but it is not known if he actually went to CERN at that time.
 The invitation was very pressing, as we can see from Touschek’s answer:
"Dear Dr. Johnsen, your threat of ringing me has been transmitted by Amman and I am looking forward to its execution.
I shall certainly come for 3 or 4 days to participate at the enthusiast’s meeting…
With many greetings and looking forward to your call…" \cite[345-347]{Pancheri:2022}. } 
which 
played a fundamental role in the progress of particle physics, with the discovery of the rise of the total proton proton cross section in 1972 \cite{Amaldi:1972uw,Amaldi:1973yv}. Confirming  a recent similar observation from   cosmic rays events \cite{Yodh:1972fv}, the rise of the total cross-section  was  signalling  the appearance of interactions between fundamental constituents of matter, not unlike the contemporary   appearance of  multihadron production in ADONE.
 The complexity  of strong interactions prevented a clear interpretation of the ISR results, 
contrary to ADONE, where 
  a well defined  initial quantum state  was  producing a virtual photon directly probing  the hadronic world and  making possible the   quark  interpretation.  Soon confirmed by  experiments at other electron-positron colliders, multihadron production was one of
the motivations to propose quantum chromodynamics as the field theory
of strong interactions. 

{After the discovery of the $J/\Psi$, new discoveries soon took place. 
New colliders were planned.  At the Deutsches Elektronen-Synchrotron (DESY), in Germany, after DORIS (The Double-Ring Storage Facility for electrons and positrons)  gave further evidence of the nature of the $J/\Psi$ \cite{DASP:1974rzt},
  the new collider, PETRA, Positron-Electron Tandem Ring Accelerator,  gave evidence for the existence of the gluon, the carrier of  strong interactions.
The W-and neutral weak bosons were discovered at a $p \bar{p}$ machine \cite{UA1:1983crd,UA2:1983tsx,UA1:1983mne,UA2:1983mlz},
while precision measurement on the weak bosons at the CERN  Large Electron Positron collider (LEP)  allowed
to show that the $Q^2$ dependence of the strong coupling constant
is in agreement with QCD and that the number of light neutrinos is
$3$, in agreement with the primordial production of the light elements, for a review of LEP result} see \cite[pp. 83-112]{AmaldiU:2023}.
The top quark was discovered at the US proton-antiproton collider at FermiLab \cite{CDF:1995wbb,D0:1995jca},  while electron-positron colliders such as BELLE from KEK in Japan, and BABAR at Stanford completed the knowledge of the Cabibbo-Kobayashi-Maskawa  matrix by measuring  CP violation in the golden channel of the neutral
$B-$meson  decay into $J/\psi \ K^S$. In the early  '90s, a new type of collider, called HERA (High Energy Ring Accelerator),
 entered into  operation at DESY,   bringing 
 important studies of QCD,   by smashing   27  GeV electrons against 920 GeV protons. HERA's  studies     
 prepared the way to the LHC discoveries,  such    the Higgs boson,  information about properties of   quark-gluon plasma,  finding of  further 
  evidence for  CP violation.
Finally, the LHC experiments extended the validity of the Standard Model up to scales of several TeVs, while precision experiments from new electron-positron collider experiments in the US and China 
led to the  discovery of new types of quark bound states, tetra and pentaquarks.

\section{Conclusion}
\label{sec:conclusion}
\begin{figure}
\centering
 \includegraphics[scale=0.355]{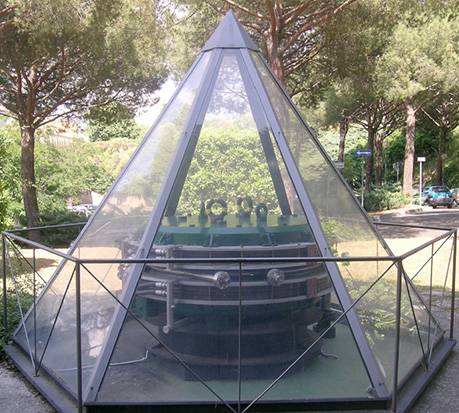}
 \includegraphics[scale=0.11]{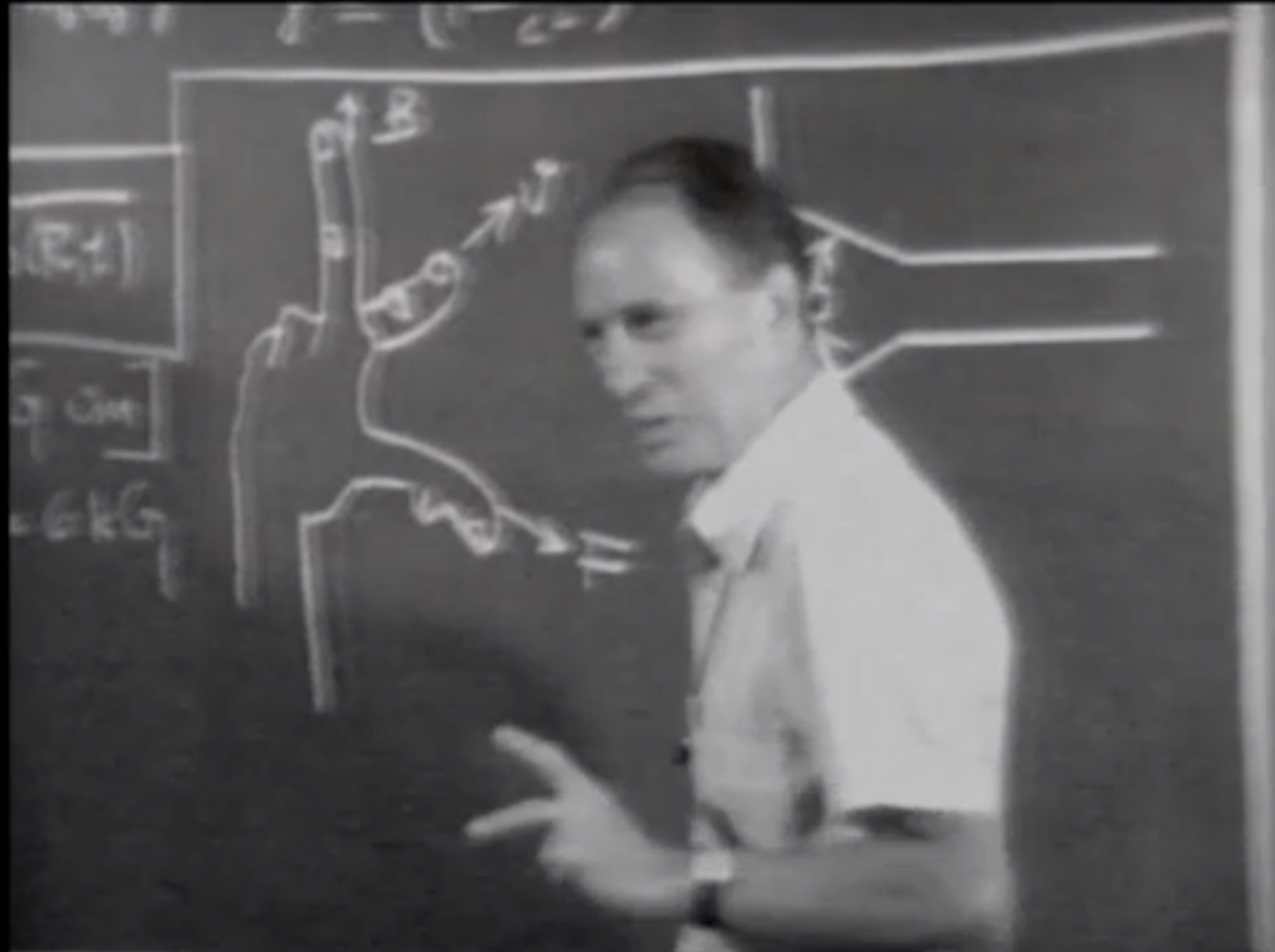}
\caption{Left: AdA  on the grounds of \LNF,  2015 photo by A. Srivastava, courtesy INFN-LNF, all rights reserved; right:  Bruno Touschek during a 1975 lecture at Accademia dei Lincei, in a  video-registration by Francis Touschek, \copyright \  Accademia dei Lincei, all rights reserved.}
\label{fig:BTvideo-AdA}
\end{figure}

The parallel but also converging accomplishments of Raoul Gatto and Bruno Touschek
in the development  of {particle} physics in Italy in the 1960s  have  been 
described.

{We have started by   recounting} the little-known story of how Gatto, in collaboration with Bruno Touschek, contributed to the theoretical grounds for the construction of the first electron-positron collider, AdA, Fig.~\ref{fig:BTvideo-AdA},   and to the ADONE  proposal.
We have  seen how these two exceptional scientists together shaped physics between Rome and Frascati from 1960 to 1964, and, in Touschek's case until 1969, when he was finally nominated University Professor and devoted himself entirely  to teaching and science communication to high school students, their teachers, and the academic  public at large \cite{Amaldi:1981}, Fig.~\ref{fig:BTvideo-AdA}.

Gatto and Touschek also shaped the further development of 
 particle physics through their students, who    were  part of an exceptional roster of pupils  and collaborators whose work  contributed to the renaissance of Italian theoretical physics after the Second World War, and to the establishment of the Standard Model of particle physics.  
 Outstanding among  them  in  shaping theoretical physics in Rome and elsewhere, there is  Nicola Cabibbo, Gatto's collaborator and one of Touschek's first students.\footnote{The other two  students who graduated with Touschek in 1958 were Francesco  Calogero and Paolo Guidoni.} 
 Cabibbo was  then Giorgio Parisi's advisor. From 1971 to 1981, Parisi   was  a researcher in Frascati, and  came to know of  Touschek's  work   on resummation. Nominated  professor in 1981, Parisi   was called to the chair of theoretical physics in Rome.

In Nicola's subsequent scientific and public life,  we can see the profound interconnection between Gatto and Touschek: after  Cabibbo became INFN president in 1983, he was instrumental in envisioning  a new electron-positron accelerator  project in Frascati, 
meant to study CP violation effects, the $\phi$-factory DAFNE, whose   construction was  approved
and fully funded by the INFN Board of Directors in June
1990 \cite{Valente:2007,Bonolis:2021}.

In the present note, the extent of Gatto and Touschek's contribution to particle  physics has only considered in detail the period during which AdA was built  
and  the $J/\Psi$ was discovered. ADONE, which confirmed the US discovery within a few days,  
 had been conceived and proposed by Fernando Amman, Carlo Bernardini, Raoul Gatto,  Giorgio Ghigo and Bruno Touschek \cite{Amman:1961}. These are  the years when  electron-positron physics started between Rome and Frascati, and the road was open for future colliders to became the major   discovery tool in high energy physics .  

\backmatter

\bmhead{Acknowledgements}
  We thank  Francis Touschek for allowing to reproduce some family documents, which had been photographed courtesy of  the late Mrs. Elspeth Yonge Touschek in 2009-2010. Among thes documents, we acknowledge a {\it cache} of unpublished drawings and letters written  by Bruno Touschek to his father, from 1939 until 1960, and then from 1969 until 1971. We thank Emilia Campochiaro from the Preparata Association for permission to reproduce material from \cite{Preparata:2020}, 
   Raoul  Gatto and Guido Altarelli's families  for permission to  reproduce their photographs. 
   We thank Gianni Battimelli, Mario Greco,     Giovanni Paoloni, Yogendra Srivastava and Galileo Violini  for exchanges and useful discussions. For  assistance in bibliographic searches we  thank the  library staff  from  \LNF, {the archivist Antonella Cotugno from the Library of the Physics Department at Sapienza University of Rome,} and   Mimma  Zaccheo  from the Physics library at University of Bari. 
  We  gratefully acknowledge   permission  from the \LNF \ to reproduce documents and photographs.

\begin{appendices}
\section{What Gatto said of Touschek in 1987}
\label{app-BTML}
\centerline{
Memories of Bruno Touschek by}

\begin{center}
Raoul Gatto\\
Department de Physique Th\'eorique, Universit\'e de Gen\`eve
\end{center}

\vskip 0.5cm

When Mario Greco called some month[s] ago, he asked for a general talk on the 
present status of electroweak theory. I accepted with pleasure and I felt honored to be 
able to present such a talk within this commemoration of Bruno Touschek, one of the 
most intelligent physicist I have ever known, and a dear friend. Later on, Greco 
informed me that the program had to be modified and that he rather expected a talk 
within the present open session. It is much harder for me to talk on things that go 
beyond present day physics, essentially because of the limitations of my personality. 
But, I consider a compelling duty to dedicate my thoughts to Bruno Touschek and to 
some of the physics to which he contributed. We are here to commemorate Bruno, who 
was a friend of most of us, a most original and profound physicist, who disappeared so 
prematurely, leaving all of us in great sadness. I think that for what he did, he deserved 
much more than the difficult times and the circumstances his life offered him {\cite{Amaldi:1981}}. 
Particularly to me, the memory of Bruno is so dear, as he was, together with Ferretti 
and Amaldi, one of my first teachers in physics. I learned a lot from him, discussing 
entire afternoons at my early times in physics during the years 1953-1956.

I never had unfortunately the chance of directly collaborating with Bruno. I must 
say that the only paper where our two names appear jointly was the internal Frascati 
report {\cite{Amman:1961}} containing the Adone proposal, written together with Giorgio Ghigo, Fernando 
Amman, and Carlo Bernardini. But I had only been asked to join a few pages on the 
theory to this proposal, which was essentially the work of Bruno and his collaborators. 
The reason for this lack of direct collaboration were mostly logistic. I was too inexpert 
in the period in Rome before I went to the United States, and, afterwards, I had to travel 
so frequently between Universities\footnote{In 1959 Gatto won the national competition for a Chair in Theoretical Physics, together with Bruno Zumino and Sergio Fubini.  His first assignement was at  the University of Cagliari, which he joined  in 1960. 
 After moving to Florence,  Padova and Rome, his  final academic destination was    University of Geneva.} that I could'nt enjoy that constant precious contact 
with Bruno that I had had before. So my most intense memories go mainly back to the 
years from 1953 to 1956 when I met Bruno almost every day and I talked with him and 
learned so much from him. Looking back at the work of that time, I see how often I felt 
I had to acknowledge his generous help and encouragement. Most important was his 
friendliness and his consideration. At that time, especially at the beginning, I felt rather 
lost and unsecure, in a career which seemed to be very competitive and where some 
people occasionally exhibited an intense pride of hierarchies. Bruno, on the opposite, 
was friendly, cordial, encouraging. I remember, I was 22, at a Conference in Cagliari. 
He was sitting at a caf\`e with Pauli,\footnote{Wolfgang Pauli had been awarded the 1945 Nobel Prize in Physics   ``for the discovery of the Exclusion Principle, also called the Pauli Principle".} who partecipated in the meeting,  as shown in Fig.~\ref{fig:Pauli-Touschek-1953-Cagliari-lowres096}, and I was passing { [by]} the side walk, rather trying to get unnoticed. He called me and wanted me to sit 
down with him and Pauli and partecipate in the discussion. Similar things happened 
many times. When a foreign visitor arrived, we often went with the visitor to Albano or 
Nemi, two small towns here in the neighbourhoods, for a walk and a glass of wine. He 
had bought at that time a strange sports car, I think it was a Triumph, an extremely 
uncomfortable convertible. He used to drive in full winter with the windshield lowered 
so that all the air would blow directly into our faces. Before returning to Rome, in the 
not very dense but totally disordered traffic of the roman fifties, he would not separate 
from the colleagues before pronouncing the historical sentence that the fighters in the 
Coliseum would tell Cesar in the old Rome: ``morituri te salutant", in his wonderful 
precise Latin. He was referring to the uncertain conditions of his car. I think we lived in 
that period a rather adventurous life, but the friendliness and generosity of Bruno were 
an uncomparable  and unforgettable compensation.

I have been instructed to try to give a view on what were the theoretical problems 
of the late fifties, which related to the yet inexisting electron-positron physics. As 
always happens when one tries to compare with older times, one cannot avoid to remark 
how different it was from nowadays, how much more limited were our problems and 
purposes. Of course, it would not be correct, historically, to judge on such a perspective. 
At the same time, comparing with all that was later done, illustrates, I think, the courage 
and vision of Bruno, with his unique combination of competences in so many different 
fields of physics.

As the older people in this audience will remember, one of the dominant problems 
of theory in the late fifties, was that of the nucleon's electromagnetic from factors. 
Measurements had been done at Stanford, a laboratory which was at that time, and still 
is, at the advancing frontiers of physics. Already since 1954, Hofstadter and 
McAllister \cite{Hofstadter:1955ae,Mcallister:1956ng} 
 first observed structure effects in the proton, and in the subsequent years 
an impressive amount of data was collected. In a very short note of remarkable 
originality, in 1957, Nambu \cite{Nambu:1957wzj} 
pointed out two main features: (i) the relevance of using 
a spectral representation, and, (ii), the possible role of mesonic resonances. Specifically 
he drew attention on the role of a possible isoscalar resonance of the type later called {$\omega$} 
(but he called it {$\rho$}). 
The isoscalar property would guarantee same sign for proton and 
neutron. On the other hand, what Nambu called the pion cloud, the isovector part, 
would change sign. The electric form factor would thus add in the proton and 
approximately cancel in the neutron. The dispersion theory approach for the nucleon 
form factors was soon later developed by Chew, Karplus, Gasiorowicz, and 
Zachariasen  {\cite{Chew:1958zjr}}, 
and by Federbush, Goldberger, and Treiman { \cite{Federbush:1958zz}}. 
Basic to the dispersion 
analysis is the knowledge of the absorptive contribution, like in optics for the Kramers-
Kroning relations. For the nucleon form factors the absorptive part starts with 
contributions which correspond to a virtual time-like photon going into two pions for 
the isovector part, and into three pions for the isoscalar part. Having an electron-
positron machine would have rapidly settled most of the problems. Nobody  however 
dared to start such a project. When $e^+e^-$  machine became operational, and it was 
essentially the merit of Bruno and of a few other courageous physicists, part of this 
particular history had already been unveiled. Frazer and Fulco {\cite{Frazer:1959gy}} 
had already proposed a 
resonant isovector pion-pion interaction. The experimental evidence came from pion-
proton inelastic collisions, preliminarly by Derado {\cite{Derado:1960}} 
and through extrapolation method 
by Anderson at al.  {\cite{Anderson:1961zz}},
by Erwin et al. {\cite{Erwin:1961ny}},
by Stonehill at al. {\cite{Stonehill:1961zz}}.
 As for the isoscalar 
resonance, that Nambu had conjectured, it was Maglic, Alvarez, Rosenfeld,  and 
Stevenson {\cite{Maglich:1961rtx}},
who discovered it in proton-antiproton. But the precision work still came 
from the electron-positron machines.

This is one particular aspect of the theoretical situation and problematics of that 
time. Another aspect had to do with the efforts to test the validity of quantum 
electrodynamics. Again at Stanford, especially Sidney Drell {\cite{Drell:1958gv}}
had pushed in this 
direction. In Europe, we had the successful g-2 experiment {\cite{Charpak:1961mz}}.

Electron-electron collisions would allow to test the photon propagator. I remember 
a conference by Professor Panofsky, at the end of 1959, reporting on the pioneering 
work of Barber, Gittelman, O'Neil, Panofsky, and Richter \cite{Barber:1959vg}, on electron rings. 
Answering to a question, Panofsky mentioned that, to test the electron (rather than the 
photon) propagator, electron-positron collisions would have been suitable, through 
observation of 2-photon annihilation, but that such a development could present 
additional technical difficulties and that for  the moment had been postponed. This also, I 
think, shows that a strong courage and optimism was required to embark in the 
direction of $e^+ e^- $collisions and, beyond any doubt, without the vision, the optimism, 
the courage of Touschek, $ e^+e^-$ physics would, at least, have suffered a delay.

The Frascati laboratory produced at that time first class physics, in a quiet and 
almost unperceptible way. The Frascati atmosphere was a typical country atmosphere. 
From the windows of our offices one could admire a large extension of vineyards and 
sometimes  hear  people singing what in America would be called country songs. It was 
a relaxed and perhaps provincial atmosphere. But it gave all of us the possibility of 
working hard and of imagining the future not only the immediate future but also what 
was, for that time, the far-away future. To imagine, for instance, the $e^+ e^-$ production 
of neutral weak vector bosons, coupled to neutral currents, or the $ e^+ e^-$ production of 
pairs of weak charged vector bosons, and the weak asymmetries which are now being 
measured. That relaxed Frascati atmosphere may have been  provincial, but certainly it 
gave all of us a feeling of doing something together, and that  something was 
worthwhile. All this we owe to Bruno, to his scientific and human qualities. The 
contribution of Touschek's direct collaborators, Giorgio Ghigo, Carlo Bernardini, 
Gianfranco Corazza, who were the initial collaborators for AdA {\cite{Bernardini:1960osh}},
of  Querzoli, Sacerdoti, Puglisi, Massarotti, Bizzarri, Di Giugno, of Marin and Lacoste at Orsay at 
those early times, was undoubtedly of the highest quality \cite{Bernardini:1962zza}.
Fernando Amman took  {[on]}
the responsibility  directing of the Adone project  \cite{Amman:1962}. As far as theory is concerned, let 
me mention the contribution of Nicola Cabibbo and the contribution of Francesco 
Calogero. Much physics was done with Adone. Much more, we all know, could and 
should have been done, were it not situations and circumstances which were essentially 
external to us physicists.

I shall not go back to those results, to which so many Italian physicist 
contributed  {[\dots]}
Although mainly concentrated on proton machines, CERN was not 
insensitive to progress on electron-electron and electron-positron physics. Already in 
June 1961, a conference on very high energy phenomena was organized at CERN and it 
was remarkable that all the three invited talks on electromagnetic interactions were on 
electron-electron and electron-positron colliding beams. One of the three talks was given 
by Bruno, who gave an exact presentation of Ada and of the Adone project. The 
report {\cite{Bell:1961gi}},
 is in the Proceedings, which were edited by John Bell et al. 

We know that Touschek had a deep respect for Pauli. His relations with Pauli 
were steady but they become more intense when Pauli got interested in what were later 
called the Pauli-Pursey transformations, a general class of rigid, that is global as 
opposite to local, transformations \cite{Pursey:1957,Pauli:1957voo}. 
This was towards the end of Pauli's life \cite{Pauli:1959kmw,Touschek:1958aa}.\footnote{Ref.  \cite{Pauli:1959kmw} is a posthumous work, 
 introduced by a note (by Touschek) which reads: ``The contribution by W. Pauli to this report  was not intended for publication. However, it was decided to publish it, in the form the talk was given, as a document of His last activities."  }
 But,
even before, Touschek always found very attractive Pauli's ideas on non-abelian gauge 
theories (Professors Enz and Jost have recently helped me in clarifying this part of 
Pauli's history). Bruno often told me of these, for that time, quite new ideas \cite{Straumann:2000zc}.\footnote{{Gatto refers here to a book by P. Gulmanelli, {\it Su una teoria dello spin isotopico}[On a theory of isotopic spin], edited by Pleion, Milan, Italy, 1957, the result of a series of seminars held at the Institute of Theoretical Physics of the University of Milan. See the English translation in \cite{hartmann2023translation}, containing an exposition of all elements of Pauli's attempt at a non Abelian Kaluza-Klein theory. See also a recent comment to  this book in \cite{Straumann:2000zc}}.}
Touschek and, I must say, also  Ferretti, during so many discussions, always showed a 
special attention to the role of gauge invariance. In a sense I am grateful for this to both 
Touschek and Ferretti, as they transmitted this interest also to their students.

What I learned from Bruno was also a sort of style. He never liked extremely long 
calculations and uninspiring formulae. He put ideas and invention before the hard 
mechanical effort. When he wrote a formula he seemed to carefully draw it, designing, 
more than just writing it down. He never would waste his time in checking hundredth 
of papers in the literature, but he would rather try to go directly to the heart of the 
problem. He first  wanted everything to be simplified and reduced to the essential. His 
loved books, in physics, were few and of classical authors, Sommerfeld, Pauli, Heitler. 
Once, he was going on vacation to the mountains, and he told me he wanted to work on 
beta-decay. The only thing he was taking  with him was a very small note-book, still 
empty. No books, no articles, no preprints. The notebook was extremely tiny. Like any 
good theoretician he always thought that right things have to be simple and not require a 
cumbersome apparatus.

Touschek had a deep classical culture, which certainly allowed him to assimilate 
the Italian culture and to adapt himself so easily to our country and our people. A deep 
side of his personality was however his relation to the Viennese culture. I always found 
remarkable how Austrian  culture, Anglo-saxon culture, and Latin culture could so well 
coexist in him. He had deeply thought and elaborated on the aspects of these apparently 
so disjoined cultures. The Jewish culture was undoubtly also part of his personality. I 
think it became manifest in his particular intelligent, sometimes critical, sense of humor, 
which reminds me of modern Yiddish theater.

In autumn 1977, already seriously ill, Bruno was at CERN. In spite of his evident 
unhealthy conditions he always was willing to discuss. He often developed typical 
particular interests, even outside physics. He liked to speculate on that explosion of 
cultural life that characterized the Vienna of Franz Joseph. For that he proposed a 
socio-economical explanation, which included elements of politics and also of urbanism. 
Unfortunately I have not been able to entirely reconstruct his arguments, which perhaps 
I never could follow completely because of my incompetence.

I had written, in the first version of these notes, additional recollections of my last 
encounters with Bruno at CERN. I think they are not really so relevant here, although 
they will remain vivid in my memory. When I learned of his death in Innsbruck I was 
so shocked that for a few days I could not do any useful physics. All those, among us, 
who had the privilege of knowing Bruno, will never forget him, and we have an 
immense debt of gratitude to his intelligence, generosity, and friendliness.
\vspace{0.3 cm} 

\noindent
{\it Reproduced with permission from  \cite{Greco:2004}, \copyright \ INFN-LNF, all rights reserved.}
\section{Raoul Gatto to Luisa Bonolis, 2004}
\label{app-LB}
In this appendix,  we reproduce  a letter by Gatto, in which he describes the   period he spent in Rome before he left for the United States, and his relation to Touschek.



\vskip 0.5 cm

\begin{flushright}Geneva, January 15, 2004
\end{flushright}
\vskip 0.3 cm

Dear Mrs. Bonolis,

I am sorry for the delay of this reply, due to a family accident ( not serious ).

During the last years of my stay in Rome before I left for the United States (I am talking about the years '54, '55, '56), Touschek was the expert person from whom I {learned the most and to whom I turned} for discussions, doubts, ideas. Much of my work done in Rome in those early  years {bears} explicit thanks to Bruno Touschek who was my main reference for theory. In fact, my teacher had been Ferretti, who had been thinking for some time about returning to Bologna (which he did in '56). Ferretti's  various health problems and overwork did not always allow me to bother him with discussions related to my calculations, which I felt I could do with less scruples with Bruno Touschek instead.  Ferretti had brilliant and important ideas, and I will remember, for example, the role he already at that time saw in gauge theories, something that remained engrained in my education in Rome.

  The discussions I remember as most decisive from the point of view of my work in the following years took place with our group of theorists, {with} Bruno Touschek at the forefront. 

 I must also mention Zumino's visits to Rome. Bruno Zumino  would drop by Rome from time to time in those years. Zumino had also been a student of Ferretti but had continued his career abroad in Germany and the United States. Zumino informed us about the work of Pauli and especially of Gerhart L\"uders on the TCP theorem.  Zumino had discussed with L\"uders, who  had been working on the problem as early as 1952, which he then correctly  formulated  during 1953. These things were close to Touschek's interests and there were extensive discussions.  This shows how one was in a sense  at the center of the most important progress in Rome at that time, because  of Bruno's vision and his contacts abroad. 
 
Touschek was also aware of Pauli's thoughts on various topics such as gauge theories. It must be said that at that time everything was at a very preliminary and pioneering stage.

I think Schwinger was already convinced of the TCP theorem perhaps even before 1953 because he somehow followed it up with the spin-statistics theorem. I don't recall him giving an explicit statement of it, though. These issues, which are fundamental to field theory, were still not well defined in the crucial years of 1953, '54 and '55, i.e., somewhat before the discovery of non-conservation of parity, which then made
apparent their fundamental importance for particle physics. 

The discussions in Rome, which unfortunately did not lead to anything published, but 
{had the sole purpose} of understanding L\"uders' and Pauli's ideas, 
{concerned}  the role of the Lorentz group and the spin-statistical connection for local theories with at most spin 1/2 (Pauli was the one who removed this last {restriction}). For me these discussions were very important because (1) I realized the crucial role of symmetries, and (2) I realized that general questions of field theory could characterize the phenomenology of elementary particles. 

Of course {Bruno Touschek had  {perhaps} been aware of these things}  all along and he had been particularly interested in time reversal publishing various papers with Morpurgo and Radicati at the time. 

I was already in the U.S., or about to leave, when Touschek {became}  interested in the role of parity in the theory of zero-mass spinors. Touschek was one of the rare physicists at that time who was familiar with Majorana's work, and he was clear about the distinction between Dirac and Majorana neutrinos. Touschek kept a little notebook on the neutrino which he showed me from time to time, and the discussions were about how to characterize the various representations of the Dirac algebra in { terms of  space-time invariance properties}. Not only that, but also about the massive or non-massive character of the fermion. {If Bruno's {premature death}  had not robbed us of his intelligence,  there  would  probably have been, by now, his important contributions to the theory of neutrinos and their mixing}. 

As soon as I arrived in the United States, I was naturally imbued with the Roman discussions and especially with the style of problematics {mainly} originated by Bruno's fervent intelligence and curiosity. With what I had learned about TCP and with the knowledge I had of K-phenomenology (and here I must mention my long collaboration with the Rome  emulsion group, in particular with Carlo Castagnoli), I quickly realized that it made sense to  {make}  the hypothesis that CP was a good quantum number and to  {search for its}  consequences (a few years later it turned out that CP {also  was only an approximate symmetry}). On February 4, 1957 I sent  the work you mention to {\it The Physical Review}, where I used the name  {\it L} for  the new quantum number, as a tribute to  L\"uders, whom I had in the meantime met personally. In the same days a paper by Landau was sent to {\it Nuclear  Physics},  with the same proposal and a general paper by Lee,Oehme and Yang ( the one that is usually cited) was sent to Physical Review, in which however the proposal of CP as the conserved quantum number was not made.  This shows the simultaneous interest in these problems  by  {different}  groups at that time.

{Simultaneously, in various places, interest in the still poorly defined  concept of chiral symmetry, which for Touschek emerged from his mastery of neutrino theory.} 
Chiral symmetry is not only the basis of the electroweak theory, but has long played a crucial role in QCD, the theory of  strong interactions, where it is  {spontaneously} broken  in the hadronic normal phase, only to be restored at very high temperatures. Had he lived, Touschek would certainly have contributed a great deal to these theoretical developments in QCD, partly because of his {continuing} interest in thermodynamics. Touschek quickly realized that the  extension of discrete symmetry to continuous symmetry would play a role that was not yet well defined at the time. This extension was the basis for the formulation of electroweak theory in terms of continuous groups, as required for a gauge theory (both in the electroweak case and for QCD).

Bruno should also be credited with preparing excellent graduate students. On my return from the U.S. I met Nicola Cabibbo at the Institute, who had just finished his dissertation  on beta transitions between nuclei under Bruno's supervision. He was planning to continue working on weak interactions.  At that time I was working with Malvin Rudermann on the decay of the pion into neutrino electron, which was a difficulty (fortunately only apparent) for V-A theory. I gave Nicola a problem on Kaons, which he solved brilliantly. {Afterwards} I had the pleasure of collaborating with Nicola for many years. Besides Nicola, Bruno had welcomed other good students. Among them was {[Giovanni]} Gallavotti who came with me to Florence, {[Aurelio]} Grillo, who collaborated with me and Sergio Ferrara for a long time, {G.} Putzolu who interacted with me and Cabibbo at Frascati, and many others with whom I had less close scientific relations. 

With best wishes for your work and cordial greetings, 

Raoul Gatto
\section{Raoul Gatto to Giulia Pancheri, ~ 2010}
\label{app-GP}

Dear Lia,\footnote{The letter  can be dated around 2010, in response to G.P.'s  inquiries about Gatto and Touschek's role in the development of  the theory group in Frascati.}

it is with with great shock that  {I discovered, looking back at old emails,}  that I probably never responded to your email about the history of Frascati. If so, I apologize and will try to 
{help you as much as} I can by answering your questions.

{I graduated} in 1951 and it's a little complicated. As a student of the Scuola Normale, I should have done my thesis in Pisa. But the chair of theoretical physics in Pisa was vacant. From the Scuola Normale I was able to work on my thesis in Rome, except for the final discussion which was to take place in Pisa. My thesis, with Ferretti, was on a non-perturbative model of inelastic diffusion (with a view to application to diffusions on systems with excited states). Marcello Conversi, who was the chair of experimental physics in Pisa, acted as {advisor} of my work with great awareness and competence. After graduation I stayed in Rome, where I was when Bruno Touschek came. Bruno's arrival was a great fortune for the theoretical group in Rome. Ferretti moved to Bologna shortly afterwards and in fact Bruno Touschek was the person with whom I 
{discussed most during} those years and who generously offered me a lot of his time. For this, I cannot help but be grateful to him.

From the end of 1956 on, my scientific base became Berkeley, 
{where I stayed at first for a little more than a year and then returned on a regular basis for shorter periods of time}.

{In 1960 I became an Extraordinary  Professor  in Cagliari. Almost at the same time, Giorgio Salvini, always full of scientific initiatives, and the then director Italo Federico Quercia, offered me} the possibility of creating a theoretical group in Frascati through a consultancy contract. (To answer your question, there was no discussion or information about how much time this would go through once it was enacted, etc.). I cannot say 
{whether} a similar offer had previously been made to Bruno, which would have seemed {perfectly} right and natural to me. Bruno had never told me about it, and it was only now, from your email that I learned of this possibility, which would have seemed completely correct and scientifically excellent to me.

After my move to Florence I continued to deal with Frascati on a regular basis, within the limits of my increased university commitments until 1965. These last years, however, were years of serious economic difficulties for Frascati and for all of Italian physics and the discussions we had in the Senate of the laboratory were largely centered on financial and administrative issues.

The theoretical group of the laboratory, in addition to myself and Nicola Cabibbo, had as members and collaborators (I hope I remember {them all}), Bassetti, De Franceschi, Putzolu, Mosco, Altarelli, Buccella, and regular visitors from the French CNRS and various foreign universities. I don't have precise records but everyone was very active. I particularly remember Nicola's work with De Franceschi  and Da Prato on photons in crystals.

In 1973 I resumed contact with the Frascati group (the collaboration 
with Ferrara - Grillo - Parisi) and worked on the commission for SuperAdone and the preparation of the related project, which unfortunately did not come to fruition. I say ``unfortunately'' because [with 
we could have done the physics of heavy mesons in Frascati.

I'm not very tidy. 
However, among my works containing the Frascati affiliation, I have found some, which I  list below.\footnote{Works with Frascati affiliation noted by Gatto: \cite{Cabibbo:1961zza},\cite{Cabibbo:1961ab},\cite{Gatto:1963a},\cite{Ademollo:1964tta},\cite{Ademollo:1964sr},\cite{Borchi:1965sdr}.  In addition to what Gatto writes, there are other papers cited in  {\it inspirehep.net} with  affiliation to Frascati (or CNEN)  and a  number of LNF reports as well.} There were no electronic databases at that time so it is not easy to find them or to realize the impact of these works: No medium, no message. It's a miracle that they are still sometimes mentioned. {It is even worse, of course, for the articles published} in the Nuovo Cimento, a journal  which cannot be found in most libraries. I think that Nicola.
 who contributed substantially to the first important works, can provide any comments (I am sending him a copy of this e-mail).
 

I'm sorry for the unexpected delay. I remember that I had put your email in a separate file telling myself that I had to give you all my help. I hope to be  useful to you anyway.

With warm greetings, 

Raoul Gatto

\end{appendices}

\vskip 0.5 cm
\bibliography{Touschek_book_resummation_12jan2022-oct2023-LB-nov26GPMay}

  \end{document}